\def\mycolor{black}

 \documentclass[final,1p,times]{amsart}

\usepackage{epsfig}

\usepackage{amssymb}
\usepackage{amsthm}
\usepackage{amsmath}
\usepackage{color}
\usepackage{hyperref}
\usepackage{url}
\usepackage{cite}
\usepackage{nccmath}

\newcommand{\fref}[1]{Fig.~\ref{#1}}  
\newcommand{\eref}[1]{Eq.~\eqref{#1}} 
\newcommand{\sref}[1]{Sect.~\ref{#1}} 
\newcommand{\PBS}[1]{\let\temp=\\#1\let\\=\temp}

\newcommand{\ci}{\mathrm{i}}
\newcommand{\supp}{\operatorname{supp}}
\newcommand{\Sset}{\mathbb{S}}
\newcommand{\id}{\mathrm{d}}
\newcommand{\iD}{\mathrm{D}}
\newcommand{\iP}{\mathrm{P}}
\newcommand{\iS}{\mathrm{S}}
\newcommand{\itr}{{\sf T}}
\newcommand{\ygj}{y}
\newcommand{\zgj}{z}
\newcommand{\ugj}{u}
\newcommand{\vgj}{v}

\newcommand{\yg}{{\bf\ygj}}
\newcommand{\zg}{{\bf\zgj}}
\newcommand{\hzg}{\hat{\zg}}
\newcommand{\ug}{{\bf\ugj}}
\newcommand{\vg}{{\bf\vgj}}

\newcommand{\speci}{w}
\newcommand{\specij}{w}
\newcommand{\speciv}{{\bf\speci}}
\newcommand{\cel}{c}
\newcommand{\vcel}{{\bf\cel}}
\newcommand{\eigl}{\omega}
\newcommand{\jeig}{\alpha}
\newcommand{\keig}{\beta}
\newcommand{\II}{\mathbf{I}}

\newcommand{\Hamil}{{\mathcal \gamma}}
\newcommand{\proj}{\boldsymbol{\Pi}}
\newcommand{\Mode}{R}
\newcommand{\mode}{r}
\newcommand{\Go}{\mathrm{O}}
\newcommand{\po}{\mathrm{o}}
\newcommand{\varXset}{{\mathcal X}}

\newcommand{\vyj}{y}
\newcommand{\vy}{\mathbf{\vyj}}
\newcommand{\xg}{\mathbf{x}}
\newcommand{\vqj}{q}
\newcommand{\vp}{{\bf p}}
\newcommand{\vq}{{\bf\vqj}}
\newcommand{\hvq}{\hat{\vq}}
\newcommand{\bzero}{{\bf 0}}					
\newcommand{\eigvec}{{\bf p}}					
\newcommand{\kgj}{k}
\newcommand{\kg}{{\bf\kgj}}
\newcommand{\hkg}{{\hat\kg}}

\newcommand{\eib}{\hat{\mathbf{ e}}}       

\newcommand{\wg}{\omega}					
\newcommand{\strain}{\boldsymbol\epsilon}		
\newcommand{\stress}{\boldsymbol\sigma}		
\newcommand{\tenselas}{{\rm {\large C}}}			
\newcommand{\TA}{\boldsymbol{\Gamma}}		
\newcommand{\roi}{\varrho}					
\newcommand{\domain}{{\mathcal O}}			
\newcommand{\pala}{\lambda}					
\newcommand{\palb}{\mu}					

\newcommand{\dd}{\mathrm{d}}				
\newcommand{\bnabla}{\boldsymbol\nabla}		
\newcommand{\Dx}{\mathrm{\bf D}}				
\newcommand{\Dxx}{\Dx_\xg}
\newcommand{\Dxy}{\Dx_\vy}	
\newcommand{\Divx}{\operatorname{{\bf Div}}}	
\newcommand{\Wigner}{{\bf W}}				
\newcommand{\TF}[1]{\widehat{#1}}				
\newcommand{\iexp}{\operatorname{e}}			
\newcommand{\trace}{\operatorname{Tr}}		

\newcommand{\WaveEq}{\mathcal{L}}

\newcommand{\cjg}[1]{\overline{#1}}
\newcommand{\adj}[1]{{#1}^*}

\newcommand{\Rset}{\mathbb{R}}			

\newcommand{\SyH}{\mathbf{H}}                      
\newcommand{\SyHin}{\mathrm{H}}                  

\newcommand{\diag}{\operatorname{diag}}              

\newcommand{\vM}{\mathbf{M}}
\newcommand{\vC}{\mathbf{C}}
\newcommand{\vL}{\mathbf{L}}

\newcommand{\NCi}{\mathrm{i}}            
\newcommand{\demi}{\frac{1}{2}}	      
\newcommand{\tiers}{\frac{1}{3}}

\newcommand{\ccub}{\mathrm{c}}               

\newcommand{\xgj}{x}

\newcommand{\comC}{\tenselas}             
\newcommand{\coefC}{\mathrm{d}} 

\newcommand{\fu}{\ug}
\newcommand{\fv}{\mathbf{v}}
\newcommand{\eigv}{\omega}                    
\newcommand{\saj}{a}  
\newcommand{\sa}{{\bf\saj}}                
\newcommand{\dir}{\delta}                       
\newcommand{\obsj}{\varphi}
\newcommand{\obsg}{\boldsymbol{\obsj}}
\newcommand{\av}{{\bf a}}
\newcommand{\bv}{{\bf b}}

\newcommand{\coroij}{\varrho}
\newcommand{\coro}{\mathrm{R}}        

\newcommand{\indi}{\alpha}             
\newcommand{\indj}{\beta}                

\newcommand{\inda}{i}        
\newcommand{\indb}{j}        

\newcommand{\indL}{2}
\newcommand{\indLT}{{2}}
\newcommand{\isym}{\mathrm{s}}
\newcommand{\iasym}{\mathrm{a}}

\newcommand{\escale}{\varepsilon}

\newcommand{\vinttem}{s}                   
\newcommand{\Wignere}{\Wigner_\escale}
\newcommand{\fue}{\fu_\escale}
\newcommand{\fve}{\fv_\escale}

\newcommand{\sequence}[1]{(#1)}

\newcommand{\sig}[1]{\hat{#1}}
\newcommand{\matN}{{\bf N}}
\newcommand{\matM}{{\bf K}}
\newcommand{\matGij}{G}
\newcommand{\matG}{{\bf \matGij}}
\newcommand{\matTij}{T}
\newcommand{\matT}{{\bf \matTij}}
\newcommand{\dscatij}{\sigma}
\newcommand{\dscat}{{\boldsymbol\dscatij}}
\newcommand{\tscati}{\Sigma}
\newcommand{\tscat}{{\boldsymbol\tscati}}
\newcommand{\lcor}{a}
\newcommand{\lcorp}{\overline{\lcor}}

\newcommand{\faniso}{\mathcal{A}}

\begin{document}




\title[Kinetic model of elastic waves in anisotropic media]{Kinetic modeling of multiple scattering of elastic waves in heterogeneous anisotropic media}


\author[I. Baydoun]{Ibrahim Baydoun}
\address[I. Baydoun]{Laboratoire MSS-Mat, \'Ecole Centrale Paris, UMR 8579 CNRS, France}
\email{ibrahim.baydoun@ecp.fr}

\author[\'E. Savin]{\'Eric Savin}
\address[\'E. Savin]{ONERA--The French Aerospace Lab, France}
\thanks{Corresponding author: \'E. Savin, ONERA--The French Aerospace Lab, 29 avenue de la Division Leclerc, F-92322 Ch\^atillon cedex, France (Eric.Savin@onera.fr).}
\email{eric.savin@onera.fr}

\author[R. Cottereau]{R\'egis Cottereau}
\address[R. Cottereau]{Laboratoire MSS-Mat, \'Ecole Centrale Paris, UMR 8579 CNRS, France}
\email{regis.cottereau@ecp.fr}

\author[D. Clouteau]{Didier Clouteau}
\address[D. Clouteau]{Laboratoire MSS-Mat, \'Ecole Centrale Paris, UMR 8579 CNRS, France}
\email{didier.clouteau@ecp.fr}

\author[J. Guilleminot]{Johann Guilleminot}
\address[J. Guilleminot]{Laboratoire Mod\'elisation et Simulation Multi-\'Echelle, Universit\'e Paris-Est, UMR 8208 CNRS, France}
\email{johann.guilleminot@univ-paris-est.fr}


\begin{abstract}
In this paper we develop a multiple scattering model for elastic waves in random anisotropic media. It relies on a kinetic approach of wave propagation phenomena pertaining to the situation whereby the wavelength is comparable to the correlation length of the weak random inhomogeneities--the so-called weak coupling limit. The waves are described in terms of their associated energy densities in the phase space position~$\times$~wave vector. They satisfy radiative transfer equations in this scaling, characterized by collision operators depending on the correlation structure of the heterogeneities. The derivation is based on a multi-scale asymptotic analysis using spatio-temporal Wigner transforms and their interpretation in terms of semiclassical operators, along the same lines as Bal [Wave Motion {\bf 43}, 132-157 (2005)]. The model accounts for all possible polarizations of waves in anisotropic elastic media and their interactions, as well as for the degeneracy directions of propagation when two phase speeds possibly coincide. Thus it embodies isotropic elasticity which was considered in several previous publications. Some particular anisotropic cases of engineering interest are derived in detail.
\end{abstract}


\date{\today}


\maketitle

\section{\textcolor{\mycolor}{Introduction and summary}}

\subsection{Modeling of wave propagation phenomena in random media}

The study of multiply-scattered elastic waves in heterogeneous, anisotropic media has relevance to non destructive evaluation of materials and structures, seismic waves characterization, acoustic emission and backscattered echo analyses, with possible applications in geophysical prospection, biomedical imaging, or structural health monitoring, among others. In this respect, the use of ultrasound to infer the microstructure of polycrystalline materials has been widely considered in the past since the earlier work of Mason \& McSkimin~\cite{MAS47}. The nondestructive techniques elaborated afterwards are based on the measurement of exponential rates of spatial decay (attenuations) and speeds of averaged plane waves, \emph{i.e.} coherent fields. The difficulty raised by this approach is the impossibility to distinguish the various sources of potential decays between scattering, geometrical spreading, internal absorption, or the influence of the reflections at the opposite faces of the sample. These shortcomings have prompted the development of probing techniques based on the measurement of the evolution of the incoherent part of ultrasonic waves, \emph{i.e.} multiply scattered, possibly diffusive fields~\textcolor{\mycolor}{\cite{GOE94}}. The earlier attempts in this direction can be tracked back to the works of Guo \emph{et al.}~\cite{GUO85} and Weaver~\cite{WEA90}. This alternative approach has received a considerable attention in the last decade since it was observed that the empirical cross-correlations of such diffuse fields could be directly related to the Green function of the propagation medium~\cite{DUV93,WEA01,CAM03}. The numerous developments and applications in geophysics which have followed are described in \emph{e.g.}~\cite{CAM11} and references therein.

These advances call for accurate analytical models of ultrasonic wave propagation phenomena in unstructured or structured heterogeneous media. Iterative perturbation expansions for weakly random polycrystalline materials were considered in~\cite{STA84,HIR82,HIR85,VAR85} in the spirit of the seminal developments of Karal \& Keller~\cite{KAR64}. Other approaches are based upon diagrammatic expansions in which the mean response is governed by a Dyson equation, and the mean square response is governed by a Bethe-Salpeter equation~\cite{FRI68,VAN99}. Both are analytically intractable unless low-order truncations are enforced, typically a so-called first-order smoothing approximation (FOSA) for the Dyson equation, and a so-called ladder approximation for the Bethe-Salpeter equation. These approximations have been used in~\cite{WEA90,TUR94a,TUR94b,TUR99,TUR01} for the derivation of (i) scattering-based attenuation coefficients on one hand, and of (ii) radiative transfer equations for the (renormalized) mean square wave fields in the limit of small wavelengths (high frequencies) with respect to the macroscopic features of the medium on the other hand.
The works evoked above have been mainly confined to untextured or textured aggregates of cubic-symmetry crystallites. Besides, radiative transfer models of high-frequency wave propagation in heterogeneous media have a broad range of validity and were derived in many fields from either phenomenological principles~\cite{CHA60,ISH78,SHE06,SAT12} or, more recently, systematic formal multiple-scale asymptotic expansions~\cite{RYZ96,GUO99,BRA02,BAL05,BAL10}. 

Transport and radiative transfer equations describe the mesoscopic regime of wave propagation when the wavelength is comparable to the characteristic length of the heterogeneities, typically a correlation length in a random medium \textcolor{\mycolor}{(hereafter referred to as the fast scale)}. It corresponds to a situation of strong interaction between waves and random heterogeneities which cannot be addressed by usual homogenization and multi-scale techniques. Also it considers large propagation distances compared to the wavelength, and weak amplitudes of the random perturbations of the material parameters with respect to a bare, possibly heterogeneous, background medium varying at a length scale \textcolor{\mycolor}{(the slow scale)} one order of magnitude larger than the wave/correlation lengths. This corresponds to the so-called weak coupling limit as defined in the dedicated literature, whereby an explicit separation of scales can be invoked. The analysis developed in~\cite{RYZ96,GUO99,BRA02,BAL05,BAL10} is based on the use of a Wigner transform of the wave field, of which high-frequency, non-negative limit captures the angularly resolved energy density in time and space. It can be made mathematically rigorous as in~\cite{LIO93,GER97,AKI12} ignoring however the influence of random inhomogeneities, except for some particular situations~\cite{ERD00,LUK07}.

The purpose of the research presented in this paper is to assess the influence of material \emph{full anisotropy} on the radiative transfer regime of elastic waves in randomly heterogeneous media. Anisotropy is considered at two levels. The first one is related to the constitutive law of random materials. The second level is related to the correlation structure of these random materials, referred to as anisomery in the dedicated literature~\cite{MAR06}. More specifically we have developed formal models for the consideration of anisotropy in the collision kernels of the radiative transfer equations pertaining to multiply-scattered elastic waves. These models describe the evolution of their energy density in the phase space position $\times$ wave vector in terms of the Wigner measure of the wave fields. The analysis follows to a great extent the techniques used by Bal~\cite{BAL05} and Akian~\cite{AKI12} in that it handles a second-order wave equation and introduces the spatio-temporal Wigner transform of the elastic waves. However, as opposed to~\cite{BAL05} it considers \emph{vector} wave fields, and as opposed to~\cite{AKI12} it considers the influence of random perturbations in the weak coupling regime. Therefore our derivation generalizes those previous works and the classical reference~\cite{RYZ96} to fully anisotropic bare elastic media with fully anisotropic random perturbations.

\subsection{\textcolor{\mycolor}{Summary of the main results}}

\textcolor{\mycolor}{We now summarize our main results. We aim at describing elastic waves in a random medium taking into account the non-uniformity of the background medium and their scattering by random inhomogeneities, with due consideration of the effects of coupling between their different polarizations. We also consider the regime where the leading wavelength is comparable to the (small) correlation length of the heterogeneities, in order to ensure maximum interactions with the waves. This is a necessary condition if we want to probe the medium and its fluctuations. It defines the high-frequency range terminology we shall use throughout the paper. At last, our fundamentally new results are that this objective is achieved for \emph{arbitrary anisotropy} of the random medium. As a \emph{scalar} wave propagates in a random medium with an incident wave vector $\vq$, it can be scattered at any time $t$ and position $\xg$ into any direction $\hkg$ and wave vector $\kg$ (such that $\hkg=\kg/|\kg|$). Therefore it is relevant to consider an angularly resolved scalar energy density $\saj(t,\xg,\kg)$ for this wave, defined for all positions $(\xg,\kg)$ in phase space. In~\cite{RYZ96,BAL05} it is shown that energy conservation takes the form of a scalar radiative transfer equation:}
\textcolor{\mycolor}{
\begin{multline}\label{eq:RTE-scal}
\partial_t\saj(t,\xg,\kg)+\bnabla_\kg\eigl(\xg,\kg)\cdot\bnabla_\xg\saj(t,\xg,\kg)-\bnabla_\xg\eigl(\xg,\kg)\cdot\bnabla_\kg\saj(t,\xg,\kg) \\
=\int\dscatij(\xg,\kg,\vq)\saj(t,\xg,\vq)\dd\vq-\tscati(\xg,\kg)\saj(t,\xg,\kg)\,,
\end{multline}
where $\eigl(\xg,\kg)$ is the frequency of the waves at $\xg$ with wave vector $\kg$, and $\dscatij(\xg,\kg,\vq)$ is the rate of conversion of energy with wave vector $\vq$ into energy with wave vector $\kg$ at position $\xg$--the so-called scattering cross-section. The total scattering cross-section $\tscati$ is:
\begin{displaymath}
\tscati(\xg,\kg)=\int\dscatij(\xg,\kg,\vq)\dd\vq\,,
\end{displaymath}
such that the transport equation is conservative because the former relationship yields:
\begin{displaymath}
\iint\saj(t,\xg,\kg)\dd\kg\dd\xg=\operatorname{Const}
\end{displaymath}
for all times. The scattering cross-section is explicitly determined by the power spectral density of the inhomogeneities~\cite{RYZ96,BAL05}. The transport equation (\ref{eq:RTE-scal}) also holds when the waves are scattered by randomly distributed discrete scatterers, in which case the scattering cross-section is the cross-section of a single scatterer multiplied by their density. Here we only consider continuous random inhomogeneities.}

\textcolor{\mycolor}{For \emph{vector} waves we must in addition keep track of their state of polarization. In a three-dimensional anisotropic elastic medium three orthogonal polarization directions exist, corresponding to at most three different directionally-dependent phase velocities: one for quasi-longitudinal compressional wave, and two for quasi-transverse shear waves. Labeling the polarization states by $\indi=1,2$ or $3$, each one has its own energy density $\saj_\indi(t,\xg,\vq)$ but it may be converted to any other state and any other direction $\hkg$ at any position $\xg$ when scattered by the inhomogeneities. Conservation of energy is now expressed in terms of \emph{coupled} radiative transfer equations for the energy densities of the different polarizations:
\begin{multline}\label{eq:RTE-vect}
\partial_t\saj_\indi(t,\xg,\kg)+\bnabla_\kg\eigl_\indi(\xg,\kg)\cdot\bnabla_\xg\saj_\indi(t,\xg,\kg)-\bnabla_\xg\eigl_\indi(\xg,\kg)\cdot\bnabla_\kg\saj_\indi(t,\xg,\kg) \\
=\sum_{\indj=1}^3\int\dscatij_{\indi\indj}(\xg,\kg,\vq)\saj_\indj(t,\xg,\vq)\dd\vq-\tscati_\indi(\xg,\kg)\saj_\indi(t,\xg,\kg)\,,\quad\indi=1,2,3\,,
\end{multline}
where $\eigl_\indi(\xg,\kg)$ is the frequency of the waves at $\xg$ with wave vector $\kg$ and polarization state $\indi$, and $\dscatij_{\indi\indj}(\xg,\kg,\vq)$ is the rate of conversion of energy with wave vector $\vq$ and polarization state $\indj$ into energy with wave vector $\kg$ and polarization state $\indi$, at position $\xg$. The total scattering cross-sections $\tscati_\indi$ are:
\begin{displaymath}
\tscati_\indi(\xg,\kg)=\sum_{\indj=1}^3\int\dscatij_{\indi\indj}(\xg,\kg,\vq)\dd\vq\,,\quad\indi=1,2,3\,,
\end{displaymath}
such that the coupled transport equations are conservative since:
\begin{displaymath}
\sum_{\indi=1}^3\iint\saj_\indi(t,\xg,\kg)\dd\kg\dd\xg=\operatorname{Const}
\end{displaymath}
in the vector case. The scattering cross-sections $\dscatij_{\indi\indj}$ are explicitly determined in terms of the power spectral density of the inhomogeneities--which may be characterized by up to $21$ coefficients in a triclinic medium. They were originally derived in~\cite{RYZ96} for isotropic media solely, such that the quasi-transverse shear velocities are the same everywhere and independent of the propagation direction. Here we obtain generalized formulas for them accounting for all possible symmetry classes of anisotropic media, namely~\eref{eq:dscat} together with the definitions of \eref{symb-H} and \eref{ela-ten-fluc}. In this latter equation the elasticity tensor $\tenselas_0$ of the background medium and the elasticity tensor $\tenselas_1$ of the random inhomogeneities are formally allowed to belong to different symmetry classes, and to vary at different scales. This implies that they have different contributions to the wave dynamics:
\begin{itemize}
\item The variations of $\tenselas_0$ at the slow scale contribute only to the left-hand side of the radiative transfer equations (\ref{eq:RTE-scal}) and (\ref{eq:RTE-vect}). They basically characterize the phase velocities $\eigl_\indi(\xg,\hkg)$ and polarizations $\indi$ in this regime.
\item The variations of $\tenselas_1$ contribute only to the right-hand side of these radiative transfer equations in terms of the power spectral densities of its $21$ elasticity coefficients. It describes how high-frequency waves are continuously scattered by the material inhomogeneities at the fast scale, which is also their (small) wavelength. These collision operators account for both the elastic (change of direction without changing polarization) and inelastic (with a change of polarization) processes.
\end{itemize}
In \eref{eq:RTE-vect} the energy densities are scalars as long as the phase velocities are all distinct. Our derivation considers the most general case when they may coincide for some modes, though. This is the case for isotropic media for example, for which \eref{eq:RTE-vect} becomes a matrix system. This situation is fully addressed in the subsequent analyses}.

\subsection{\textcolor{\mycolor}{Outline}}

The rest of the paper is organized as follows. In \sref{sec:HFwaves} we introduce the basic framework and notations used throughout. We more particularly focus on the characterization of anisotropy in terms of the acoustic, or Christoffel tensor of the bare medium, and the relevance of using a Wigner transform and its non-negative limit measure for the analysis of multiple scattering phenomena in the high-frequency range. \textcolor{\mycolor}{This limit is simply the energy density $\saj_\indi$ introduced above for each polarization state $\indi$}. The corresponding transport model is derived in detail in \sref{sec:Transport} ignoring the influence of random inhomogeneities in a first step. The spatio-temporal Wigner transform and the formal mathematical tools used for the subsequent analyses are also introduced there. The main contribution of the paper is \sref{sec:RTE} which outlines the extension of the previous transport model to account for anisotropic random inhomogeneities. Matrix radiative transfer equations are obtained in the most general case. Their collision operators are explicitly described in terms of the correlation structure of the heterogeneities for all possible symmetries arising in elastic constitutive relations. In this respect, it should be noted that the proposed theory requires a full characterization of the power spectral densities of the tensorial random fluctuations (assumed to be statistically homogeneous at the small wavelength scale), but no other statistical information. These data may be obtained from the random matrix theory of the elasticity tensor developed recently (see~\cite{TA10,GUI13} and references therein) for example. We will however not pursue the analysis presented here in that direction, considering that simplified correlation models as classically encountered in the literature are sufficient for the purpose and scope of this paper. Using these models, the collision kernels (the scattering cross-sections) for some selected material symmetries are plotted in \sref{sec:examples} in order to illustrate our results. Some conclusions and perspectives are finally drawn in~\sref{sec:CL}.

\section{Elastic bulk wave propagation in a high-frequency setting}\label{sec:HFwaves}

In this section we establish the high-frequency setting we are interested in for the derivation of the multiple-scattering kinetic (transport) model that will be detailed in the subsequent parts of this paper. The material below is essentially adapted from~\cite{SAV13}. The primary objective is to introduce the main notations that will be used throughout the paper.

\subsection{Elastic wave equation}\label{sec:HFwaves-basics}

We first recall how the vector wave equation arises in an elastic medium occupying an open domain $\domain\subseteq\smash{\Rset^3}$, where $\Rset^3$ stands for the usual three-dimensional Euclidean space. That medium is constituted by a heterogeneous linear viscoelastic material, of which density is denoted by $\roi(\xg)$ and fourth-order relaxation, or elasticity tensor is denoted by $\tenselas(\xg)$, for $\xg\in\domain$. Its displacement field about a quasi-static equilibrium considered as the reference configuration is denoted by $\ug_\escale(\xg,t)\in\Rset^3$, and its second-order Cauchy stress tensor is denoted by $\stress_\escale(\xg,t)$. The subscript $\escale$ stands for the (small) spatial scale of variation of the initial conditions that will be imposed to the materials and will be propagated to the displacement and stress fields at later times by hyperbolicity. Then the balance of momentum in a fixed reference frame ignoring the action of body forces reads:
\begin{equation}\label{eq:Navier}
\roi\partial_t^2\ug_\escale(\xg,t)=\Divx\,\stress_\escale(\xg,t)\,,\quad\xg,t\in\domain\times\Rset\,.
\end{equation}
Here the divergence of a second-order tensor ${\bf A}$ is defined by $\smash{(\Divx{\bf A},\bv)}=\smash{\bnabla_\xg}\cdot({\bf A}^\itr\bv)$ for any constant vector $\bv$, and $\smash{\bnabla_\xg}$ is the gradient vector with respect to $\xg$. At last $(\cdot)^\itr$ stands for the matrix transpose. The initial conditions for~\eref{eq:Navier} read:
\begin{equation}\label{eq:CI}
\ug_\escale(\xg,0)=\ug_0(\xg;\escale)\,,\quad\partial_t\ug_\escale(\xg,0)=\vg_0(\xg;\escale)\,.
\end{equation}
They are parameterized by the small parameter $\escale$, which quantifies the rate of change of $\xg\mapsto\smash{\ug_0}(\xg)$ and $\xg\mapsto\smash{\vg_0}(\xg)$ with respect to the dimensions of $\domain$ or the propagation/observation distances. Since high-frequency waves will be generated by an initial vibrational energy oscillating at a scale $\escale\ll 1$, the functions $\smash{\bnabla_\xg\otimes\ug_0}$ and $\smash{\vg_0}$ shall be considered as strongly $\escale$-oscillating functions in the sense of G\'erard \emph{et al}.~\cite{GER97}. The plane waves $\smash{\ug_0(\xg;\escale)}=\smash{\escale{\bf A}(\xg)\iexp^{\ci\kg\cdot\xg/\escale}}$ and $\smash{\vg_0(\xg;\escale)}=\smash{{\bf B}(\xg)\iexp^{\ci\kg\cdot\xg/\escale}}$ for a given wave vector $\kg\in\smash{\Rset^3}$ and $\smash{\ci=\sqrt{-1}}$, typically fulfill this condition. 

In addition, the stress field $\smash{\stress_\escale}$ is given as a function of the linearized strain tensor $\smash{\strain_\escale}$ by the material constitutive equation:
\begin{equation}\label{eq:comportement}
\stress_\escale(\xg,t)=\tenselas(\xg)\strain_\escale(\xg,t)\,,\quad\strain_\escale(\xg,t)=\bnabla_\xg\otimes_s\ug_\escale(\xg,t)\,.
\end{equation}
Both $\smash{\stress_\escale}$ and $\smash{\strain_\escale}$ are parameterized by the scale $\escale$ of the applied loads since $\smash{\ug_\escale}$ also is. Here $\smash{\otimes_s}$ is the symmetrised tensor product of two vectors $\smash{\av\otimes_s\bv}=\smash{\operatorname{sym}(\av\otimes\bv)}$. The elastic wave equation for $\smash{\ug_\escale}$ is derived plugging this relation into~\eref{eq:Navier} and thus reads:
\begin{equation}\label{eq:elastic-waves}
\WaveEq(\ug_\escale) = \roi\partial_t^2\ug_\escale-\bnabla_\xg\cdot(\tenselas:\bnabla_\xg\otimes\ug_\escale)=\bzero\,,\quad\xg,t\in\domain\times\Rset\,.
\end{equation}
Indeed, since $\smash{\stress_\escale}$ and $\smash{\strain_\escale}$ are symmetric the elasticity tensor $\tenselas$ shall satisfy the minor symmetries $\smash{\tenselas^{ijkl}}=\smash{\tenselas^{jikl}}= \smash{\tenselas^{ijlk}}$. Invoking a thermodynamical reversibility argument, it also satisfies the major symmetry $\smash{\tenselas^{ijkl}}=\smash{\tenselas^{klij}}$.

\subsection{The Christoffel tensor}

Let us define the $9\times 3$ matrix $\vM(\kg)$ by:
\begin{equation}\label{eq:defM}
\vM(\kg)=\NCi\begin{bmatrix} \kg & \bzero & \bzero \\  \bzero & \kg & \bzero \\  \bzero & \bzero & \kg \end{bmatrix}\,,\quad\kg\in\Rset^3\,,
\end{equation}
and the $9\times 9$ symmetric, positive semi-definite matrix $\vC(\xg)$ which is constituted by the $21$ independent coefficients of the elasticity tensor $\tenselas(\xg)$. More precisely, $\vC$ is a $3\times 3$ block matrix of which block $(i,k)$ is the $3\times 3$ matrix with elements $\smash{\tenselas^{ijkl}}$. Then the second-order acoustic, or Christoffel tensor $\TA(\xg,\kg)$ of the propagation medium is defined by:
\begin{equation}\label{eq:TA}
\TA(\xg,\kg)=\roi(\xg)^{-1}\adj{\vM}(\kg)\vC(\xg)\vM(\kg)\,,\quad\xg\in\domain\,,\;\kg\in\Rset^3\,,
\end{equation}
where $\smash{\adj{\vM}}=\smash{\cjg{\vM}^\itr}$ stands for the conjugate transpose matrix. It is symmetric, real and positive definite in $\domain\times\Rset^3\setminus\{\kg=\bzero\}$. So for a given direction $\hkg:=\kg/|\kg|$ on the unit sphere $\Sset^2$ of $\Rset^3$ it has three real positive (possibly multiple) eigenvalues $\eigl_\jeig^2(\xg,\kg)$ for $\jeig=1,2$ or $3$, and the associated eigenvectors $\eigvec_\jeig(\xg,\kg)$ can be chosen real and orthogonal. They can also be normalized such that:
\begin{equation}\label{eq:TAspectral}
\TA(\xg,\kg)=\sum_{\jeig=1}^3\eigl_\jeig^2(\xg,\kg)\eigvec_\jeig(\xg,\kg)\otimes\eigvec_\jeig(\xg,\kg)\,,\quad\II=\sum_{\jeig=1}^3\eigvec_\jeig(\xg,\kg)\otimes\eigvec_\jeig(\xg,\kg)\,,
\end{equation}
where $\II$ is the identity matrix of $\Rset^3$. The eigenvalues and eigenvectors of the Christoffel tensor correspond to the phase velocities and polarizations of plane waves propagating along the direction $\smash{\hkg}$ in the medium $\domain$. Here we do not assume any particular ordering of the eigenvalues with indices $1$, $2$ and $3$. The polarization with the closest direction to $\smash{\hkg}$ is called the quasi-longitudinal wave, the other two components being the quasi-transversal waves. Also in view of~\eref{eq:TA}, the eigenvalues $\smash{\eigl_\jeig^2(\xg,\kg)}$ have the form $\smash{\eigl_\jeig^2(\xg,\kg)}=\smash{\eigl_\jeig^2(\xg,\hkg)|\kg|^2}:=\smash{\cel_\jeig^2(\xg,\hkg)|\kg|^2}$ where the $\smash{\cel_\jeig}$'s have dimension of celerities. One has in addition $\smash{\eigvec_\jeig(\xg,\kg)}=\smash{\eigvec_\jeig(\xg,\hkg)}$. The above spectral expansion of the Christoffel tensor is valid for all directions but the so-called acoustic axes~\cite{ALS04b,BOU98b,NOR04}, along which two eigenvalues may coincide. For such a direction $\smash{\hkg=\hkg_a}$ of degeneracy where, say $\smash{\eigl_1(\xg,\hkg_a)=\eigl_2(\xg,\hkg_a)}$, the expansion reduces to:
\begin{equation}\label{eq:TA-degenerated}
\TA(\xg,\hkg_a)=\eigl_1^2(\xg,\hkg_a)\II+\left(\eigl_3^2(\xg,\hkg_a)-\eigl_1^2(\xg,\hkg_a)\right)\eigvec_3(\xg,\hkg_a)\otimes\eigvec_3(\xg,\hkg_a)\,.
\end{equation}
It follows from~\eref{eq:TA-degenerated} that any vector which is orthogonal to $\smash{\eigvec_3(\xg,\hkg_a)}$ is an eigenvector of the Christoffel tensor $\smash{\TA(\xg,\hkg_a)}$, \emph{i.e.} it is an allowed polarization for a wave propagating along $\smash{\hkg_a}$ with a wave celerity $\smash{\eigl_1(\xg,\hkg_a)}$. A common example is elastic isotropy, when $\tenselas=\lambda\II\otimes\II+2\mu\II\boxtimes\II$ where $\lambda$ and $\mu$ are the Lam\'e parameters and $({\bf A}\boxtimes{\bf A}){\bf B}:=\smash{{\bf A}{\bf B}^\isym{\bf A}}$, for $\smash{{\bf B}^\isym}$ standing for the symmetric part of a square matrix ${\bf B}$. Then the Christoffel tensor reads:
 \begin{equation}\label{eq:TA-isotropy}
\TA(\xg,\hkg)=\cel_\iS^2(\xg)\II+\left(\cel_\iP^2(\xg)-\cel_\iS^2(\xg)\right)\hkg\otimes\hkg\,,\quad\forall\hkg\in\Sset^2\,,
\end{equation}
where $\smash{\cel_\iP}=\smash{\sqrt{(\lambda+2\mu)/\roi}}$ and $\smash{\cel_\iS}=\smash{\sqrt{\mu/\roi}}$ are the velocities for compressive and shear waves, respectively. The various properties of the Christoffel tensor for anisotropic media are discussed in \emph{e.g.}~\cite{ALS04b,ALS04a,BOU98a,ITI13} and references therein. 

\subsection{High-frequency setting}

\begin{figure}[ht]
\centering{\includegraphics[scale=0.45]{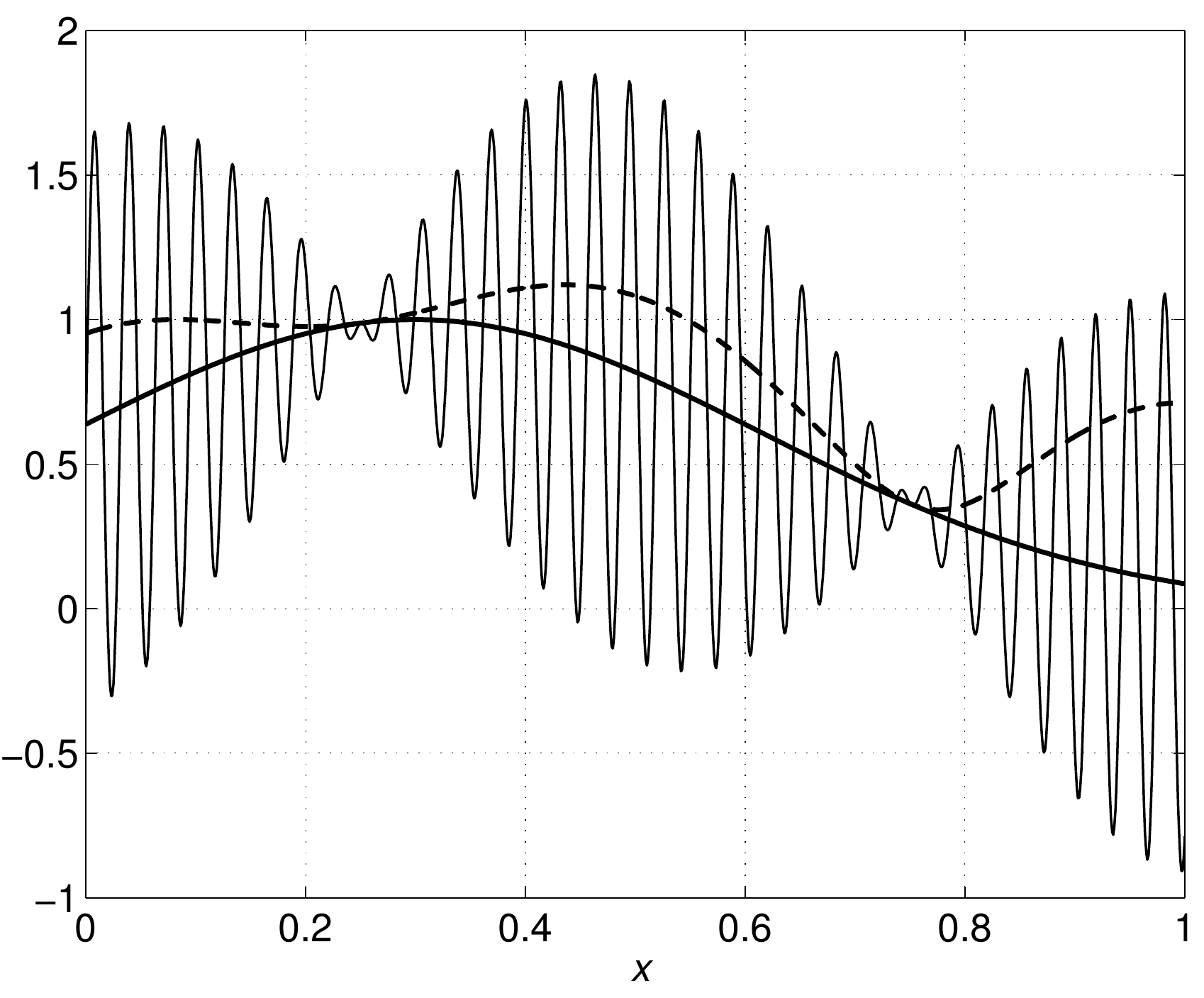}}
\caption{The energy limit of a strongly oscillating sequence: the oscillating function $\ugj_\escale(x)$ (thin line), the mean function $\underline{\ugj}(x)$ (thick line), and the square-root limit $\smash{(\underline{\ugj}(x)^2+\demi a(x)^2)^\demi}$ (thick dashed line). After~\cite{SAV13}.}\label{fg:Wigner}
\end{figure}

The high-frequency limit $\escale\rightarrow 0$ in the previous setting shall be considered for quadratic observables of the wave displacement field $\smash{\ug_\escale}$ as explained now. We rely on the simple following example which was already discussed in~\cite{SAV12,SAV13}.  The real function $x\mapsto\ugj_\escale(x)$ oscillating with an amplitude $a(x)$ about its mean $\underline{\ugj}(x)$:
\begin{equation}
\ugj_\escale(x)=\underline{\ugj}(x)+a(x)\sin\frac{x}{\escale}\,,\quad0<\escale\ll 1\,,
\end{equation}
has no strong limit when $\escale\rightarrow 0$, although the functions $a$ and $\underline{\ugj}$ vary slowly. However for any smooth function $\obsj$ with compact support on $\Rset$, one can obtain a vague limit as:
\begin{equation}\label{eq:vaguelim}
\lim_{\escale\rightarrow 0}\int_\Rset\obsj(x)\left(\ugj_\escale(x)\right)^2\id x=\int_\Rset\obsj(x)\left((\underline{\ugj}(x))^2+\demi(a(x))^2\right)\id x\,.
\end{equation}
Thus the purpose of the observation function $\obsj$ is to compute a smoothened version of the "energy" of $\ugj_\escale(x)$ as given by $\underline{\ugj}(x)^2+\demi a(x)^2$. It allows to estimate the deviation of oscillations with amplitude $a(x)$ about the mean at any point $x$ selected by the support of $\obsj$. This feature is illustrated on~\fref{fg:Wigner}.

A mathematical generalization of this idea is possibly given by the notion of Wigner measure~\cite{RYZ96,GER97,MAR02,BAL10,AKI12}. Let us now consider that the observable function $\obsg$ is an arbitrary $3\times 3$, compactly supported matrix function of both the space variable $\xg$ and the wave vector $\kg\in\Rset^3$. For a vector field $\ug\in\smash{[L^2(\Rset^3)]^3}$, the functional space of $\Rset^3$-valued, square integrable functions equipped with the scalar product $\smash{(\ug,\vg)_{L^2}}=\smash{\int_{\Rset^3}\ug(\xg)\cdot\vg(\xg)\,\id\xg}$, consider the (semiclassical) operator:
\begin{equation}
\obsg^\theta(\xg,\escale\Dx)\ug(\xg)=\frac{1}{(2\pi)^3}\int_{\Rset^3\times\Rset^3}\iexp^{\ci\kg\cdot(\xg-\yg)}\obsg((1-\theta)\xg+\theta\yg,\escale\kg)\ug(\yg)\,\id\yg\id\kg\,,
\end{equation}
for $\theta\in[0,1]$. This parameter defines the so-called quantization of the operator. The case $\theta=0$ corresponds to the standard quantization. It is simply denoted by $\smash{\obsg(\xg,\escale\Dx)}$ such that:
\begin{equation}\label{eq:PDO}
\obsg(\xg,\escale\Dx)\ug(\xg)=\frac{1}{(2\pi)^3}\int_{\Rset^3}\iexp^{\ci\kg\cdot\xg}\obsg(\xg,\escale\kg)\TF{\ug}(\kg)\,\id\kg\,,
\end{equation}
where $\smash{\TF{\ug}(\kg)}:=\smash{\int_{\Rset^3}\iexp^{-\ci\kg\cdot\xg}\ug(\xg)\id\xg}$ stands for the Fourier transform of $\ug(\xg)$. The case $\smash{\theta=\demi}$ corresponds to the Weyl quantization, which is usually denoted by $\smash{\obsg^W(\xg,\escale\Dx)}$. Then for a sequence $\smash{\sequence{\ug_\escale}}$ uniformly bounded in $\smash{[L^2(\Rset^3)]^3}$, there exists a positive, Hermitian measure $\Wigner[\ug_\escale]$ such that, up to extracting a subsequence if need be:
\begin{equation}\label{eq:mes-semi-classique}
\lim_{\escale\rightarrow 0}(\obsg^\theta(\xg,\escale\Dx)\ug_\escale,\ug_\escale)_{L^2}=\trace\int_{\Rset^3\times\Rset^3}\obsg(\xg,\kg)\Wigner[\ug_\escale](\id\xg,\id\kg)\,,\quad\forall\obsg\,,
\end{equation}
independently of the quantization $\theta$. $\Wigner[\ug_\escale]$ is the so-called Wigner measure of the sequence $\smash{\sequence{\ug_\escale}}$ because it can also be interpreted as the weak limit of its Wigner transform $\smash{\Wignere[\ug_\escale,\ug_\escale]:=\Wignere[\ug_\escale]}$. Indeed, if the latter is defined for $\smash{\Rset^3}$-valued temperate distributions $\ug,\vg$ by:
\begin{equation}\label{eq:TWigner}
\Wignere[\ug,\vg](\xg,\kg)=\frac{1}{(2\pi)^3}\int_{\Rset^3}\iexp^{\ci\kg\cdot\vy}\ug\left(\xg-\frac{\escale\yg}{2}\right)\otimes\cjg{\vg\left(\xg+\frac{\escale\yg}{2}\right)}\,\id\vy\,,
\end{equation}
then one has:
\begin{displaymath}
(\obsg^W(\xg,\escale\Dx)\ug,\vg)_{L^2}=\trace\int_{\Rset^3\times\Rset^3}\obsg(\xg,\kg)\Wigner_\escale[\ug,\vg](\id\xg,\id\kg)\,.
\end{displaymath}
Thus $\smash{\Wigner[\ug_\escale]}$ describes the limit energy of the sequence $\smash{\sequence{\ug_\escale}}$ in the phase space $\smash{\Rset^3_\xg\times\Rset^3_\kg}$. As in~\eref{eq:vaguelim}, the matrix function $\obsg(\xg,\kg)$ is used to select any quadratic observable or quantity of interest associated to this energy: the kinetic energy, or the free energy, or the power flow, etc. For example, the high-frequency strain energy $\smash{{\mathcal V}_\escale(t)}:=\smash{\demi\int_\domain\tenselas\strain_\escale:\strain_\escale\,\id\xg}$ in $\domain$ may be estimated with $\obsg(\xg,\kg)\equiv\roi(\xg)\TA(\xg,\kg)$:
\begin{equation}\label{eq:strain-energy}
\lim_{\escale\rightarrow 0}{\mathcal V}_\escale(t)=\demi\int_{\domain\times\Rset^3}\roi(\xg)\TA(\xg,\kg):\Wigner[\ug_\escale(\cdot,t)](\id\xg,\id\kg)\,,
\end{equation}
up to some possible boundary effects on $\partial\domain$. Similarly, the kinetic energy $\smash{{\mathcal T}_\escale(t)}:=\smash{\demi\int_\domain\roi|\partial_t\ug_\escale|^2\,\id\xg}$ is estimated by:
\begin{equation}\label{eq:kinetic-energy}
\lim_{\escale\rightarrow 0}{\mathcal T}_\escale(t)=\demi\int_{\domain\times\Rset^3}\roi(\xg)\trace\Wigner[\escale\partial_t\ug_\escale(\cdot,t)](\id\xg,\id\kg)\,.
\end{equation}
The vibrational energy density ${\mathcal E}_\escale:={\mathcal V}_\escale+{\mathcal T}_\escale$ does not solve a closed-form equation in the high-frequency limit $\escale\rightarrow 0$. However the Wigner measure, which provides a decomposition of these quantities in the phase space, does so as explained in the subsequent derivations. This is another reason why we shall now focus on such limit measure rather than $\smash{\ug_\escale}$ directly or quadratic quantities of $\smash{\ug_\escale}$.

\section{Wigner measure of high-frequency elastic waves}\label{sec:Transport}

In this section we show how to obtain explicitly the Wigner measure~\eqref{eq:mes-semi-classique} of the high-frequency solutions of the elastic wave equation~\eqref{eq:PDOwave} in the setting outlined in the foregoing section. Indeed, it is needed in~(\ref{eq:strain-energy}) and~(\ref{eq:kinetic-energy}) for the computation of the evolution of the strain and kinetic energies, for example. The objectives are also to outline its main properties for a slowly varying medium, as well as the (formal) mathematical tools used for its derivation. Both will prove useful in the subsequent~\sref{sec:RTE} where elastic waves in a rapidly varying random medium with correlation lengths comparable to the small wavelength $\escale$ are considered. The analysis presented here is derived from~\cite{GER97}, where first-order hyperbolic systems with constant and slowly varying coefficients are addressed, and~\cite{AKI12}, where arbitrary order hyperbolic systems with slowly varying coefficients are addressed. The dispersion properties of the elastic Wigner measure are derived in~\sref{sec:dispersion}, and its evolution properties are derived in~\sref{sec:evolution}. Here it is shown that its components in the eigenspaces of the Christoffel tensor, the so-called specific intensities, satisfy transport equations of the Liouville type. The latter states that the energy densities in these eigenspaces are transported in phase space with celerities corresponding to the associated eigenvalues. Before doing so, we shall need some formal mathematical tools in order to compute the Wigner transform and its limit for high-frequency elastic waves. They are introduced in \sref{sec:comp-adj} and \sref{sec:space-time-WT} below. Now in order to hopefully clarify the subsequent derivations, we start by reformulating the elastic wave equation in a form that is adapted to the analysis developed in the remaining of the paper. 

\subsection{Elastic wave equation as a semiclassical operator}\label{sec:PDO-wave-eq}

Here we write~\eref{eq:elastic-waves} in a more convenient form for the derivation of the high-frequency regime $\escale\ll 1$. We shall first consider a slowly fluctuating medium characterized by an elastic tensor $\tenselas(\xg)$ which is independent of the small parameter $\escale$. The corresponding Christoffel tensor being given by~\eref{eq:TA} as $\TA(\xg,\kg)=\smash{\roi(\xg)^{-1}\adj{\vM}(\kg)\vC(\xg)\vM(\kg)}$ where $\vM$ has been defined in~\eref{eq:defM}, the elastic wave equation (\ref{eq:elastic-waves}) then reads~\cite{AKI12}:
\begin{equation}\label{eq:PDOwave}
(\ci\escale)^2\WaveEq(\ug_\escale) = \left(\roi(\xg)(\escale\iD_t)^2\ug_\escale-\vL_\escale(\xg,\escale\Dxx)\right)\ug_\escale=\bzero\,,\quad\xg,t\in\domain\times\Rset\,,
\end{equation}
with $\smash{\Dxx=\frac{1}{\ci}\bnabla_\xg}$, $\smash{\iD_t=\frac{1}{\ci}\partial_t}$, and the operator $\vL_\escale(\xg,\escale\Dxx)=\adj{\vM}(\escale\Dxx)\circ\vC(\xg)\circ\vM(\escale\Dxx)$. The latter may be expanded as:
\begin{equation}\label{eq:PDO-L}
\vL_\escale=\vL_0+\frac{\escale}{\ci}\vL_\indL\,, \\
\end{equation}
where:
\begin{equation}
\begin{split}
\vL_0(\xg,\kg) &=\roi(\xg)\TA(\xg,\kg)\,,\\
\vL_\indL(\xg,\kg) &=\roi(\xg)\TA_\indL(\xg,\kg)=\bnabla_\kg\adj{\vM}(\kg)\cdot\bnabla_\xg\vC(\xg)\vM(\kg)\,,
\end{split}
\end{equation}
with the notation $\smash{\bnabla_\kg{\bf A}\cdot\bnabla_\xg{\bf B}}=\smash{\sum_{j=1}^3(\partial_{\kgj_j}{\bf A})(\partial_{\xgj_j}{\bf B})}$ for two matrices ${\bf A}$ and ${\bf B}$. It should be observed that $\smash{\vL_\indL(\xg,\escale\Dxx)}$ is a first-order partial differential operator (independent of time), and that $\smash{\vL_\indL\equiv\bzero}$ in a homogeneous medium.

\subsection{Some formal rules of pseudo-differential calculus}\label{sec:comp-adj}

Let $\obsg(\kg)$ be a smooth matrix-valued observable defined on $\smash{\Rset^3}$, we recall the notation of \eref{eq:PDO} for the homogeneous semiclassical operator $\obsg(\escale\Dxx)$ in the standard quantization. We then have (see \emph{e.g.}~\cite{BAL05}):
\begin{equation}\label{reg3-cal}
\begin{split}   
\Wignere[\obsg(  \escale \Dxx ) \fue ,\fve] &= \obsg\left ( \kg +\frac{ \escale\Dxx }{2} \right) \Wignere[  \fue,\fve]\,, \\
\Wignere[\fue, \obsg(\escale\Dxx ) \fve] &=\Wignere[\fue,\fve]\adj{\obsg} \left( \kg -\frac{ \escale \Dxx}{2} \right)\,. 
\end{split}  
\end{equation}
Here we assume that the differential operator $\smash{\Dxx}$ within the observable $\obsg$ acts on $\smash{\Wignere[\fue,\fve]}$ so that, for instance, $\smash{\Wignere[\fue,\fve]\adj{\obsg}(\kg - \frac{\escale\Dxx}{2})}$ should be interpreted as the component-wise inverse Fourier transform of the matrix $\smash{\TF{\Wigner}_\escale[\fue,\fve]}\cdot\smash{\adj{\obsg}(\kg - \frac{\escale\vp}{2})}$. These relations can be extended for a non homogeneous semiclassical operator $\obsg(\xg,\escale\Dxx)$ (as in~\eref{eq:PDO} again) in the form~\cite{GER97,BAL05}:
\begin{multline}\label{reg1-cal}   
\Wignere[  \obsg( \xg, \escale \Dxx ) \fu_{\escale},\fv_{\escale} ] =  \obsg(\xg,\kg) \Wignere[\fue,\fve] +  \frac{\escale}{2\ci}  \{\obsg,\Wignere[\fue,\fve]\}  \\
 - \frac{\escale}{2\ci}\bnabla  _{\xg} \cdot \bnabla _{\kg}\obsg( \xg,\kg)\Wignere[\fue, \fve]  + \Go(\escale^2)\,,
\end{multline}
and:
\begin{multline}\label{reg2-cal}   
\Wignere[\fue, \obsg( \xg, \escale \Dxx) \fve ] =   \Wignere[\fue,\fve]\adj{\obsg}(\xg,\kg) + \frac{\escale}{2\ci}\{\Wignere[\fue ,\fve],\adj{\obsg}\} \\ 
+ \frac{\escale}{2\ci}\Wignere[\fue,\fve]\bnabla_{\xg} \cdot \bnabla _{\kg}\adj{\obsg}(\xg,\kg)+  \Go(\escale^{2})\,,
\end{multline}
since $\Wignere[\fue,\fve]=\adj{\Wignere[\fve,\fue]}$. Here $\{\mathbf{A},\mathbf{B}\}:= \bnabla _{\kg}\mathbf{A}\cdot \bnabla  _{\xg}\mathbf{B}  -\bnabla  _{\xg}\mathbf{A} \cdot \bnabla _{\kg}\mathbf{B}$ stands for the usual Poisson bracket such that $\adj{\{\mathbf{A},\mathbf{B}\}}=-\{\adj{\mathbf{B}},\adj{\mathbf{A}}\}$.

\subsection{Spatio-temporal Wigner transform}\label{sec:space-time-WT}

In the sequel, the Wigner measure of the solutions of \eref{eq:PDOwave} shall be obtained using a spatio-temporal Wigner transform of that equation and its high-frequency limit as $\escale\rightarrow 0$. This is because the spatial and temporal scales in the wave equation (\ref{eq:PDOwave}) play a symmetric role, and their oscillations should be accounted for altogether. Therefore a larger phase space than the one considered in the definition (\ref{eq:TWigner}) has to be introduced. For two sequences $\smash{\sequence{\fue}}$ and $\smash{\sequence{\fve}}$ of square-integrable functions, the spatio-temporal Wigner transform $\Wignere[\fue,\fve]$ is defined as:
\begin{multline}\label{eq:TWigner-xt}
\Wignere[\fue,\fve](t,\omega,\xg,\kg):= \\
\frac{1}{(2\pi)^4}\int_{\Rset^4}\iexp^{\ci(\yg\cdot\kg+\vinttem\omega)}\fue\left(t-\frac{ \escale \vinttem}{2}, \xg-\frac{\escale \vy}{2}\right) \otimes \cjg{\fve\left(t+\frac{\escale \vinttem}{2},\xg+\frac{\escale \vy}{2}\right)}\,\id\vy  \id\vinttem\,.
\end{multline}
Note that if it is applied to $\fue$ and $\fve\equiv\fue$, $\Wignere[\fue,\fue]$ will be denoted by $\Wignere[\fue]$ as implicitly done in~\eref{eq:mes-semi-classique}. The spatio-temporal Wigner transform (\ref{eq:TWigner-xt}) is related to the (time-dependent) spatial Wigner transform (\ref{eq:TWigner}) by $\Wignere[\fue](t,\xg,\kg)=\int_\Rset\Wignere[\fue](t,\omega,\xg,\kg)\id\omega$, where $\omega$ arises as the dual variable of $t$. Besides, one should observe that the rules expounded in \sref{sec:comp-adj} above can be extended further on to a time-dependent observable $\obsg(t,\omega)$ or a space-time observable $\obsg(t,\omega,\xg,\kg)$. It suffices, for example, to redefine the Poisson bracket in \eref{reg1-cal} as a differential operator with respect to the space-time variable $(t,\xg)\in\Rset^4$ and the impulse variable $(\omega,\kg)\in\Rset^4$: $\{\mathbf{A},\mathbf{B}\}:= \bnabla _{\omega,\kg}\mathbf{A}\cdot \bnabla  _{t,\xg}\mathbf{B}  -\bnabla  _{t,\xg}\mathbf{A} \cdot \bnabla _{\omega,\kg}\mathbf{B}$.


\subsection{Dispersion properties}\label{sec:dispersion}

The pseudo-differential calculus and the spatio-temporal Wigner transform are now used for the wave equation (\ref{eq:PDOwave}). Computing the space-time Wigner transform~\eqref{eq:TWigner-xt} of $\WaveEq(\fue)$ and $\fue$, yields:
\begin{equation*}
\Wignere\left[ \left((\escale\iD_t)^2\II-\TA(\xg,\escale\Dxx)-\frac{\escale}{\ci}\TA_\indL(\xg,\escale\Dxx)\right)\fue,\fue\right] = 0\,.
\end{equation*}
But the partial derivative with respect to time reads:  
\begin{equation}\label{def-der-tem}
\iD_t^{2} f (t):=\left(\frac{\partial_t}{\ci}\right)^2 f (t)=    \int_{\Rset}  \frac{\dd \wg}{2\pi} \iexp^{\ci\wg t}Q(\wg)\TF{f}(\wg)\,,
\end{equation}
where $Q(\wg):=\wg^2$. Thus invoking the rules of calculus of the previous section, we get:
\begin{multline}\label{eq:step0}
\left (\wg+\frac{\escale\iD_t}{2}\right )^2\Wignere[\fue] = \TA(\xg,\kg)\Wignere[\fue] + \frac{\escale}{2\ci}\{\TA,\Wignere[\fue]\}  \\
- \frac{\escale}{2\ci}\bnabla_\xg \cdot \bnabla_\kg\TA(\xg,\kg)\Wignere[\fue] + \frac{\escale}{\ci}\TA_\indL(\xg,\kg)\Wignere[\fue]+\Go(\escale^2)\,.
\end{multline}
Considering the leading-order term, one obtains:
\begin{equation}\label{eq:step01}
(\wg^2\II-\TA(\xg,\kg))\Wigner[\fue]=\bzero\,,
\end{equation}
for the Wigner measure $\Wigner[\fue]$ of the sequence $\smash{\sequence{\fue}}$ given by~\eref{eq:mes-semi-classique}. Owing to the properties~(\ref{eq:TAspectral}) of the Christoffel tensor, one thus has:
\begin{equation}\label{eq:step02}
\sum_{\jeig=1}^\Mode\Hamil_\jeig\proj_\jeig\Wigner[\ug_\escale]=\bzero\quad\text{on}\;\varXset:=\Rset_\wg\times\domain\times\Rset^*_\kg\,,
\end{equation}
where $\Hamil_\jeig(\wg,\xg,\kg)=\wg^2-\eigl_\jeig^2(\xg,\kg)$ and:
\begin{equation*}
\begin{split}
\proj_\jeig(\xg,\kg) &=\eigvec_\jeig(\xg,\kg)\otimes\eigvec_\jeig(\xg,\kg) \\
&=\eigvec_\jeig(\xg,\kg)\adj{\eigvec}_\jeig(\xg,\kg)\,.
\end{split}
\end{equation*}
Here $\Mode\leq 3$ is the number of different eigenvalues of the Christoffel tensor $\TA$, and $\smash{\eigvec_\jeig=\{\eigvec_{\jeig,\mode}\}_{1\leq\mode\leq r_\jeig}}$ is the family of eigenvectors associated to the positive eigenvalue $\eigl_\jeig^2$ of which order of algebraic multiplicity is $r_\jeig$. It is assumed in the remaining that these multiplicities remain constant in phase space. We shall however see in \sref{sec:examples}, following~\cite{BOU98b,NOR04}, that this is not always true for the classes of symmetry considered there and that the eigenvalues of the Christoffel tensor may coincide at some points or lines. The analysis of such crossings in terms of Wigner measures is a difficult task which has been addressed mathematically in~\cite{FER02,FER03a}. An explicit transition coefficient can be obtained in terms of the gap between the eigenvalues thanks to a proper rescaling of the crossing process. However, the extension of the results presented in these works to elasticity is not straightforward and requires further analyses out of the scope of the present publication. 
Going back to~\eref{eq:step02}, the $\Mode$ families of eigenvectors $\eigvec_\jeig$, $1\leq\jeig\leq\Mode$, form an orthonormal basis of $\Rset^3$: $\smash{\eigvec_\jeig^*\eigvec_\keig}=\smash{\eigvec_\jeig\cdot\eigvec_\keig}=\smash{\dir_{\jeig\keig}\II_\jeig}$, where $\smash{\II_\jeig}$ is the $\smash{r_\jeig\times r_\jeig}$ identity matrix and $\smash{\dir_{\jeig\keig}}$ stands for the Kronecker symbol. Then the $\proj_\jeig$'s are projectors, implying that each term in the sum above cancels. Thus:
\begin{displaymath}
\Hamil_\jeig\proj_\jeig\Wigner[\fue]\proj_\keig=\bzero\quad\text{on}\;\varXset\,,\quad\forall\jeig,\keig\,.
\end{displaymath}
Likewise, $\smash{\Wigner[\fue]}$ being hermitian, $\smash{\Hamil_\keig\Wigner[\ug_\escale]\proj_\keig}=\bzero$ on $\varXset$ and consequently one has $\smash{\Hamil_\keig\proj_\jeig\Wigner[\ug_\escale]\proj_\keig}=\bzero$ on $\varXset$ for all $\jeig,\keig$. Taking the difference of these equalities yields:
\begin{displaymath}
(\Hamil_\jeig-\Hamil_\keig)\proj_\jeig\Wigner[\ug_\escale]\proj_\keig=\bzero\quad\text{on}\;\varXset\,,\quad\forall\jeig,\keig\,.
\end{displaymath}
But $\smash{\Hamil_\jeig\neq\Hamil_\keig}$ on $\varXset$ as soon as $\jeig\neq\keig$, and $\smash{\proj_\jeig\Wigner[\ug_\escale]\proj_\keig}=\bzero$ in this case. Expanding $\smash{\Wigner[\ug_\escale]}$ on $\varXset$ on the basis $\smash{\{\proj_\jeig\}_{1\leq\jeig\leq\Mode}}$ such that $\smash{\sum_{\jeig=1}^\Mode\proj_\jeig}=\II$:
\begin{displaymath}
\Wigner[\ug_\escale]=\sum_{\jeig=1}^\Mode\proj_\jeig\Wigner[\fue]=\sum_{\jeig.\keig=1}^\Mode\proj_\jeig\Wigner[\fue]\proj_\keig\,,
\end{displaymath}
the expansion on $\varXset$ finally reduces to the diagonal terms $\jeig=\keig$ solely:
\begin{equation}\label{eq:Wigner-spectral}
\Wigner[\fue]=\sum_{\jeig=1}^\Mode\Wigner_\jeig[\fue]\,,\quad\Wigner_\jeig[\ug_\escale]=\proj_\jeig\Wigner[\fue]\proj_\jeig\,,
\end{equation}
where $\supp\smash{\Wigner_\jeig[\fue]}\subset\smash{\{(\wg,\xg,\kg)\in\varXset;\,\Hamil_\jeig(\wg,\xg,\kg)=0\}}$. So the Wigner measure $\Wigner[\ug_\escale]$ of the $\escale$-oscillating elastic wave fields $\smash{\sequence{\fue}}$ is expanded into the finite sum of its orthogonal projections onto the different energy paths of the propagation operator with symbol $\smash{\wg^2\II-\TA(\xg,\kg)}$. These paths are determined by the equation $\smash{\Hamil_\jeig(\wg,\xg,\kg)}=0$, $1\leq\jeig\leq\Mode$, in phase space. They correspond to the rays for the medium arising in classical Hamiltonian dynamics, \textcolor{\mycolor}{as shown in the subsequent section}. 

\subsection{Evolution properties}\label{sec:evolution}

On the other hand, considering the Wigner transform of the adjoint wave equation $\adj{\WaveEq}(\fue)$ and $\fue$ we have:
\begin{equation*}
\Wignere\left[\fue,\left((\escale\iD_t)^2\II-\TA(\xg,\escale\Dxx)-\frac{\escale}{\ci}\TA_\indL(\xg,\escale\Dxx)\right)\fue\right] = 0\,,
\end{equation*}
since the wave operator is actually formally self-adjoint, $\adj{\WaveEq}\equiv\WaveEq$. This yields:
\begin{multline}\label{eq:step0-star}
\left (\wg-\frac{\escale\iD_t}{2}\right )^2\Wignere[\fue] = \Wignere[\fue]\TA(\xg,\kg ) + \frac{\escale}{2\ci}\{\Wignere[\fue],\TA\} \\
+ \frac{\escale}{2\ci}\Wignere[\fue]\bnabla_\xg \cdot \bnabla_\kg\TA(\xg,\kg) - \frac{\escale}{\ci}\Wignere[\fue]\adj{\TA}_\indL(\xg,\kg)+\Go(\escale^2)\,,
\end{multline}
or, upon substracting \eref{eq:step0} and \eref{eq:step0-star}, multiplying by $\smash{\frac{\ci}{\escale}}$, and observing that $\smash{\adj{\TA}_\indL}=\smash{\bnabla_\xg \cdot \bnabla_\kg\TA-\TA_\indL}$:
\begin{multline}\label{eq:Wigner-equation}
2\wg\partial_t\Wignere[\fue] = \frac{\ci}{\escale}\Big(\TA\Wignere[\fue]-\Wignere[\fue]\TA\Big) + \demi\Big(\{\TA,\Wignere[\fue]\}-\{\Wignere[\fue],\TA\}\Big) \\
+\Big[\TA_\indL-\demi\bnabla_\xg \cdot \bnabla_\kg\TA,\Wignere[\fue]\Big]+\Go(\escale)\,.
\end{multline}
Here $[\mathbf{A},\mathbf{B}]:= \mathbf{A} \mathbf{B}-\mathbf{B}\mathbf{A}$ stands for the Lie bracket. One should observe that the matrix $\matM:=\smash{\TA_\indL-\demi\bnabla_\xg \cdot \bnabla_\kg\TA}$ above is skew-symmetric. Following~\cite{RYZ96}, we then introduce the $\smash{r_\jeig\times r_\jeig}$ matrices $\smash{\speciv_\jeig}=\smash{\adj{\eigvec}_\jeig\Wigner[\fue]\eigvec_\jeig}$ for $1\leq\jeig\leq\Mode$ such that $\smash{\Wigner[\fue]}=\smash{\sum_{\jeig=1}^\Mode\eigvec_\jeig\speciv_\jeig\adj{\eigvec}_\jeig}$ with $\supp\smash{\Wigner_\jeig[\fue]}\subset\smash{\{(\wg,\xg,\kg)\in\varXset;\,\Hamil_\jeig(\wg,\xg,\kg)=0\}}$, and compute the projection of~\eref{eq:Wigner-equation} on the eigen directions $\smash{\eigvec_\jeig}$. Thus multiplying~\eref{eq:Wigner-equation} by $\adj{\eigvec}_\jeig$ on the left side and by $\eigvec_\jeig$ on the right side, one first obtains:
\begin{displaymath}
\adj{\eigvec}_\jeig\matM\eigvec_\jeig=\adj{\eigvec}_\jeig\TA_\indL\eigvec_\jeig+\matM_\jeig^\isym-\demi(\bnabla_\xg\cdot\bnabla_\kg\eigl_\jeig^2)\II_\jeig\,,
\end{displaymath}
where $\smash{\matM_\jeig}=\smash{\bnabla_\kg\adj{\eigvec}_\jeig(\eigl_\jeig^2\II-\TA)\cdot\bnabla_\xg\eigvec_\jeig}$. Indeed, the normalization condition $\adj{\eigvec}_\jeig\eigvec_\keig=\dir_{\jeig\keig}\II_\jeig$ yields $(\partial_{\kgj_j}\adj{\eigvec}_\jeig)\eigvec_\keig=-\adj{\eigvec}_\jeig(\partial_{\kgj_j}\eigvec_\keig)$ and a similar relation for the partial derivatives $\partial_{\xgj_j}$. We then use this relation and its derivatives recursively in the computation of the projection of $\smash{\bnabla_\xg\cdot\bnabla_\kg\TA}$. Secondly, we consider the projection of the Poisson's brackets in~\eref{eq:Wigner-equation}. It is derived from the following identity:
\begin{equation*}
\begin{split}
\adj{\eigvec}_\jeig(\partial_{\kgj_j}\TA) &=\partial_{\kgj_j}(\adj{\eigvec}_\jeig\TA)-(\partial_{\kgj_j}\adj{\eigvec}_\jeig)\TA \\
&=\partial_{\kgj_j}(\eigl_\jeig^2\adj{\eigvec}_\jeig)-(\partial_{\kgj_j}\adj{\eigvec}_\jeig)\TA \\
&=(\partial_{\kgj_j}\eigl_\jeig^2)\adj{\eigvec}_\jeig+(\partial_{\kgj_j}\adj{\eigvec}_\jeig)(\eigl_\jeig^2\II-\TA)\,,
\end{split}
\end{equation*}
and a similar result for the partial derivatives $\adj{\eigvec}_\jeig(\partial_{\xgj_j}\TA)$. Likewise:
\begin{equation*}
\begin{split}
(\partial_{\xgj_j}\Wigner[\fue])\eigvec_\jeig &=\partial_{\xgj_j}(\Wigner[\fue]\eigvec_\jeig)-\Wigner[\fue]\partial_{\xgj_j}\eigvec_\jeig \\
&=\partial_{\xgj_j}(\eigvec_\jeig\speciv_\jeig)-\Wigner[\fue]\partial_{\xgj_j}\eigvec_\jeig \\
&=\eigvec_\jeig\partial_{\xgj_j}\speciv_\jeig+(\partial_{\xgj_j}\eigvec_\jeig)\speciv_\jeig-\Wigner[\fue]\partial_{\xgj_j}\eigvec_\jeig\,,
\end{split}
\end{equation*}
and a similar result for the partial derivatives $(\partial_{\kgj_j}\Wigner[\fue])\eigvec_\jeig$. Therefore one has:
\begin{equation*}
\begin{split}
\adj{\eigvec}_\jeig\{\TA,\Wigner[\fue]\}\eigvec_\jeig &=\{\eigl_\jeig^2,\speciv_\jeig\}+\adj{\eigvec}_\jeig\{\eigl_\jeig^2,\eigvec_\jeig\}\speciv_\jeig+\speciv_\jeig\{\eigl_\jeig^2,\adj{\eigvec}_\jeig\}\eigvec_\jeig+2\matM_\jeig^\iasym\speciv_\jeig\,, \\
\adj{\eigvec}_\jeig\{\Wigner[\fue],\TA\}\eigvec_\jeig &=\{\speciv_\jeig,\eigl_\jeig^2\}+\speciv_\jeig\adj{\eigvec}_\jeig\{\eigl_\jeig^2,\eigvec_\jeig\}+\{\eigl_\jeig^2,\adj{\eigvec}_\jeig\}\eigvec_\jeig\speciv_\jeig+2\speciv_\jeig\matM_\jeig^\iasym\,,
\end{split} 
\end{equation*}
where ${\bf A}^\iasym$ stands for the skew-symmetric part of a square matrix ${\bf A}$. Here one should observe that $\adj{\eigvec}_\jeig\{\eigl_\jeig^2,\eigvec_\jeig\}=-\{\eigl_\jeig^2,\adj{\eigvec}_\jeig\}\eigvec_\jeig$ owing to the normalization condition. Now combining all these results in~\eref{eq:Wigner-equation} and passing to the limit $\escale\rightarrow 0$ one obtains the transport equations:
\begin{equation}\label{eq:Wigner-equationb}
2\wg\partial_t\speciv_\jeig = \{\eigl_\jeig^2,\speciv_\jeig\}+[\matN_\jeig,\speciv_\jeig]\,,\quad1\leq\jeig\leq\Mode\,,
\end{equation}
where $\smash{\matN_\jeig}:=\smash{\matM_\jeig+\adj{\eigvec}_\jeig\{\eigl_\jeig^2,\eigvec_\jeig\}+\adj{\eigvec}_\jeig\TA_\indL\eigvec_\jeig}$ is skew-symmetric. Note that the matrix $\smash{\matN_\jeig}$ vanishes in an homogeneous medium, and that the Lie bracket in~\eref{eq:Wigner-equationb} does so whenever $r_\jeig=1$.

\textcolor{\mycolor}{Now we show how the transport equations above localize the energy on rays as described by classical Hamiltonian dynamics. Indeed, introducing the following system of Hamiltonian equations:
\begin{equation}\label{eq:syst-Hamiltonian}
\begin{array}{rlrl}
\displaystyle\frac{\id\xg}{\id\tau} &\!\!\!\!=\bnabla_\kg\Hamil_\jeig(\wg(\tau),\xg(\tau),\kg(\tau))\,,\quad&\displaystyle\frac{\id t}{\id\tau} &\!\!\!\!=\bnabla_\wg\Hamil_\jeig(\wg(\tau),\xg(\tau),\kg(\tau))\,, \\
\displaystyle\frac{\id\kg}{\id\tau} &\!\!\!\!=-\bnabla_\xg\Hamil_\jeig(\wg(\tau),\xg(\tau),\kg(\tau))\,,\quad&\displaystyle\frac{\id\wg}{\id\tau} &\!\!\!\!=-\bnabla_t\Hamil_\jeig(\wg(\tau),\xg(\tau),\kg(\tau))=0\,, \\
\end{array}
\end{equation}
with initial conditions satisfying $\Hamil_\jeig(\wg(0),\xg(0),\kg(0))=0$ and $t(0)=0$, then its solutions $\tau\mapsto\smash{(t(\tau)=2\wg\tau,\wg(\tau)=\wg,\xg_\jeig(\tau),\kg_\jeig(\tau))}$ are the so-called null bicharacteristics such that $\smash{\Hamil_\jeig(\wg,\xg_\jeig,\kg_\jeig)}$ remains constant (and null) since one observes by a straightforward application of the chain rule that $\smash{\frac{\id\Hamil_\jeig}{\id\tau}=0}$. Thus the energy rays supporting the Wigner measures $\smash{\Wigner_\jeig[\fue]}$ may be constructed by solving the ordinary differential equations~(\ref{eq:syst-Hamiltonian}) (provided that the usual conditions for the local existence, uniqueness and smoothness of its solutions with respect to the initial conditions are fulfilled, see \emph{e.g.}~\cite{HAL80}. This issue is however much beyond the scope of this paper)}. From the Hamiltonian system~(\ref{eq:syst-Hamiltonian}) and the definition of $\Hamil_\jeig$ one can notice that the above equations read:
\begin{equation*}
\frac{\id}{\id\tau}\speciv_\jeig(t(\tau),\wg,\xg_\jeig(\tau),\kg_\jeig(\tau))=[\matN_\jeig,\speciv_\jeig]\,,\quad1\leq\jeig\leq\Mode\,,
\end{equation*}
with $t=2\wg\tau$. As in~\cite[Remark~6.2]{GER97}, if one introduces the matrix ${\bf U}_\jeig$ satisfying:
\begin{equation*}
\frac{\id{\bf U}_\jeig}{\id\tau}=\matN_\jeig{\bf U}_\jeig\,,\quad{\bf U}_\jeig(\tau=0)=\II_\jeig\,,
\end{equation*}
and $\speciv_\jeig={\bf U}_\jeig\tilde{\speciv}_\jeig\adj{{\bf U}}_\jeig$, then~\eref{eq:Wigner-equationb} reduces to:
\begin{equation}
\frac{\id}{\id\tau}\tilde{\speciv}_\jeig(t(\tau),\wg,\xg_\jeig(\tau),\kg_\jeig(\tau))=\bzero\,,\quad1\leq\jeig\leq\Mode\,.
\end{equation}

We finally conclude this section by observing that from~\eref{eq:syst-Hamiltonian} we further obtain that $\smash{\frac{\id\xg_\jeig}{\id t}}=\smash{\mp\vcel_\jeig}$ whenever $\wg=\smash{\pm\eigl_\jeig}$, where $\smash{\vcel_\jeig}=\smash{\bnabla_\kg\eigl_\jeig}$ is the group velocity for the mode $\jeig$. This shows that the specific intensity $\smash{\speciv_\jeig}$ propagates in the "forward" direction $\smash{\hat{\vcel}_\jeig}$ on the energy path $\smash{\xg_\jeig}$ for $\smash{\wg=-\eigl_\jeig}$, and in the "backward" direction $-\smash{\hat{\vcel}_\jeig}$ for $\smash{\wg=\eigl_\jeig}$. Formally writing it $\smash{\speciv_\jeig}=\smash{\sa_\jeig^+\dir(\wg+\eigl_\jeig)+\sa_\jeig^-\dir(\wg-\eigl_\jeig)}$ for the forward and backward traveling components, the so-called specific intensities $\sa_\jeig^\pm$, respectively, the transport equations~\eqref{eq:Wigner-equationb} yield:
\begin{equation}\label{eq:Liouville-equation}
\partial_t\sa_\jeig^\pm\pm\{\eigl_\jeig,\sa_\jeig^\pm\}+[\matN_\jeig^\pm,\sa_\jeig^\pm]=\bzero\,,\quad1\leq\jeig\leq\Mode\,,
\end{equation}
where $\smash{\pm2\eigl_\jeig\matN_\jeig^\pm=\matN_\jeig}$. The Liouville transport equations (\ref{eq:Liouville-equation}) above generalize Eqs.~(3.99) and (3.100) of \cite{RYZ96} to an arbitrary anisotropy of the elastic medium. The isotropic case considered in this latter publication is recovered as briefly explained below.

\subsection{The isotropic case}\label{sec:Wigner-isotropic}

For isotropic elasticity for example, $\Mode=2$ with $\jeig=\iP$ or $\jeig=\iS$ and $r_\iP=1$, $r_\iS=2$. The eigenvalues of the Christoffel tensor are $\smash{\eigl_\iP(\xg,\kg)}=\smash{\cel_\iP(\xg)|\kg|}$ and $\smash{\eigl_\iS(\xg,\kg)}=\smash{\cel_\iS(\xg)|\kg|}$, with the velocities $\cel_\iP$ and $\cel_\iS$ for the compressional and shear waves are as in~\eref{eq:TA-isotropy}. Then $\smash{\eigvec_\iP(\xg,\kg)=\hkg}$ and $\smash{\eigvec_\iS(\xg,\kg)=[\hzg_1(\kg),\hzg_2(\kg)]}$ such that $\smash{(\hkg,\hzg_1,\hzg_2)}$ forms an orthonormal triplet, and the projectors are $\smash{\proj_\iP}=\smash{\hkg\otimes\hkg}$ and $\smash{\proj_\iS}=\smash{\II-\hkg\otimes\hkg}$. The Wigner measure is then expanded as: 
\begin{equation}
\Wigner[\ug_\escale] = \specij_\iP\eigvec_\iP\adj{\eigvec}_\iP+\eigvec_\iS\speciv_\iS\adj{\eigvec}_\iS\,,
\end{equation}
where $\specij_\iP$ is a scalar and $\speciv_\iS$ is a $2\times 2$ matrix. Thus the multiply-scattered wave energy in an elastic medium may be characterized by five parameters, four for the transverse waves and one for the longitudinal wave. They correspond to the elastic Stokes parameters introduced in \cite{TUR94a,TUR94b}. Other particular anisotropies shall be described later on in the \sref{sec:examples}.

\section{Radiative transport equations}\label{sec:RTE}

We now turn to the weak coupling regime of high-frequency waves in a random anisotropic medium. The weak coupling regime denotes the situation whereby (i) propagation distances are large compared to the wavelength $\escale$, and (ii) the perturbations of the elasticity tensor of the background medium are weak and vary at the same scales as the wavelength (meaning that their correlation lengths scale as $\escale$). The subsequent analysis is derived from~\cite{BAL05} where scalar (acoustic) waves are considered, and~\cite{BRA02} where general first-order anti-selfadjoint systems are considered. The main result of this section is the (matrix) radiative transfer equation (\ref{eq:RTE-anisotropic}) which couples all wave polarizations in an arbitrarily anisotropic elastic medium. Radiative transfer equations are linear Boltzmann equations which describe the kinetics of particles in a lattice of randomly distributed inclusions, for example. Thus high-frequency wave propagation phenomena may be very well understood in terms of a gas kinetics analogy. It involves collisional processes characterized in terms of differential and total scattering cross-sections, of which expressions are precisely given by \eref{eq:dscat} and \eref{eq:tscat}, respectively, for the present case of arbitrarily anisotropic random elastic media. The different steps for this derivation are the following. The mathematical form chosen for modeling such inhomogeneities is first given in the next \sref{sec:C1}. The random perturbations of the elasticity tensor of the bare anisotropic medium considered in the previous part are assumed to vary at the fast scale $\smash{\frac{\xg}{\escale}}$ as opposed to the slow scale of variation $\xg$ of the latter. Therefore one has to introduce a two-scale expansion of the Wigner transform of the wave fields in this situation (\sref{sec:multiple-scale-expansion}), and a dedicated rule of pseudo-differential calculus accounting for both scales and generalizing those of the previous part (\sref{pse-dif-cal}). A major consequence of this separation of scales and of the scaling of the amplitudes of the random inhomogeneities in \sref{sec:C1} is that the fast scale does not modify the spectral (dispersion) properties of the Wigner measure already derived in \sref{sec:dispersion}. \sref{dis-rel-sec} shows why this property holds. However, the fast scale modifies the next-order contribution to the Wigner transform and consequently the evolution properties of the Wigner measure. The contribution of the fast scale of variations of the random inhomogeneities to the two-scale expansion of the Wigner transform is given explicitly in \sref{fir-ord-cor-sec}. This correction actually gives rise to the collision operator characterizing the multiple scattering process of high-frequency waves on the random inhomogeneities. It is therefore responsible for the modification of the transport equations of \sref{sec:evolution} for the bare elastic medium into radiative transfer equations for the randomly perturbed elastic medium. The final \sref{second-corre-sec} outlines how this modification arises.

\subsection{Elasticity tensor of a randomly perturbed anisotropic medium}\label{sec:C1}
 
In the setting invoked above it is assumed that the elasticity tensor now reads:
\begin{eqnarray}\label{ela-ten-fluc}
\tenselas(\xg)=  \tenselas_0 \left(\xg\right)+\sqrt{ \escale}\tenselas_1\left(\frac{\xg}{\escale}\right) ,
\end{eqnarray}
where $\smash{\tenselas_0}$ is the elasticity tensor of the anisotropic background medium, and $\smash{\tenselas_1}$ is its fluctuation with amplitude $\sqrt{\escale}$. This fluctuation is modeled by a tensor-valued, second-order stochastic field:
\begin{displaymath}
\left\{\tenselas_1(\vy)\,;\,\vy\in\Rset^3\right \},
\end{displaymath}
which has mean zero and is mean square homogeneous (stationary). The latter property means that the cross-correlations of the perturbations at two different locations $\vy_1$ and $\vy_2$ depend on $\vy_1-\vy_2$ solely; it is referred to as anisomery in the geophysical literature (see~\cite{MAR06}). If the cross-correlations depend on $|\vy_1-\vy_2|$ the medium is statistically isotropic, but this does not preclude it from being anisotropic. At last, the inhomogeneities are small as expressed by their $\smash{\Go(\escale^\demi)}$ amplitude. This size is the unique scaling which allows them to significantly modify the energy spreading in the transport regime at long propagation distances; see \emph{e.g.} \cite{RYZ96,BAL05}. Larger fluctuations could lead to localization of the waves, a situation beyond the scope of kinetic models. The fluctuation tensor $\tenselas_1$ introduced in \eqref{ela-ten-fluc} does not necessarily have the same symmetry as the mean elasticity tensor, as exemplified in \emph{e.g.}~\cite{STA84,HIR85,TUR99}. Thus it depends on twenty one coefficients $\smash{\{\coefC_\inda\}_{1\leqslant\inda\leqslant 21}}$ in the general case. The model of correlation between these coefficients is given as follows:
\begin{equation}\label{def-aver}
\mathbb{E}\left\{\smash{\TF{\coefC}_\inda(\vq)\TF{\coefC}_\indb(\vp)}\right\}:= (2\pi)^{3}\dir(\vq+\vp) \TF{\coro}_{\inda\indb}(\vq)\,,
\end{equation}
where the correlation function is $\smash{\coro_{\inda\indb}(\vy_1-\vy_2) := \mathbb{E}\{\coefC_{\inda}(\vy_1) \coefC_{\indb}(\vy_2)\}}$ and:
\begin{equation*}
\TF{\coro}_{\inda\indb}(\vq) := \int_{\Rset^{3}} \frac{\dd\vy}{(2\pi)^{3}}\iexp^{\ci\vy\cdot\vq}\coro_{\inda\indb}(\vy)\,.
\end{equation*}
In the above $\mathbb{E}\{\cdot\}$ stands for the mathematical expectation (average). We stress that the phase function $\vq\mapsto\smash{\TF{\coro}_{\inda\indb}(\vq)}$ is even, such that $\smash{\TF{\coro}_{\inda\indb}(-\vq)}=\smash{\TF{\coro}_{\inda\indb}(\vq)}$.

We note at this stage that a randomly perturbed density may be accounted for as well in the subsequent developments. However we ignore that possibility in the remaining of the paper for clarity purposes. The analysis could be carried on, though, along the same lines as in~\cite[Sect. 7]{BAL05}.

\subsection{Multiple scale expansion of the Wigner transform of the elastic wave equation}\label{sec:multiple-scale-expansion}

Having introduced the random fluctuations of the elasticity tensor, we can write the elastic wave equation accounting for these inhomogeneities in a similar form of \eref{eq:PDOwave} as follows. Let us introduce the Christoffel tensor $\smash{\TA_0}$ of the slowly varying background, such that $\smash{\TA_0(\xg,\kg)}=\smash{\roi^{-1}(\xg)\adj{\vM}(\kg)\vC_0(\xg)\vM(\kg)}$. We also introduce the second order tensor $\TA_1$ corresponding to the random fluctuations $\smash{\TA_1(\xg,\yg,\kg)}:=\smash{\roi^{-1}(\xg)\adj{\vM}(\kg)\vC_1(\yg)\vM(\kg)}$. The elastic wave equation (\ref{eq:PDOwave}) is now considered with the operator $\vL_\escale$ defined by:
\begin{equation}\label{eq:PDO-L-RTE}
\vL_\escale=\vL_0+\escale^\demi\vL_1+\frac{\escale}{\ci}\vL_\indLT+\Go(\escale^\frac{3}{2})\,,
\end{equation}
where:
\begin{equation}\label{eq:L-operators}
\begin{split}
\vL_0(\xg,\kg) &=\roi(\xg)\TA_0(\xg,\kg)\,,\\
\vL_1\left(\xg,\frac{\xg}{\escale},\kg\right) &=\roi(\xg)\TA_1\left(\xg,\frac{\xg}{\escale},\kg\right)\,,\\
\vL_\indLT(\xg,\kg) &=\roi(\xg)\TA_\indLT(\xg,\kg)=\bnabla_\kg\adj{\vM}(\kg)\cdot\bnabla_\xg\vC_0(\xg)\vM(\kg)\,.
\end{split}
\end{equation}
Then by applying the space-time Wigner transforms $\smash{\Wignere[\cdot,\fue]}$ and $\smash{\Wignere[\fue,\cdot]}$  to \eref{eq:PDOwave} with $(\escale\iD_t)^{2}\fue = Q(\escale\iD_t) \fue$ of~\eref{def-der-tem} and $\vL_\escale$ given by~\eref{eq:PDO-L-RTE}, we obtain respectively:
\begin{multline}\label{sys-wig-dr-ga}
\Wignere[Q(\escale \iD_t) \fue,\fue]=  \Wignere[\TA_0( \xg,\escale \Dxx) \fue,\fue] 
+ \sqrt{\escale} \Wignere\left[\TA_1\left(\xg,\frac{\xg}{\escale},\escale \Dxx\right)\fue,\fue\right] \\
+ \frac{\escale}{\ci}  \Wignere[\TA_\indLT( \xg,\escale\Dxx) \fue,\fue] +\Go(\escale^\frac{3}{2})\,,
\end{multline}
and:
\begin{multline}\label{sys-wig-dr-ga-adj}
 \Wignere[\fue, Q(\escale \iD_t) \fue] =  \Wignere[\fue,\TA_0( \xg,\escale \Dxx) \fue]+\sqrt{\escale} \Wignere\left[\fue , \TA_1\left( \xg,\frac{\xg}{\escale},\escale \Dxx\right)\fue\right]  \\
- \frac{\escale}{\ci}\Wignere[\fue,\TA_\indLT( \xg,\escale \Dxx) \fue]+\Go(\escale^\frac{3}{2})\,.
\end{multline}
Taking the difference of \eref{sys-wig-dr-ga} and \eref{sys-wig-dr-ga-adj} and recalling the rule (\ref{reg3-cal}) for $Q (\escale \iD_t)$, yields:
\begin{multline}\label{firs-equ-subS}
2 \omega (\escale\iD_t)\Wignere[\fue]  =  \Wignere[  \TA_0 ( \xg,\escale\Dxx) \fue,\fue ] -  \Wignere[\fue , \TA_0 ( \xg,\escale\Dxx) \fue    ]  \\
+ \sqrt{\escale} \Wignere\left[  \TA_1\left( \xg,\frac{\xg}{\escale},\escale  \Dxx\right)    \fue,\fue\right]
- \sqrt{\escale} \Wignere\left[ \fue,  \TA_1 \left( \xg,,\frac{\xg}{\escale},\escale\Dxx\right)\fue\right] \\
+ \frac{\escale}{\ci} \Wignere[ \TA_\indLT ( \xg,\escale \Dxx) \fue,\fue ]  + \frac{\escale}{\ci}  \Wignere[\fue , \TA_\indLT ( \xg,\escale \Dxx) \fue ] +\Go(\escale^\frac{3}{2})\,.          
\end{multline}
\eref{firs-equ-subS} for the case of a randomly inhomogeneous medium is the counterpart of the Wigner equation (\ref{eq:Wigner-equation}) for a slowly varying medium, before the rules (\ref{reg1-cal}) and (\ref{reg2-cal}) introduced in the~\sref{sec:comp-adj} are applied. The main difference lies in the terms involving $\smash{ \TA_1}$, which must be carefully evaluated in an asymptotic analysis since they contain both scales $\xg$ and $\smash{\frac{\xg}{\escale}}$. We may then make use of the aforementioned rules of calculus, and an additional one for pseudo-differential calculus with oscillating coefficients (see~\sref{pse-dif-cal} below). 
In view of these considerations, we also introduce a two-scale version of $\smash{\Wignere[\fue]}$ as follows:
\begin{equation*}
\Wignere[\fue](t,\omega, \xg,\kg)=\tilde{\Wigner}_\escale\left(t,\omega,\xg,\frac{ \xg}{\escale},\kg\right)\,.
\end{equation*}
Moreover, letting $\vy:=\smash{\frac{\xg}{\escale}}$, the differential operator $\smash{\Dxx}$ acting on the spatial variables should be replaced by $\smash{\Dxx+\frac{1}{\escale}\Dxy}$ in \eref{firs-equ-subS}, such that the previous asymptotics in this new set of variables can now account for the fast oscillations of the medium. \eref{firs-equ-subS} thus reads:
\begin{multline}\label{second-equ}
2 \omega(\escale \iD_t)\tilde{\Wigner}_\escale = \Wignere[ \TA_0( \xg,\escale\Dxx +\Dxy) \fue,\fue] -\Wignere[\fue, \TA_0( \xg,\escale\Dxx+\Dxy) \fue]  \\
+ \sqrt{\escale}  \Big( \Wignere[ \TA_1( \xg,\vy,\escale\Dxx+\Dxy)\fue,\fue]-\Wignere[ \fue,  \TA_1(\xg,\vy,\escale\Dxx+\Dxy)\fue]\Big) \\
+\frac{\escale}{\ci}   \Big(\Wignere[\TA_\indLT( \xg,\escale\Dxx+\Dxy) \fue,\fue] + \Wignere[\fue , \TA_\indLT(\xg,\escale\Dxx+\Dxy)\fue]\Big)+\Go(\escale^\frac{3}{2}) \,.                 
\end{multline}
Using finally an asymptotic expansion of $\smash{\tilde{\Wigner}_\escale}(t,\omega, \xg, \vy,\kg)$ as:
\begin{multline}\label{asym-expan}
\tilde{\Wigner}_\escale(t,\omega, \xg, \vy,\kg) \\
= \Wigner_{0}(t,\omega, \xg,\kg) + \sqrt{\escale} \Wigner_{1}(t,\omega, \xg, \vy,\kg)    + \escale  \Wigner_{2}(t,\omega, \xg, \vy,\kg) + \po(\escale)\,,
\end{multline}
we equate like-powers of $\escale$ in \eref{second-equ} to obtain a sequence of three equations for the orders $\smash{\Go(\escale^0)}$, $\smash{\Go(\escale^\demi)}$ and $\smash{\Go(\escale)}$. This procedure follows~\cite[Sect.~7]{BAL05} for the scalar case but is extended in the following to a vector wave equation with matrix coefficients. Thus we can follow the analysis developed in the~\sref{sec:Transport} to account for the influence of the random perturbations characterized by the operator $\smash{\vL_1}$ above. The $\smash{\Go(\escale^0)}$ terms yield the dispersion properties of $\smash{\Wigner_0}$ as in \sref{sec:dispersion}, while the $\smash{\Go(\escale)}$ terms yield its evolution properties as in~\sref{sec:evolution}. The $\smash{\Go(\escale^\demi)}$ terms yield a linear relation between $\smash{\Wigner_1}$ and $\smash{\Wigner_0}$ that explicit the contribution of the random inhomogeneities on the evolution properties of the latter. To obtain it, a formal rule of pseudo-differential calculus with oscillating coefficients is needed, as already noticed above. It is given in the next section.

\subsection{A rule of pseudo-differential calculus with oscillating coefficients} \label{pse-dif-cal}

Let $\mathbf{ V}(\xg,\vy )$ be a (real) matrix-valued function. Then, we have that~\cite{BAL05}: 
\begin{equation*}\label{reg4-cal}
\begin{split}
\Wigner_{\escale}\left[ \mathbf{ V}\left(\xg,\frac{\xg}{\escale} \right) \fu_\escale ,\fv_\escale \right] &=  
\int_{\Rset^{3}} \frac{\dd \vp}{(2\pi)^{3}}  \iexp^{\NCi \frac{\xg}{\escale} \cdot\vp} \TF{ \mathbf{ V} } (\xg,\vp)  \Wigner_{\escale}[ \fu_\escale,\fv_\escale ]    \left(\xg,\kg- \frac{\vp}{2}\right) + \Go(\escale)\,,   \\  
\Wigner_{\escale}\left[  \fu_\escale , \mathbf{V}\left(\xg,\frac{\xg}{\escale} \right)\fv_\escale \right] &=  
\int_{\Rset^{3}} \frac{\dd \vp}{(2\pi)^{3}}  \iexp^{\NCi \frac{\xg}{\escale} \cdot\vp} \Wigner_{\escale}[ \fu_\escale,\fv_\escale ]    \left(\xg,\kg+ \frac{\vp}{2}\right)\TF{\mathbf{V}}^* (\xg,\vp) + \Go(\escale)\,,
\end{split}
\end{equation*}
where $ \TF{ \mathbf{ V} } (\xg,\vp)$ is the component-wise Fourier transform of $\mathbf{ V}(\xg,\vy)$ with respect to the second variable. In the above we have dropped the dependency of $\Wignere[ \fue,\fve]$ with respect to $t$ and $\wg$ for clarity purposes. Applying the above formula for highly oscillatory fluctuations in random media with $\yg\equiv\frac{\xg}{\escale}$ yields:
\begin{multline}\label{reg5-cal}
\Wignere[\TA_1\left( \xg,\yg, \escale\Dxx \right)\fue,\fve](\xg,\vy,\kg)= \\
\int_{\Rset^3} \frac{\dd \vp}{(2\pi)^{3}}\iexp^{\ci\yg\cdot\vp}\SyH\left(\xg,\kg+\frac{\Dxy}{2},\vp,\kg-\frac{\vp}{2}+\frac{\Dxy}{2}\right)\Wignere[ \fue,\fve] \left(\xg,\vy,\kg-\frac{\vp}{2}\right) + \Go(\escale) \,,
\end{multline}
and:
\begin{multline}\label{reg5-cal-adj}
\Wignere[\fue,\TA_1\left( \xg,\yg, \escale\Dxx \right)\fve](\xg,\vy,\kg)= \\
\int_{\Rset^3} \frac{\dd \vp}{(2\pi)^{3}}\iexp^{\ci\yg\cdot\vp}\Wignere[ \fue,\fve] \left(\xg,\vy,\kg+\frac{\vp}{2}\right)\SyH\left(\xg,\kg+\frac{\vp}{2}-\frac{\Dxy}{2},\vp,\kg-\frac{\Dxy}{2}\right) + \Go(\escale) \,,
\end{multline}
where the $3\times 3$ matrix $\SyH$ is defined by:   
\begin{equation}\label{symb-H}
\SyH(\xg, \kg, \vp,\vq ):=\roi^{-1}(\xg) \adj{\vM}( \kg)  \TF{\vC}_1(\vp) \vM ( \vq)\,. 
\end{equation}
Note that it is useful for the sequel to observe that from (\ref{symb-H}) this matrix verifies the following property: 
\begin{equation}\label{decomp-H-mat-conj}
\SyH(\xg,  \kg, \vp,\vq )= \adj{\SyH}(\xg,  \vq, \vp,\kg )\,.
\end{equation}    
These formulas will be used in the subsequent derivation of the evolution properties of the Wigner measure accounting for a randomly perturbed elasticity tensor. Similar results can be established if one also considers random perturbations of the material density at the small lengthscale $\escale$.






\subsection{Dispersion properties}\label{dis-rel-sec}

We start by establishing the connection between the temporal and spatial oscillations of the waves in the high-frequency limit, the so called dispersion relation. It is given by the leading order terms $\smash{\Go(\escale^0)}$ in~\eref{second-equ}. Since the symbol of the operator $\vL_0$ of \eref{eq:L-operators} is identical with the Christoffel tensor (\ref{eq:TA}) for the case $ \tenselas\equiv\tenselas_0$ up to $\roi^{-1}$, we adopt the same notations for the eigenvectors and eigenvalues of $\TA _0(\xg, \kg)$ as in~\eref{eq:TAspectral}. Thus we denote the latter by $\smash{\{\eigvec_\indi(\xg,\kg)\}_{1\leq\indi\leq\Mode}}$ and $\smash{\{ \eigl_\indi(\xg,\kg)\}_{1\leq\indi\leq\Mode}}$, respectively, with $\smash{r_\indi}$ the order of multiplicity of the eigenvalue $\smash{\eigl_\indi}$. Then:
\begin{equation}\label{disper-mat-from-chri-eig-vec}
\TA_0(\xg, \kg)  = \sum_{\indi=1}^\Mode \eigv^2_{\indi}(\xg,\kg )\eigvec _{\indi}(\xg,\kg)\adj{\eigvec} _{\indi}(\xg,\kg )\,,
\end{equation} 
with $\adj{\eigvec}_{\indi}(\xg,\kg)\eigvec_{\indj}(\xg,\kg)=\dir_{\indi \indj}\II_\indi$ (the family $\smash{\{\eigvec_{\indi}(\xg,\kg)\}_{1\leq\indi\leq\Mode}}$ forms an orthonormal basis of $\Rset^{3}$ for any $(\xg,\kg)\in\domain\times\Rset^3$). Consequently, any $3\times 3$ real matrix ${\bf A}(\xg,\kg)$ can be expanded on this new basis by a spectral decomposition as follows:
\begin{equation*}\label{matrice-base}
{\bf A}(\xg,\kg)=\sum_{\indi,\indj=1}^\Mode\eigvec_{\indi} (\xg,\kg){\bf A}_{\indi\indj} (\xg,\kg)\adj{\eigvec}_\indj(\xg,\kg)\,,
\end{equation*}
where $\smash{{\bf A}_{\indi \indj} (\xg,\kg)}=\smash{\adj{\eigvec}_{\indi} (\xg,\kg){\bf A}(\xg,\kg)\eigvec_{\indj} (\xg,\kg)}$. Now the matrix-valued Wigner measure $\Wigner_{0}(t,\omega,\xg ,\kg) $ being Hermitian and positive definite, it is expanded as:
\begin{equation}\label{deve-w}
\Wigner_{0}(t,\omega,\xg ,\kg)  = \sum_{\indi=1}^\Mode\eigvec_\indi(\xg,\kg  )\speciv_\indi(t,\omega,\xg,\kg)\adj{\eigvec}_\indi(\xg,\kg)\,.   
\end{equation}
On the other hand, letting $\escale\rightarrow 0$ in \eref{sys-wig-dr-ga}, and invoking the rule \eqref{reg1-cal} and the first line of \eqref{reg3-cal}, we deduce the following eigenvalue equation:
\begin{equation}\label{dis-rel-1}
\wg^2 \Wigner_0(t,\omega,\xg ,\kg) = \TA_0(\xg,\kg) \Wigner_0(t,\omega,\xg ,\kg)\,,
\end{equation}
which is of course the same as \eref{eq:step01}. Therefore, multiplying (\ref{dis-rel-1}) by $\adj{\eigvec}_{\indi} (\xg,\kg)$ on the left side and by $\eigvec_{\indi} (\xg,\kg)$ on the right side, and using the spectral decomposition \eqref{deve-w} of $\Wigner_{0}$, we get that the eigenvalues for the system \eqref{dis-rel-1} are identical to $\smash{\{\eigl^2_\indi(\xg,\kg)\}_{1\leq\indi\leq\Mode}}$. Consequently, let $(\wg,\xg,\kg)$ be such that $\wg^2= \eigl^2_\indi(\xg ,\kg)$, which uniquely defines the polarization mode $\indi$ for this choice of frequency, position and wave vector. Then the coefficients of the decomposition (\ref{deve-w}) can be rewritten as: 
\begin{equation}\label{dis-rel-3}
\speciv_\indi(t,\omega,\xg,\kg) = \sa_\indi(t,\xg,\kg)\dir \left(\wg^2-\eigl^2_\indi(\xg,\kg)\right)\,.
\end{equation}
The $\sa_{\indi}$'s are the (possibly matrix-valued) specific intensities for high-frequency waves in a randomly varying elastic medium. They are Hermitian and positive definite since the limiting Wigner measure $\smash{\Wigner_0}$ is Hermitian and positive definite.


\subsection{Half-order correction $\Go(\escale^\demi)$}\label{fir-ord-cor-sec}

From now on, we drop the $(t,\omega,\xg)$ dependence for clarity purposes, it being understood that we will come back to this dependence once the derivation has been completed. By considering the $\smash{\Go(\escale^\demi)}$ terms in~\eref{second-equ} we can calculate $\smash{\TF{\Wigner}_{1}(\vp,\kg)}$, the Fourier transform of $ \Wigner_{1} (\vy,\kg)$ with respect to $\vy$, in terms of $\Wigner_{0}(\kg )$. This expression will be used in the sequel for the derivation of the evolution properties of $\smash{\Wigner_0}$. So, by inserting the asymptotic expansion (\ref{asym-expan}) in (\ref{second-equ}), and making use of \eqref{reg1-cal}, \eqref{reg2-cal}, \eqref{reg5-cal} and \eqref{reg5-cal-adj}, we obtain the following $\smash{\Go(\escale^\demi)}$ terms:
\begin{equation} \label{cal-1-s-ins}
\begin{split}
\bzero &=\TA_0 \left(\kg+\frac{\Dxy}{2}\right) \Wigner_1(\vy,\kg)- \Wigner_1(\vy,\kg)\TA_0 \left( \xg,\kg-\frac{\Dxy}{2}\right)  \\
&\quad+\int_{\Rset^3} \frac{\dd \vp}{(2\pi)^{3}}\iexp^{\ci\yg\cdot\vp}\SyH\left(\kg+\frac{\Dxy}{2},\vp,\kg-\frac{\vp}{2}+\frac{\Dxy}{2}\right)\Wigner_0\left(\kg-\frac{\vp}{2}\right) \\
&\quad-\int_{\Rset^3} \frac{\dd \vp}{(2\pi)^{3}}\iexp^{\ci\yg\cdot\vp}\Wigner_0\left(\kg+\frac{\vp}{2}\right)\adj{\SyH}\left(\kg-\frac{\Dxy}{2},\vp,\kg+\frac{\vp}{2}-\frac{\Dxy}{2}\right)\,.   
\end{split}
\end{equation}
Taking the Fourier transform of \eqref{cal-1-s-ins} with respect to $\yg$ and using the above definition, yields:
\begin{multline}\label{four-tran-1-cor}
\bzero = \TA_0\left(\kg+  \frac{\vp}{2} \right)\TF{\Wigner}_{1}(\vp,\kg)  - \TF{\Wigner}_{1}(\vp,\kg)\TA_0\left(\kg-  \frac{\vp}{2} \right)+\ci\theta \TF{\Wigner}_{1}(\vp,\kg) \\
\quad + \SyH\left(\kg+\frac{\vp}{2}, \vp,\kg-  \frac{\vp}{2} \right)  \Wigner_0\left(\kg- \frac{\vp}{2}\right) - \Wigner_0\left(\kg+ \frac{\vp}{2}\right)\adj{\SyH}\left(\kg- \frac{\vp}{2}, \vp,\kg+\frac{\vp}{2} \right)\,,
\end{multline}  
since $\Wigner_{0}$ is independent of $ \vy$ so that $\smash{\TF{\Wigner}_0}=\smash{(2\pi)^{3} \dir(\bzero)\Wigner_0}$. Here $\theta$ is a regularization (limiting absorption) parameter as in~\cite{BAL05} that will be sent to $0$ at the end of the derivation. As $\smash{\{ \eigvec_{\indi}(\kg)\}_{1\leq\indi\leq\Mode}}$ form a complete basis of $\Rset^3$ for all $\kg$ (and for all $\xg$), the following expansions of $\smash{\TF{ \Wigner}_1}$ and $\SyH$ hold: 
\begin{equation} \label{deve-w1-H-equ}
\begin{split}
\TF{ \Wigner}_{1}(\vp,\kg) & =  \sum_{\indi,\indj=1}^\Mode\eigvec_\indi\left(\kg+\frac{\vp}{2}\right)\TF{\speciv}_{\indi\indj}(\vp,\kg)\adj{\eigvec}_\indj\left(\kg-\frac{\vp}{2}\right)\,,    \\
\SyH (\kg,\vp, \vq )  &=  \sum_{\indi,\indj=1}^\Mode \eigvec_\indi(\kg)\SyH_{\indi \indj}(\kg,\vp, \vq )\adj{\eigvec}_\indj(\vq) \\
 &=\adj{\SyH}(\vq,\vp,\kg  ) \,,
\end{split}
\end{equation}
where:
\begin{equation}  \label{def-h-sym-ave-new}
\begin{split}
\TF{\speciv}_{\indi\indj}(\vp,\kg) &:=\adj{\vp}_\indi\left(\kg+\frac{\vp}{2}\right )\TF{ \Wigner}_{1}(\vp,\kg)\vp_\indj\left(\kg-\frac{\vp}{2}\right)\,, \\
\SyH_{\indi \indj}(\kg,\vp, \vq ) &:=  \adj{\vp}_{\indi} (\kg)\SyH(\kg,\vp, \vq )\vp_{\indj} (\vq)\,.
\end{split}
\end{equation}        
Injecting the two equations of (\ref{deve-w1-H-equ}) into (\ref{four-tran-1-cor}), with the expansion $\smash{\TA_0(\kg\pm\frac{\vp}{2})}=\smash{\sum_\indi\eigv^2_\indi(\kg\pm\frac{\vp}{2})}\times\smash{\proj_\indi(\kg\pm\frac{\vp}{2})}$ obtained from \eref{deve-w}, and multiplying by $\smash{\adj{\eigvec}_{\indi}(\kg +\frac{\vp}{2})}$ on the left side and by $\smash{\eigvec_{\indj}(\kg -\frac{\vp}{2})}$ on the right side, we deduce that:  
\begin{equation*}
\begin{split}
\left[  \eigv^2_{\indi}\left(\kg+ \frac{\vp}{2}\right) -  \eigv^{2}_{\indj}\left(\kg- \frac{\vp}{2}\right)+\ci \theta  \right] 
\vp_{\indi } \left(\kg+\frac{\vp}{2}\right) \TF{\speciv}_{\indi\indj}(\vp,\kg) \adj{\vp}_{\indj} \left(\kg-\frac{\vp}{2}\right ) &\\
+\vp_{\indi } \left(\kg+\frac{\vp}{2}\right)\SyH_{\indi \indj} \left(\kg+\frac{\vp}{2}, \vp, \kg-\frac{\vp}{2}\right )\speciv_{\indj}\left(\kg-\frac{\vp}{2}\right)  \adj{\vp}_{\indj} \left(\kg-\frac{\vp}{2}\right ) 
& \\
-\vp_{\indi } \left(\kg+\frac{\vp}{2}\right)\speciv_{\indi}\left(\kg+\frac{\vp}{2}\right)\SyH_{\indi \indj} \left(\kg+\frac{\vp}{2}, \vp, \kg-\frac{\vp}{2}\right ) \adj{\vp}_{\indj} \left(\kg-\frac{\vp}{2}\right) &=0
\,,
\end{split}
\end{equation*}
so that:
\begin{multline}\label{coef-w1}
\TF{\speciv}_{\indi \indj}(\vp,\kg)= \\
\frac{\speciv_{\indi}\left(\kg+\frac{\vp}{2}\right) \SyH_{\indi \indj} \left(\kg+\frac{\vp}{2}, \vp, \kg-\frac{\vp}{2}\right )-\SyH_{\indi \indj} \left(\kg+\frac{\vp}{2}, \vp, \kg-\frac{\vp}{2}\right ) \speciv_{\indj}\left(\kg-\frac{\vp}{2}\right)}{ \eigv^2_{\indi}\left(\kg+ \frac{\vp}{2}\right) -  \eigv^{2}_{\indj}\left(\kg- \frac{\vp}{2}\right)+\ci\theta}\,.
\end{multline}


\subsection{Evolution properties}  \label{second-corre-sec}

The evolution equation for ${\Wigner_0}$ is finally obtained from the $\Go(\escale)$ terms in \eref{second-equ}. It is:
\begin{multline}\label{second-correc-1}
2 \omega  \iD_t\Wigner_{0}(\kg)= \TA_0(\kg)\Wigner_{2}(\vy,\kg)-\Wigner_{2}(\vy,\kg)\TA_0(\kg) \\
+\frac{1}{2\ci}\Big(\left\{\TA_0(\kg) ,\Wigner_{0}(\kg)\right\}-\left\{\Wigner_{0}(\kg),\TA_0(\kg)\right\}\Big) +\frac{1}{\ci}\Big[\TA_\indLT(\kg)-\demi\bnabla_\xg \cdot \bnabla_\kg\TA_0(\kg),\Wigner_0(\kg)\Big]  \\
+\int_{\Rset^{3}}\frac{\dd\vp}{(2\pi)^{3}}\iexp^{\ci\vy\cdot\vp}\SyH\left(\kg+\frac{\Dx_{\vy}}{2},\vp,\kg-\frac{\vp}{2}+\frac{\Dx_{ \vy}}{2}\right)\Wigner_{1}\left(\vy,\kg-\frac{\vp}{2}\right)  \\
-\int_{\Rset^{3}}\frac{\dd\vp}{(2\pi)^{3}}\iexp^{\ci\vy\cdot\vp}\Wigner_{1}\left(\vy,\kg+\frac{\vp}{2}\right)\adj{\SyH}\left(\kg-\frac{\Dx_{\vy}}{2},\vp,\kg+\frac{\vp}{2}-\frac{\Dx_{\vy}}{2}\right)\,.        
\end{multline}
As is usual in homogenization approaches, we may assume that $\smash{\Wigner_2}$ is orthogonal to $\smash{\Wigner_0}$ in an averaged sense in order to justify the asymptotic expansion \eqref{asym-expan} with respect to $\escale$, so that we can drop the terms with $\smash{\Wigner_2}$ in \eref{second-correc-1}. On the other hand, since $\vp$ is a dummy variable in the integrals of \eref{second-correc-1}, we can interchange $\vp$ and $\vq$ so that it becomes (ignoring the $\smash{\Wigner_2}$ terms for the reason invoked just above):
\begin{multline}\label{second-correc-2}
\begin{split}
2 \omega  \iD_t\Wigner_{0}(\kg)= & \;\frac{1}{2\ci} \Big(\left \{\TA_0(\kg) ,\Wigner_{0}(\kg)\right \}-\left \{\Wigner_{0}(\kg),\TA_0(\kg)\right \}\Big) \\
&+\frac{1}{\ci}  \Big[ \TA_\indLT(\kg)-\demi\bnabla_\xg \cdot \bnabla_\kg\TA_0(\kg),\Wigner_0(\kg)  \Big] \\
&+\underbrace{\int_{\Rset^{3}}\frac{\dd \vq}{(2\pi)^{3}}\iexp^{\ci\vy\cdot\vq}\SyH\left(\kg+ \frac{\Dx_{ \vy}}{2},\vq,\kg-\frac{\vq}{2}+ \frac{\Dx_{ \vy}}{2} \right)\Wigner_1\left(\vy,\kg- \frac{\vq}{2}\right)}_{\mathcal{I}_1}    \\
&-   \underbrace{ \int_{\Rset^{3}}   \frac{\dd \vq}{(2\pi)^{3}} \iexp^{\ci\vy\cdot\vq}\Wigner_1\left(\vy,\kg+ \frac{\vq}{2}\right) 
\adj{\SyH}\left(\kg- \frac{\Dx_{ \vy}}{2}  , \vq  ,\kg + \frac{\vq}{2}- \frac{\Dx_{ \vy}}{2}\right)}_{\mathcal{I} _2}\,.              
\end{split}
\end{multline}
The next step consists in computing the integrals $\smash{\mathcal{I}_1}$ and $\smash{\mathcal{I}_2}$ above using the results obtained in the previous \sref{fir-ord-cor-sec} for $\smash{\TF{\Wigner}_1}$ in terms of $\smash{\Wigner_0}$. This closure, together with averaging in~\eref{second-correc-2}, gives rise to a collisional linear radiative transfer equation for the average of $\smash{\Wigner_0}$. Its collision operator is shown to depend on the phase functions of the random inhomogeneities, $\smash{\TF{\coro}_{\inda\indb}(\vq)}$ in~\eref{def-aver}. We obtain here a general form of the collision kernel accounting for all possible symmetry classes of elasticity tensors, including the isotropy class which was already considered in~\cite{RYZ96}. This result is the main contribution of this paper. We detail in the next two subsections how it is derived.


\subsubsection{Averaging \eref{second-correc-2}}

The averaging of \eref{second-correc-2} requires more specifically the computation of the averages $\mathbb{E}\{\mathcal{I}_{1}\}$ and $\mathbb{E}\{\mathcal{I}_{2}\}$. We first consider $\mathbb{E}\{\mathcal{I}_{1}\}$. Injecting the inverse Fourier transform $\vp\rightarrow \vy$ of $ \Wigner_{1}(\vy,\kg- \frac{\vq}{2})$, we deduce that: 
 \begin{equation}\label{i1-1}
\mathcal{I}_{1} =   \int_{\Rset^{6}} \frac{\dd \vq \dd \vp}{(2\pi)^{6}}\iexp^{\ci\vy\cdot(\vq+\vp)}\SyH \left (\kg + \frac{\Dx_{ \vy}}{2}, \vq , \kg - \frac{\vq - \vp}{2}\right)\TF{\Wigner}_{1}\left(\vp,\kg- \frac{\vq}{2}\right) .
\end{equation}
We inject the expression of $\smash{\TF{\Wigner}_{1}(\vp,\kg- \frac{\vq}{2})}$ obtained from \eref{deve-w1-H-equ} with \eref{coef-w1}:
\begin{multline}\label{dev-w1-av-1}
\TF{\Wigner}_{1}\left(\vp,\kg-\frac{\vq}{2}\right) = \sum_{\indi,\indj=1}^{\Mode}\eigvec_\indi\left(\kg-\frac{\vq-\vp}{2}\right)\TF{\speciv}_{\indi\indj}\left(\vp,\kg - \frac{\vq}{2}\right)\adj{\eigvec}_\indj \left(\kg- \frac{\vq+\vp}{2}\right) \\
=\sum_{\indi,\indj=1}^\Mode\Bigg[
\frac{\vp_{\indi }\left(\kg-\frac{\vq  -\vp   }{2} \right) \speciv_{\indi}\left(\kg-\frac{  \vq-\vp}{2}\right)\SyH_{\indi \indj}\left(\kg-\frac{\vq-\vp}{2},\vp,\kg- \frac{\vq+\vp}{2}\right) \adj{\vp} _{\indj }\left( \kg- \frac{\vq  +\vp   }{2}\right)}{\eigv^2_{\indi}\left(\kg-\frac{ \vq-\vp}{2}\right) - \eigv^{2}_{\indj}\left(\kg- \frac{ \vq+\vp}{2}\right)+\ci\theta} \\
\quad\quad\quad -\frac{\vp_{\indi }\left(\kg-\frac{\vq  -\vp   }{2} \right) \SyH_{\indi \indj}\left(\kg-\frac{\vq-\vp}{2},\vp,\kg- \frac{\vq+\vp}{2}\right)\speciv_{\indj}\left(\kg-\frac{\vq+\vp}{2}\right) \adj{\vp} _{\indj }\left( \kg- \frac{\vq  +\vp   }{2}\right)}{\eigv^2_{\indi}\left(\kg-\frac{ \vq-\vp}{2}\right) - \eigv^{2}_{\indj}\left(\kg- \frac{ \vq+\vp}{2}\right)+\ci\theta}\Bigg]\,.
\end{multline}
Now because of \eqref{def-aver}, we see that the average of $\mathcal{I}_{1}$ in \eref{i1-1} with \eref{dev-w1-av-1} will give rise to the Dirac factor $\dir(\vq+\vp)$. Thus introducing the change of variables:
\begin{displaymath}
\frac{\vq-\vp}{2}\rightarrow\vq\,,\quad\frac{\vq+\vp}{2}\rightarrow\bzero\,,
\end{displaymath} 
allows us to write:
\begin{multline*}
\dir(\bzero) \mathbb{E}\{\mathcal{I}_{1}  \} =  \sum_{\indi,\indj=1}^{\Mode} \mathbb{E} \Bigg\{\frac{1}{(2\pi)^6}\int_{\Rset^3} \frac{\dd \vq}{ \eigv^2_{\indi}(\kg-\vq) -\eigv^{2}_{\indj}(\kg)+\ci\theta } \times \\
\SyH(\kg,\vq,\kg-\vq)\eigvec_\indi(\kg-\vq)\left[\speciv_{\indi}(\kg-\vq)\SyH_{\indi \indj}(\kg - \vq,-\vq, \kg)\right. \\
\left.-\SyH_{\indi \indj}(\kg - \vq,-\vq, \kg) \speciv_{\indj}(\kg)\right]\adj{\eigvec}_\indj(\kg)\Bigg\} \,.
\end{multline*}
A similar calculus gives $\smash{\mathbb{E}\{\mathcal{I}_{2}\}}$ as:
\begin{multline*}
\dir(\bzero) \mathbb{E}\{\mathcal{I}_{2} \} = \sum_{\indi,\indj=1}^{\Mode}\mathbb{E} \Bigg\{\frac{1}{(2\pi)^6}\int_{\Rset^{3}} \frac{\dd \vq} { \eigv^2_{\indi}(\kg) -\eigv^{2}_{\indj}(\kg + \vq)+\ci\theta}\times \\
\eigvec_\indi (\kg)\left[\speciv_{\indi}(\kg )\SyH_{\indi \indj}(\kg , -\vq, \kg +\vq )\right. \\
\left.-\SyH_{\indi \indj}(\kg , -\vq, \kg +\vq )\speciv_{\indj}(\kg + \vq)\right]\adj{\eigvec}_\indj (\kg+\vq)\adj{\SyH}(\kg, \vq,\kg+\vq)\Bigg\}\,.
\end{multline*}    
          
        
\subsubsection{Matrix radiative transfer equation}
      
Now we insert in \eref{second-correc-2} the foregoing expressions of $\smash{\mathbb{E}\{\mathcal{I}_{1}\}}$ and $\smash{\mathbb{E}\{\mathcal{I}_{2}\}}$. Multiplying \eref{second-correc-2} on the left side by $\adj{\eigvec}_\indi(\kg)$ and on the right side by $\eigvec_\indi(\kg)$, then using the results of \sref{sec:evolution} for the projection of the Poisson and Lie brackets (see~\eref{eq:Wigner-equationb}), we have:
\begin{multline}\label{fina-matrix-equ-1}
\dir(\bzero)2\omega\partial_t{\mathbb E}\{\speciv_\indi\}(\kg) = \dir(\bzero)\left\{\eigl_\indi^2(\kg),{\mathbb E}\{\speciv_\indi\}(\kg)\right\}+ \dir(\bzero)[\matN_\indi(\kg),{\mathbb E}\{\speciv_\indi\}(\kg)] \\
-\ci\sum_{\indj=1}^\Mode{\mathbb E}\Bigg\{\int_{\Rset^3} \frac{\dd \vq}{(2\pi)^{6}} \frac{\SyH_{\indi\indj}(\kg,\vq, \kg-\vq)\SyH_{\indj\indi}(\kg- \vq ,-\vq, \kg )\speciv_\indi(\kg)}{\eigv^2_\indj(\kg-\vq) - \eigv^2_\indi(\kg)+\ci\theta}   \\ 
+\int_{\Rset^3} \frac{\dd \vq}{(2\pi)^{6}}\frac{\speciv_{\indi}(\kg) \SyH_{\indi \indj }(\kg ,-\vq, \kg + \vq )\SyH_{\indj \indi }(\kg+\vq,\vq, \kg) }{\eigv^2_{\indi}(\kg) -\eigv^{2}_{\indj}( \kg + \vq) +\ci \theta}  \\  
- \int_{\Rset^3} \frac{\dd \vq}{(2\pi)^{6}}
\frac{\SyH_{\indi\indj}(\kg,\vq, \kg- \vq) \speciv_{\indj} (\kg-\vq) \SyH_{ \indj \indi}(\kg-  \vq ,-\vq, \kg ) } { \eigv^2_{\indj}(\kg-\vq) -\eigv^{2}_{\indi}(\kg)  +\ci \theta}  \\  
- \int_{\Rset^3} \frac{\dd \vq}{(2\pi)^{6}}\frac{ \SyH_{\indi \indj}(\kg ,-\vq, \kg + \vq)\speciv_{\indj}(\kg + \vq) \SyH_{\indj\indi}(\kg+\vq,\vq, \kg) }  {\eigv^2_{\indi}(\kg)-\eigv^{2}_{\indj}(\kg + \vq) +\ci \theta}\Bigg\} \,, 
\end{multline}
where we recall that $\partial_t\equiv\ci\iD_t$. Here we have used the fact that $\smash{\SyH_{\indi\indj}(\kg ,\vp, \vq )}=\smash{\adj{\SyH}_{\indj\indi}(\vq ,\vp, \kg )}$ from \eref{decomp-H-mat-conj} and \eref{deve-w1-H-equ}. On the other hand, we have in the sense of distribution $\smash{\frac{1}{\ci x+\theta}}\rightarrow\smash{\frac{1}{\ci x}+\sig{\theta}\pi \dir(x)}$ as $\theta\rightarrow 0$, where $\sig{\theta}$ stands for the sign of $\theta$. Consequently, the previous equation implies by letting $\theta\rightarrow 0 $ and changing properly the variables $\vq+\kg\rightarrow  \vq$ and $\kg-\vq\rightarrow \vq$ in their respective integrals, that:
\begin{multline}\label{fina-equ-3}
\dir(\bzero)2\omega\partial_t{\mathbb E}\{\speciv_\indi(\kg)\} = \dir(\bzero)\left \{\eigl_\indi^2(\kg),{\mathbb E}\{\speciv_\indi(\kg)\}\right \}+ \dir(\bzero) \left[\matN_\indi(\kg),{\mathbb E}\{\speciv_\indi(\kg)\} \right] \\
 - \sig{\theta}\sum_{\indj=1}^\Mode \mathbb{E}  \Bigg\{\int_{\Rset^3} \frac{\dd \vq}{(2\pi)^6}
\left(\pi\dir\left(\eigv^2_{\indi}(\kg) -\eigv^2_{\indj}(\vq)\right)+\frac{\ci\sig{\theta}}{\eigv^2_{\indj}(\vq)-\eigv^{2}_{\indi}(\kg) }\right)\times \\
\SyH_{\indi\indj}(\kg, \kg- \vq , \vq)\big[\SyH_{\indj\indi}(\vq ,\vq-\kg, \kg ) \speciv_{\indi}(\kg) - \speciv_{\indj}(\vq)  \SyH_{\indj \indi }(\vq ,\vq-\kg,  \kg)\big]  \\ 
 + \int_{\Rset^3} \frac{\dd \vq}{(2\pi)^6}  \left(\pi\dir \left( \eigv^2_{\indj}(\vq)-\eigv^2_{\indi}(\kg)  \right)  +\frac{\ci \sig{\theta}}{\eigv^2_{\indi}(\kg)-\eigv^2_{\indj}(\vq) }\right)\times \\
 \big[\speciv_{\indi}(\kg)  \SyH_{\indi \indj }(\kg ,\kg-\vq, \vq ) - \SyH_{\indi \indj}(\kg ,\kg-\vq, \vq)\speciv_{\indj}(\vq)\big]\SyH_{\indj\indi}(\vq,\vq- \kg,\kg) \Bigg\}\,.
\end{multline}
We finally proceed as in \eref{eq:Liouville-equation}. At first, we observe that:
\begin{displaymath}
\dir(\eigv^2_{\indj}(\vq) -\eigv^2_{\indi}(\kg))  = \frac{1}{2  \eigv_{\indi}(\kg)}\dir(\eigv_{\indj}(\vq)- \eigv_{\indi}(\kg))\,.
\end{displaymath}
\textcolor{\mycolor}{Second, as for \eref{eq:Liouville-equation}, we split $\smash{\speciv_\indi}$ into its "forward" traveling components $\smash{\sa^+_\indi}$ for which $\wg=-\smash{\eigl_\indi}$ and its "backward" traveling components $\smash{\sa^-_\indi}$ for which $\wg=\smash{\eigl_\indi}$: $\speciv_\indi=\smash{\sa^+_\indi\dir(\wg+\eigl_\jeig)+\sa^-_\indi\dir(\wg-\eigl_\jeig)}$}. Third, $\theta$ has to be chosen negative to preserve causality, so that we obtain with~\eref{fina-equ-3}:
\begin{multline}\label{fina-equ-4}
\dir(\bzero)\left(\partial_t{\mathbb E}\{\sa^+_\indi(\kg)\}+\left\{\eigl_\indi(\kg),{\mathbb E}\{\sa^+_\indi(\kg)\}\right\}+[\matN^+_\indi(\kg),{\mathbb E}\{\sa^+_\indi(\kg)\}]\right)= \\
\sum_{\indj=1}^\Mode\int_{\Rset^3} \frac{\dd \vq}{(2\pi)^5}\frac{\dir(\eigv _{\indj}(\vq) -\eigv_{\indi}(\kg))}{4\eigl_\jeig(\kg)\eigl_\keig(\vq)}\mathbb{E}\{\SyH_{\indi \indj}(\kg ,\kg-\vq, \vq)\sa^+_{\indj}(\vq)\SyH_{\indj\indi}(\vq,\vq- \kg,\kg)\} \\
-\demi\left[\int_{\Rset^3} \frac{\dd \vq}{(2\pi)^5}\frac{\dir(\eigv_{\indi}(\kg) -\eigv_{\indj}(\vq))}{4\eigl_\jeig(\kg)\eigl_\keig(\vq)}{\mathbb E}\{\SyH_{\indi\indj}(\kg, \kg- \vq , \vq)\SyH_{\indj\indi}(\vq ,\vq-\kg, \kg )\}\right] {\mathbb E}\{\sa^+_\indi(\kg)\} \\
-\demi{\mathbb E}\{\sa^+_\indi(\kg)\}\left[\int_{\Rset^3} \frac{\dd \vq}{(2\pi)^5}\frac{\dir(\eigv_{\indi}(\kg) -\eigv_{\indj}(\vq))}{4\eigl_\jeig(\kg)\eigl_\keig(\vq)}{\mathbb E}\{\SyH_{\indi\indj}(\kg, \kg- \vq , \vq)\SyH_{\indj\indi}(\vq ,\vq-\kg, \kg )\}\right] \\  
+\left[\ci\int_{\Rset^3} \frac{\dd \vq}{(2\pi)^6}\Bigg(\frac{{\mathbb E}\{\SyH_{\indi\indj}(\kg, \kg- \vq , \vq)\SyH_{\indj\indi}(\vq ,\vq-\kg, \kg )\}}{2\eigl_\jeig(\kg)(\eigv^2_{\indj}(\vq)-\eigv^{2}_{\indi}(\kg)) }\Bigg)
\right]{\mathbb E}\{\sa^+_\indi(\kg)\}   \\ 
+{\mathbb E}\{\sa^+_\indi(\kg)\} \left[\ci\int_{\Rset^3} \frac{\dd \vq}{(2\pi)^6}  \Bigg(\frac{{\mathbb E}\{\SyH_{\indi \indj }(\kg ,\kg-\vq, \vq )\SyH_{\indj\indi}(\vq,\vq- \kg, \kg)\}}{2\eigl_\jeig(\kg)(\eigv^2_{\indi}(\kg)-\eigv^2_{\indj}(\vq) )}\Bigg) \right]\,. \\ 
 \end{multline}
In the above derivation we have invoked a crucial mixing assumption as in~\cite{RYZ96,BAL05}: indeed, it is expected that ${\mathbb E}\{\SyH_{\indi\indj}\SyH_{\indj\indi}\sa^+_{\indi}\}\simeq{\mathbb E}\{\SyH_{\indi\indj}\SyH_{\indj\indi}\}{\mathbb E}\{\sa^+_{\indi}\}$ since both quantities $\SyH$ and $\sa^+_{\indi}$ vary on different scales. On the other hand, we have from the correlation model \eqref{def-aver} that $\smash{\mathbb{E}\{\TF{\coefC}_{\inda}(\vp)\TF{\coefC}_{\indb}(-\vp)\}}=\smash{\dir(\bzero)(2\pi)^{3}\TF{\coro}_{\inda\indb}(\vp)}$. Therefore we shall introduce the following definition of the so-called differential scattering cross-sections $\dscat_{\indi \indj}$, $1\leq\indi,\indj\leq\Mode$:
\begin{multline}\label{eq:dscat}
\dir(\bzero)(2\pi)^{3}\dscat_{\indi \indj}(\kg,\vq)[{\bf A}(\vq)]:= \\
\frac{\pi}{2\eigv_{\indi}(\kg)\eigv_{\indj}(\vq)} \mathbb{E} \left \{\SyH_{\indi \indj}(\kg ,\kg- \vq,\vq){\bf A}(\vq)\SyH_{\indj \indi}(\vq,\vq-\kg,\kg)\right\}
\end{multline}
for any $r_\indj\times r_\indj$ square matrix ${\bf A}(\vq)$, such that $\smash{\dscat_{\indi \indj}[\sa^+_\indj]}=\smash{\dscat_{\indi \indj}[\mathbb{E}\{\sa^+_\indj\}]}$ invoking the aforementioned mixing assumption. \textcolor{\mycolor}{From this definition it can be verified straightforwardly, at least in the scalar case $\smash{r_\indj=1}$, that the differential scattering cross-sections satisfy the following general reciprocity relationships:
\begin{displaymath}
\begin{split}
\dscatij_{\indi\indj}(\kg,\vq) &=\dscatij_{\indj\indi}(-\vq,-\kg)\,, \\
\dscatij_{\indi\indj}\left(\frac{\wg\hkg}{\cel_\indi(\hkg)},\frac{\wg\hvq}{c_\beta(\hat{\bf q})}\right) &=\dscatij_{\indj\indi}\left(\frac{\wg\hvq}{\cel_\indj(\hvq)},\frac{\wg\hkg}{\cel_\indi(\hkg)}\right)\,,
\end{split}
\end{displaymath}
as expected from wave physics}. We also define the total scattering cross-section matrices $\tscat_\indi$, $1\leq\indi\leq\Mode$ as:
\begin{multline}\label{eq:tscat}
\tscat_\indi(\kg)=\frac{1}{2}\sum_{\indj=1}^\Mode\Bigg\{\int_{\Rset^3}\dscat_{\indi \indj} (\kg,\vq)[\II_\indj] \dir(\eigv _{\indj}(\vq) -\eigv_{\indi}(\kg))\frac{\dd \vq}{(2\pi)^3} \\
-\frac{\ci}{\pi}\int_{\Rset^3}\left(\frac{1}{\eigv_{\indj}(\vq)-\eigv_{\indi}(\kg)}\right)\dscat_{\indi \indj} (\kg,\vq)[\II_\indj]\frac{\dd \vq}{(2\pi)^3}\Bigg\} \,.
\end{multline}
\eref{fina-equ-4} then reads:
\begin{multline}\label{eq:RTE-anisotropic}
\partial_t\sa^+_\indi(t,\xg,\kg)+\left\{\eigl_\indi(\xg,\kg),\sa^+_\indi(t,\xg,\kg)\right\}+[\matN^+_\indi(\xg,\kg),\sa^+_\indi(t,\xg,\kg)]= \\
\sum_{\indj=1}^\Mode\int_{\Rset^3}\dscat_{\indi \indj} (\xg,\kg,\vq)[\sa^+_\indj(t,\xg,\vq)] \dir(\eigv _{\indj}(\xg,\vq) -\eigv_{\indi}(\xg,\kg))\frac{\dd \vq}{(2\pi)^3} \\
-\tscat_\indi(\xg,\kg)\sa^+_\indi(t,\xg,\kg)-\sa^+_\indi(t,\xg,\kg)\adj{\tscat}_\indi(\xg,\kg)\,,
\end{multline}
still denoting the averages $\smash{\mathbb{E}\{\sa^+_\indi\}}$ by $\smash{\sa^+_\indi}$. In the above we have re-introduced the space-time dependence of all quantities whenever applicable. Contrary to the result of \eref{eq:Liouville-equation}, the specific intensities $\sa_\indi$ for the $\Mode$ modes propagating in the medium get coupled by its random inhomogeneities. The radiative transport equations (\ref{eq:RTE-anisotropic}) above generalize Eqs.~(4.32) of \cite{RYZ96} to arbitrary anisotropy of the elastic medium, in that the differential and total scattering cross-sections we have derived embed all possible cases of elastic constitutive models.  \textcolor{\mycolor}{We note here that if the order of multiplicity of all modes is one, the specific intensities $\saj^+_\indi$ are scalars satisfying the system (\ref{eq:RTE-vect}). \eref{eq:RTE-anisotropic} above considers the most general case when some modes possibly have a multiplicity higher than $1$, as for an isotropic medium}. Now some particular classes of practical significance in engineering mechanics are discussed in the next section.
\section{\textcolor{\mycolor}{Example calculations}}\label{sec:examples}

The aim of this section is to apply the formula for the differential and total scattering cross-sections derived in \sref{second-corre-sec}, Eqs.~(\ref{eq:dscat}) and (\ref{eq:tscat}), for some usual classes of anisotropy. We consider cubic, transverse isotropic (hexagonal) and orthotropic (orthorhombic) materials. Elastic isotropy is also detailed in order to demonstrate that the analysis developed in the foregoing section is consistent with some already known results. The elasticity tensor $\tenselas=\smash{[\tenselas^{ijkl}]}$ of \eref{ela-ten-fluc} is a fourth-order tensor satisfying the minor ($\smash{\tenselas^{ijkl}}=\smash{\tenselas^{jikl}}= \smash{\tenselas^{ijlk}}$) and major ($\smash{\tenselas^{ijkl}}=\smash{\tenselas^{klij}}$) symmetries invoked in~\sref{sec:HFwaves-basics}. For clarity purposes we adopt Voigt's notation in this section. It considers the following one-to-one correspondence between a symmetric pair $(i,j)$ of three-dimensional indices and a multi-index $I$ ranging from 1 to 6:
\begin{equation*}
11\leftrightarrow 1\,,\; 22\leftrightarrow 2\,,\; 33\leftrightarrow 3\,,\; 23\leftrightarrow 4\,,\; 31\leftrightarrow 5\,,\; 12\leftrightarrow 6\,.
\end{equation*}
We can thus represent the elasticity tensor $\tenselas$ or its counterpart $\tenselas_0$ for the unperturbed background medium as a symmetric $6\times 6$ matrix with the following equivalence:
\begin{equation}\label{ela-ten-cons-matr-form}
\tenselas = {\scriptsize\begin{bmatrix}
\comC^{1111} &\comC^{1122}  &  \comC^{1133} & \comC^{1123} & \comC^{1131}& \comC^{1112}  \\
\comC^{2211} &\comC^{2222}  & \comC^{2233}  & \comC^{2223} &\comC^{2231}& \comC^{2212}  \\
\comC^{3311} &\comC^{3322}  &  \comC^{3333} &\comC^{3323} & \comC^{3331}&\comC^{3312}  \\
\comC^{2311} &\comC^{2322}  & \comC^{2333}  & \comC^{2323} &\comC^{2331}& \comC^{2312}  \\
\comC^{3111} &\comC^{3122}  & \comC^{3133}  & \comC^{3123} &\comC^{3131}& \comC^{3112}  \\
\comC^{1211} &\comC^{1222}  & \comC^{1233}  &  \comC^{1223} &\comC^{1231}&\comC^{1212}  \\
\end{bmatrix}}
\equiv
{\scriptsize\begin{bmatrix}
\comC^{11} &\comC^{12}  &  \comC^{13} & \comC^{14} & \comC^{15}& \comC^{16}  \\
& \comC^{22}  & \comC^{23}  & \comC^{24} &\comC^{25}& \comC^{26}  \\
& &  \comC^{33} &\comC^{34} & \comC^{35}&\comC^{36}  \\
& & & \comC^{44} &\comC^{45}& \comC^{46}  \\
& & & &\comC^{55}& \comC^{56}  \\
& & & & &\comC^{66}  \\
\end{bmatrix}}\,.
\end{equation}
For the following applications it is first necessary to identify the eigenvalues and eigenvectors of the Christoffel tensor $\TA_0$ of the bare medium from the components of $\tenselas_0$ corresponding to each case of anisotropy, and compute the tensor $\SyH(\xg,\kg,\vp,\vq)$ of \eref{symb-H}. Then we shall be able to deduce the differential and total scattering cross-sections, $\dscat_{\indi \indj}(\xg,\kg,\vq)$ and $\tscat_{\indi}(\xg,\kg)$ respectively, for the considered cases. Now the correlation functions $\vy\mapsto\coro_{\inda\indb}(\vy)$ of~\eref{def-aver} are normalized such that:
\begin{equation*}
\frac{4}{3}\pi\lcor_{\inda\indb}^3=\int_0^{+\infty}\vyj^2\id\vyj\int_{\Sset^2}\id\Omega(\hat{\vy})\coro_{\inda\indb}(\vyj\hat{\vy})\,,
\end{equation*}
where $\smash{\vyj=|\vy|}$, $\hat{\vy}=\smash{\vy/\vyj}$, and $\smash{\Sset^2}$ is the unit sphere of $\smash{\Rset^3}$ with the uniform probability measure $\Omega$. This normalization introduces the correlation lengths $\smash{\lcor_{\inda\indb}}$. We may assume in the following examples and without loss of generality that these parameters are all equal to a single parameter $\lcor$. Different models of the phase functions $\vq\mapsto\smash{\TF{\coro}_{\inda\indb}}(\vq)$, which are the three-dimensional Fourier transforms of these normalized correlation functions (NCF), may be invoked as discussed in \emph{e.g.}~\cite{LIU08} and references therein. We will adopt here a Markov (exponential) model, by which:
\begin{equation}\label{eq:Markov}
\TF{\coro}_{\inda\indb}(\vq)=\coroij_{\inda\indb}\times\frac{\lcorp^3}{\pi^2}\frac{1}{(1+(\lcorp|\vq|)^2)^2}\,,\quad\coro_{\inda\indb}(\vy)=\coroij_{\inda\indb}\times\exp\left(-\frac{|\vy|}{\lcorp}\right)\,,
\end{equation}
for $\lcorp=\smash{6^{-\tiers}\lcor}$. The $\coroij_{\inda\indb}$'s are scalars quantifying the amount of correlation between $\smash{\coefC_\inda}$ and $\smash{\coefC_\indb}$ for $1\leq\inda,\indb\leq 21$. Indeed, it is argued in~\cite{STA86} that this model describes fairly well the correlation structure of both continuous and discrete materials. This particular choice does not however restrict our results in any respect. At last, we assume in the following examples that the fluctuation tensor $\tenselas_1$ and the mean tensor $\tenselas_0$ belong to the same symmetry class. Our analysis, though, allows different \textcolor{\mycolor}{morphological and crystallographic textures to be considered at the slow and fast scales since $\tenselas_0$ and $\tenselas_1$ are allowed to vary independently with the position. So this simplifying choice, again, does not restrict it, its physical relevance being out of the scope of this paper in any case. Our objective here is only to demonstrate that our derivation can effectively be used in practical applications}. 

\subsection{Isotropic case}

We start by considering the case of elastic isotropy already derived in~\cite{RYZ96}. Our purpose is to show that the theory developed in~\sref{sec:RTE} embeds the existing results for that particular symmetry class. The elasticity tensor $\smash{\tenselas_0}$ of the background medium depends on the two Lam\'e's coefficients $\pala$ and $\palb$. In view of \eref{ela-ten-cons-matr-form} it thus reads:
\begin{equation}\label{ela-ten-cons-matr-form-iso}
\tenselas_0= \begin{bmatrix}
\pala+2 \palb   & \pala   &   \pala &  0 & 0& 0\\
\pala  &\pala +2 \palb   &  \pala &  0 & 0& 0 \\
 \pala   & \pala   &  \pala+2 \palb  & 0 &0& 0 \\
 0  & 0   &   0&  \palb   & 0& 0 \\
0 & 0 &  0&  0  & \palb &0 \\
0 & 0   &   0&  0 &0&  \palb  \\
\end{bmatrix}\,.
\end{equation}
The Christoffel tensor is given by~\eref{eq:TA-isotropy} as $\TA_0(\xg,\kg)= (\cel_\iP^2(\xg) -\cel_\iS^2(\xg)) \kg \otimes \kg  + \cel_\iS^2(\xg) |\kg|^2 \II$ and we have $\Mode=2$ ($\jeig=\iP$ or $\iS$) and $r_\iP=1$, $r_\iS=2$ as outlined in~\sref{sec:Wigner-isotropic}. Its eigenvalues are $\smash{\eigv^{2}_\iP(\xg,\kg})=\smash{\cel_\iP^2(\xg)|\kg|^2}$ and $\smash{\eigv^{2}_\iS(\xg,\kg)}=\smash{\cel_\iS^2(\xg)|\kg|^2}$, corresponding to the eigenvectors:
\begin{equation}\label{eig-ve-va-iso}
\eigvec_\iP(\xg, \kg)= \hkg\,,\quad\eigvec_\iS(\xg, \kg)=\hkg^\perp=[\hzg_1(\kg),\hzg_2(\kg)]\,,
\end{equation}
where $\hzg_1(\kg)$ and $\hzg_2(\kg)$ are such that $\smash{(\hkg,\hzg_1,\hzg_2)}$ forms an orthonormal triplet. On the other hand, we deduce  from \eref{symb-H} using the notations of \eref{def-aver}, that: 
\begin{equation*}\label{h-kq-iso}      
\SyH(\xg,\kg,\vp,\vq)=\roi^{-1}(\xg)\left[\TF{\coefC}_{\pala}(\vp)\kg\otimes \vq+\TF{\coefC}_{\palb}(\vp) \vq\otimes \kg + \TF{\coefC}_{\palb}(\vp)(\kg\cdot\vq) \II \right]\,.
\end{equation*}
By a straightforward calculation inserting (\ref{eig-ve-va-iso}) into \eref{def-h-sym-ave-new} we have:
\begin{equation*}\label{cal-hcp-iso-cas} 
\begin{split}
\SyHin_{\iP\iP}(\xg, \kg ,\vp, \vq ) &=\roi^{-1}( \xg) |\kg| |\vq|\left[\TF{\coefC}_\pala(\vp) +2(\hkg\cdot \hvq)^{2}\TF{\coefC}_\palb(\vp)\right]\,, \\
\SyH_{\iP\iS}(\xg, \kg ,\vp, \vq )  &=\roi^{-1}( \xg)2(\kg\cdot\vq)\TF{\coefC}_\palb(\vp)\left(\hkg\cdot\hzg_1(\vq),\hkg\cdot\hzg_2(\vq)\right)\,, \\
\SyH_{\iS\iS}(\xg, \kg ,\vp, \vq )  &=\roi^{-1}( \xg) |\kg| |\vq|\TF{\coefC}_\palb(\vp)\matG'(\kg,\vq)\,,
\end{split}
\end{equation*}
where $\matG'(\kg,\vq)=\matG(\kg,\vq)+(\hkg\cdot\hvq)\matT(\kg,\vq)$, and $\matG(\kg,\vq)$ and $\matT(\kg,\vq)$ are the $2\times 2$ matrices given as in Eqs.~(1.11) and (1.20) of~\cite{RYZ96} by:
\begin{equation*}
\matGij_{jk}(\kg,\vq)=(\hzg_j(\kg)\cdot\hvq)(\hzg_k(\vq)\cdot\hkg)\,,\quad\matTij_{jk}(\kg,\vq)=\hzg_j(\kg)\cdot\hzg_k(\vq)\,,\quad 1\leq j,k\leq 2\,.
\end{equation*}
The differential scattering cross-sections of \eref{eq:dscat} are then derived as:
\begin{equation*}
\begin{split}
\dscatij_{\iP\iP}(\kg,\vq)[\saj_\iP] &=\frac{\pi}{2}\frac{|\kg|^2}{ \roi(\pala+2\palb)}\big[\TF{\coro}_{\pala\pala} (\kg-\vq)  + 4(\hkg\cdot\hvq)^2\TF{\coro}_{\pala\palb}(\kg-\vq) \\
&\quad\quad\quad\quad\quad\quad\quad\quad+ 4(\hkg\cdot\hvq)^4\TF{\coro}_{\palb\palb}(\kg-\vq)\big]\saj_\iP(\vq)\,, \\
\dscatij_{\iP\iS}(\kg,\vq)[\sa_\iS] &=\frac{\pi}{2}\frac{4|\kg|^2}{\roi\palb}(\hkg\cdot\hvq)^2\TF{\coro}_{\palb\palb}(\kg-\vq)\matG(\vq,\kg):\sa_\iS(\vq)\,, \\ \dscat_{\iS\iS}(\kg,\vq)[\sa_\iS] &=\frac{\pi}{2}\frac{|\kg|^2}{\roi\palb}\TF{\coro}_{\palb\palb}(\kg-\vq)[\matG'(\kg,\vq)][\sa_\iS(\vq)][\matG'(\kg,\vq)]\,,\\
\dscat_{\iS\iP}(\kg,\vq)[\saj_\iP] &=\frac{\pi}{2}\frac{4|\kg|^2}{\roi\palb}(\hkg\cdot\hvq)^2\TF{\coro}_{\palb\palb}(\kg-\vq)\matG(\kg,\vq)\saj_\iP(\vq)\,,
\end{split}
\end{equation*}
in full agreement with the results of~\cite[Sect.4.5]{RYZ96}, up to a proper normalization of the fluctuation tensor and its power spectral density functions $\vp\mapsto\smash{\TF{\coro}_{\inda\indb}(\vp)}$.

\subsection{Cubic anisotropy} \label{sec:cubic}

Here, the elasticity tensor $\smash{\tenselas_0}$ of the background medium depends on three coefficients $\ccub_1$, $\ccub_2$ and $\ccub_3$ and reads:    
 \begin{equation*}\label{ela-ten-cons-matr-form-cub}
\tenselas_0= \begin{bmatrix}
\ccub_1  & \ccub_2 &   \ccub_2&  0 & 0& 0\\
\ccub_2 & \ccub_1  & \ccub_2&  0 & 0& 0 \\
\ccub_2 &\ccub_2  & \ccub_1 & 0 &0& 0 \\
 0  & 0   &   0& \ccub_3  & 0& 0 \\
0 & 0 &  0&  0  &\ccub_3&0 \\
0 & 0   &   0&  0 &0&  \ccub_3 \\
\end{bmatrix}\,.
\end{equation*}
Then the Christoffel tensor $\TA_0$ of \eref{eq:L-operators} reads:
 \begin{equation}\label{disper-matr-expli-cub-cas}
\TA_0(\xg, \kg  )= \roi^{-1}(\xg)\left[ (\ccub_2+\ccub_3 ) \kg \otimes \kg  + \ccub_3 |\kg|^2 \II +\faniso\diag(\kgj_{1}^2,\kgj_{2}^2,\kgj_{3}^2) \right]
\end{equation}
in a cartesian frame $\smash{(\eib_1,\eib_2,\eib_3)}$, with $\faniso=\ccub_1-\ccub_2-2\ccub_3$. Notice that if this anisotropy factor $\faniso=0$, the Christoffel tensor (\ref{disper-matr-expli-cub-cas}) relative to cubic anisotropy becomes identical with that of the isotropic case for $\ccub_2\equiv\pala$ and $\ccub_3\equiv\palb$. If the anisotropy factor $\faniso\neq 0$, the eigenvalues and eigenvectors of the Christoffel tensor cannot be computed explicitly. However it can be shown~\cite{BOU98b,AKI13} that cubic crystals have $7$ acoustic axes independently of the elasticity constants, provided that $\ccub_2+\ccub_3\neq 0$ (which correspond to "special" crystals). These axes are the three coordinate axes, and the axes $(\pm1,\pm1,\pm1)$ (with one change of sign at a time). 
On the other hand, we deduce from \eref{symb-H} using again the notations of \eref{def-aver}, that:
\begin{multline*}\label{h-kq-cub}
\SyH(\xg,\kg,\vp, \vq) = \roi^{-1}(\xg)\Big[\TF{\coefC}_2(\vp)  \kg \otimes \vq +\TF{\coefC}_3(\vp)  \vq\otimes \kg   + \TF{\coefC}_3(\vp)\left(\kg\cdot\vq\right) \II \\
+\left(\TF{\coefC}_1(\vp) -\TF{\coefC}_2(\vp)-2\TF{\coefC}_3(\vp)\right) \diag(\kgj_1\vqj_1,\kgj_2\vqj_2,\kgj_3\vqj_3) \Big]\,.
\end{multline*}
The $\SyHin_{\jeig\keig}$'s of \eref{def-h-sym-ave-new} are all scalars such that $\SyHin_{\jeig\keig}(\xg,\kg,\vp,\vq)=\SyHin_{\keig\jeig}(\xg,\vq,\vp,\kg)$ with:
\begin{multline*}
\SyHin_{\jeig\keig}(\xg,\kg,\vp,\vq)=\roi^{-1}( \xg) |\kg| |\vq|\Big[\TF{\coefC}_2(\vp)(\hkg\cdot\eigvec_\jeig(\xg,\kg))(\hvq\cdot\eigvec_\keig(\xg,\vq)) \\
+\TF{\coefC}_3(\vp)\left((\hvq\cdot\eigvec_\jeig(\xg,\kg))(\hkg\cdot\eigvec_\keig(\xg,\vq))+(\hkg\cdot\hvq)(\eigvec_\jeig(\xg,\kg)\cdot\eigvec_\keig(\xg,\vq))\right) \\
+\left(\TF{\coefC}_1(\vp) -\TF{\coefC}_2(\vp)-2\TF{\coefC}_3(\vp)\right) \trace(\hkg\otimes\hvq\otimes\eigvec_\jeig(\xg,\kg)\otimes\eigvec_\keig(\xg,\vq))\Big]\,.
\end{multline*}

\begin{figure}
\includegraphics[scale=0.23]{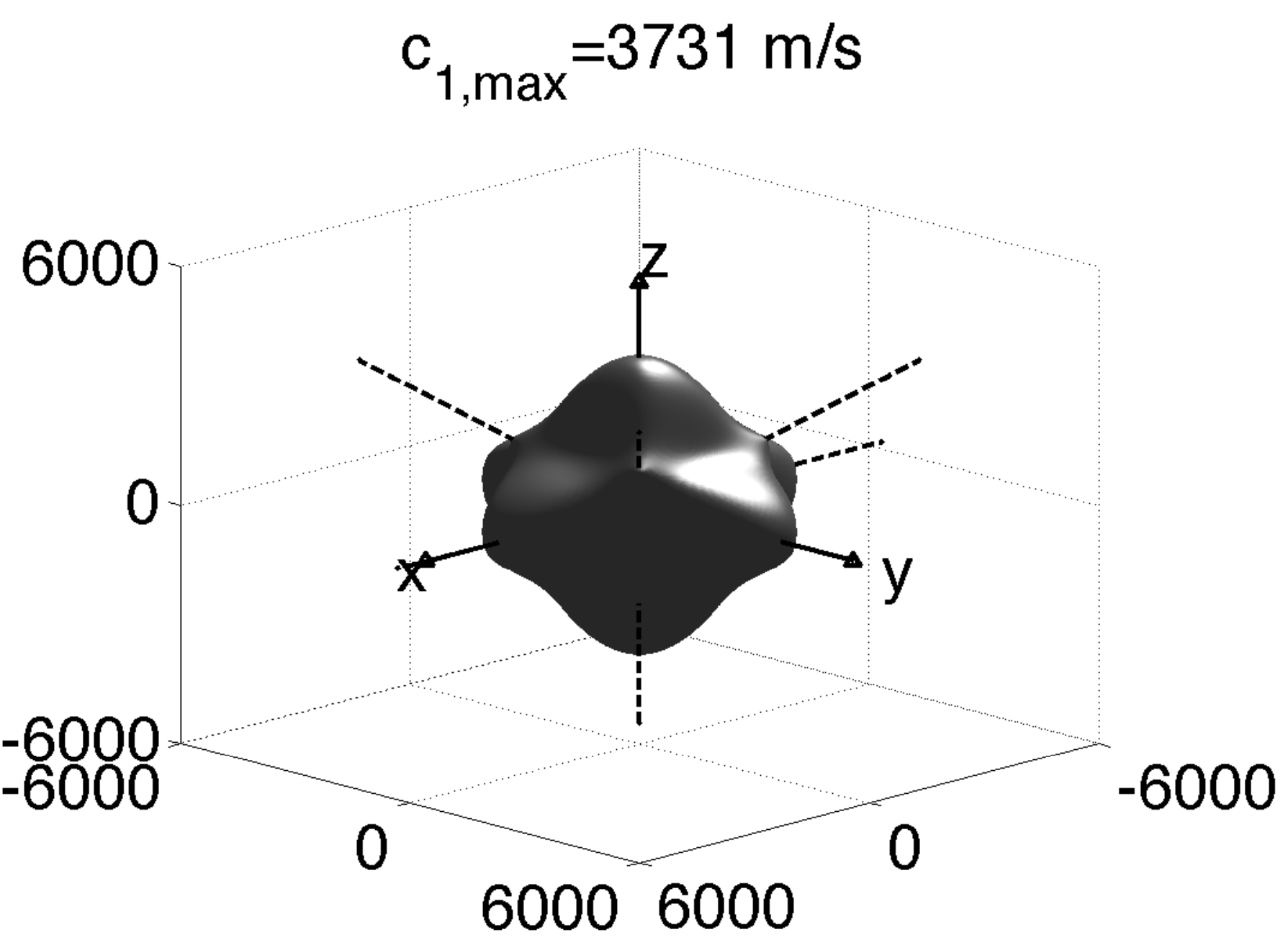}\hfill
\includegraphics[scale=0.23]{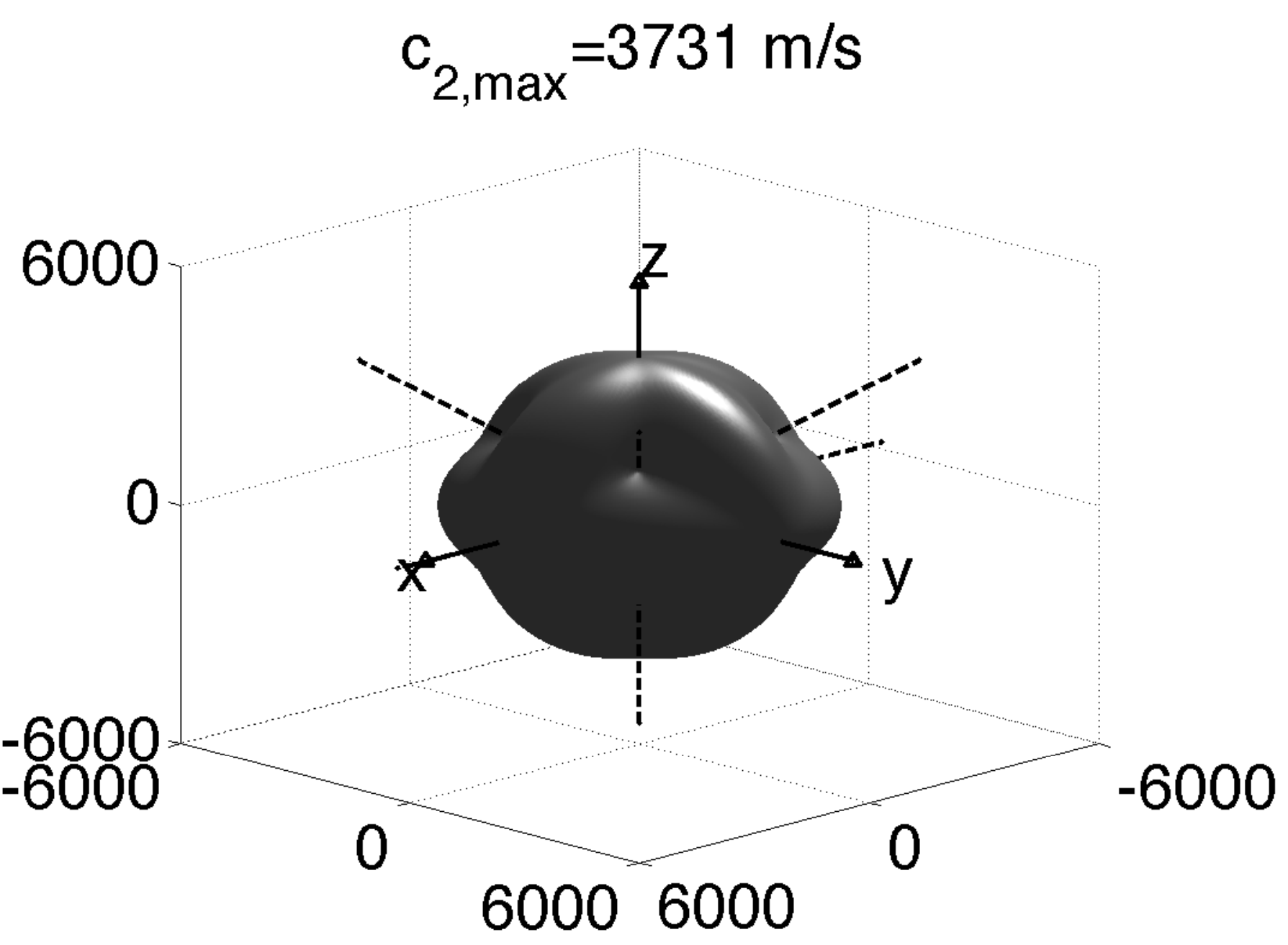}\hfill
\includegraphics[scale=0.23]{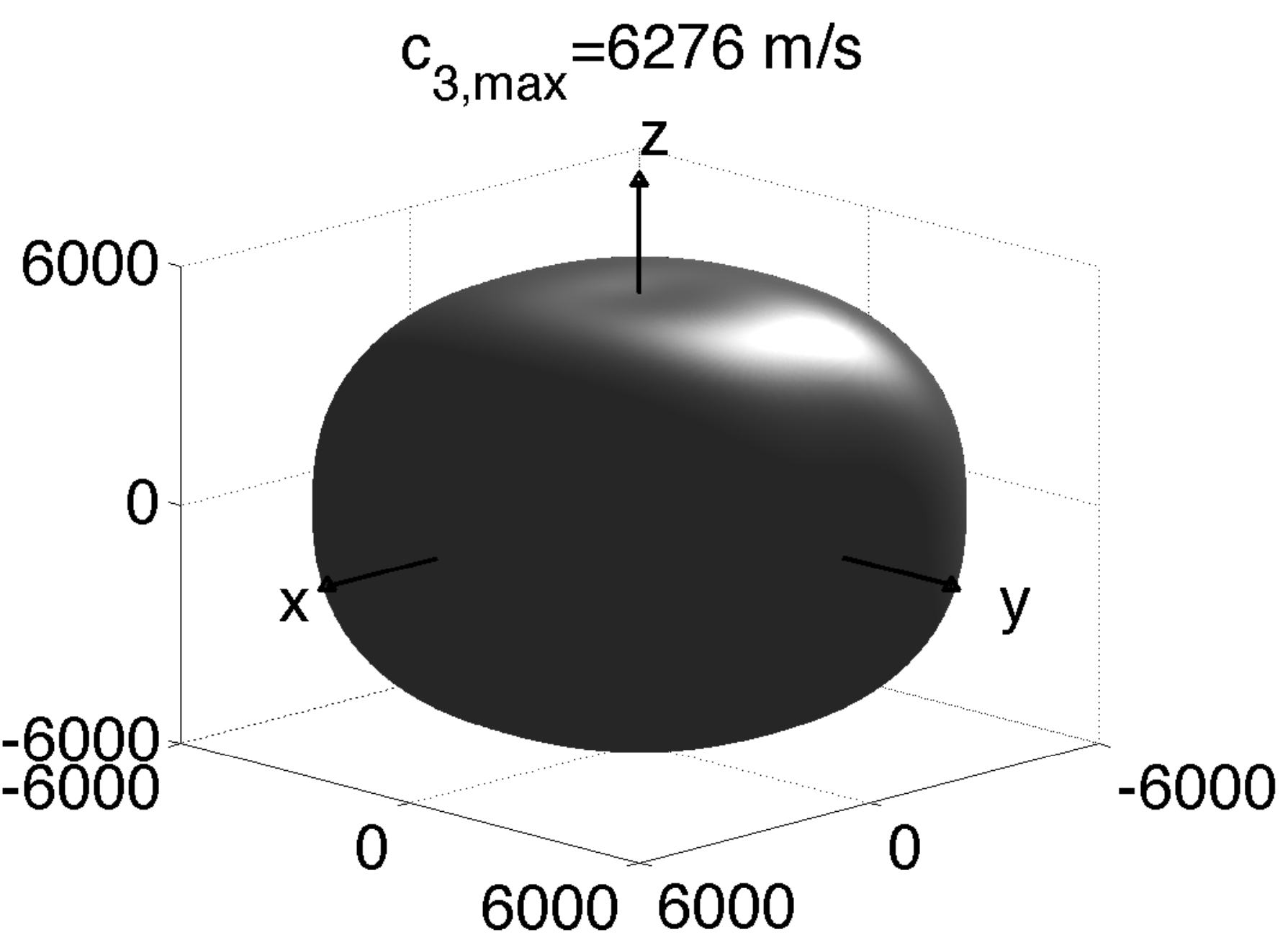}
\caption{Velocity surfaces $\smash{\hkg\mapsto\cel_\jeig(\hkg)}$ for \textcolor{\mycolor}{single crystal} nickel. Left: pseudo-transverse mode $\jeig=1$, middle: pseudo-transverse mode $\jeig=2$, right: pseudo-longitudinal mode $\jeig=3$. The dashed lines display the $4$ acoustic axes out of coordinate planes.}\label{fg:nickel-c}
\end{figure}

\begin{figure}
\centering\includegraphics[scale=0.6]{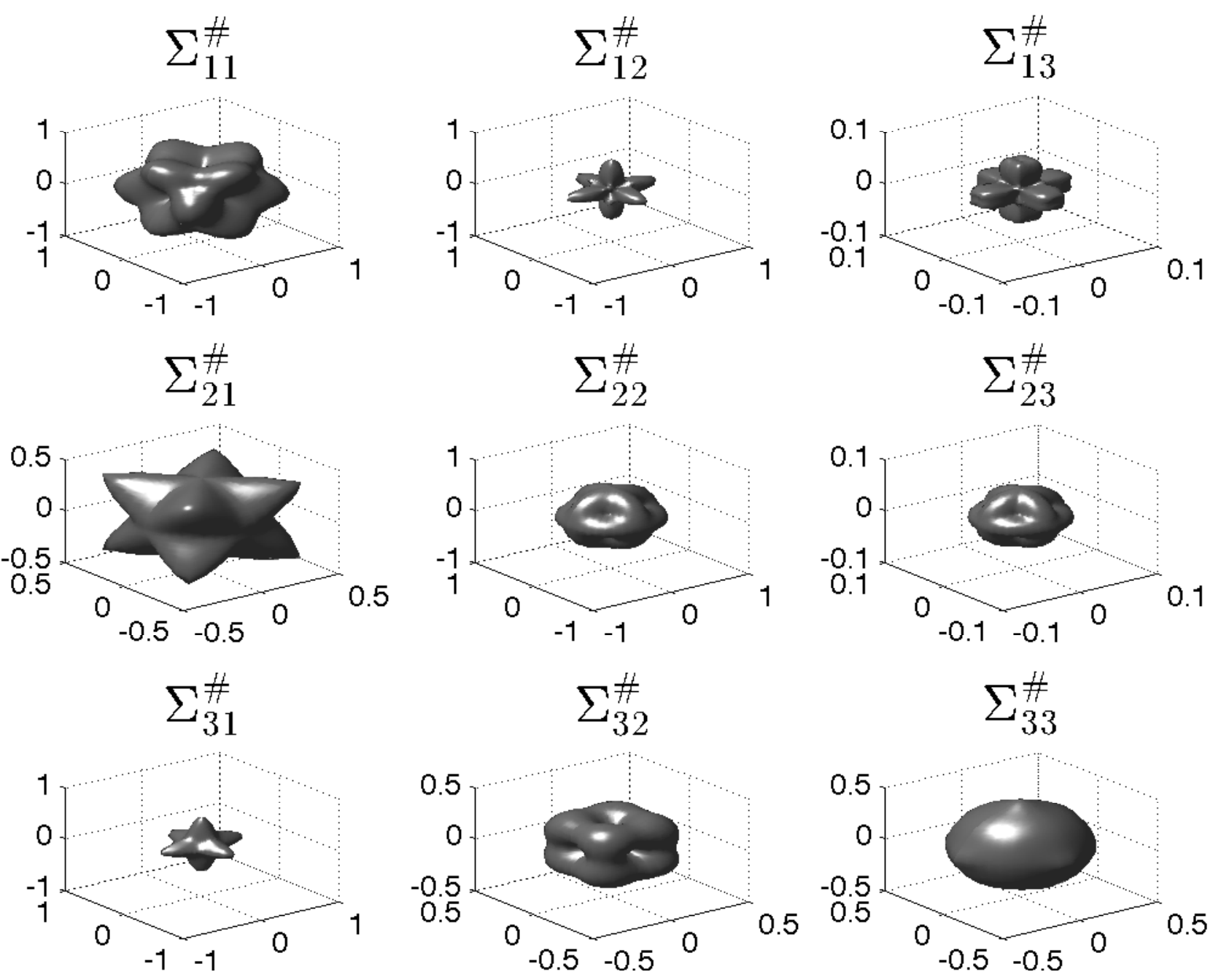}
\caption{Normalized total scattering cross-sections $\smash{\hkg\mapsto\tscati_{\indi\indj}^\#(\hkg)}$ for \textcolor{\mycolor}{single crystal} nickel with fixed frequency parameter $\lcor|\kg|=1$ and Markov model for the NCF.}\label{fg:Ni-Sigma}
\end{figure}

We apply our results to nickel (Ni), which according to the results in~\cite{NEI52} as cited in~\cite[Table 6]{LED73}, has mean elasticity constants $\ccub_1=253.0$, $\ccub_2=152.0$, $\ccub_3=124.0$ (in GPa) and density $\roi=8910$ kg/m$^3$. We first plot on~\fref{fg:nickel-c} the three velocity surfaces $\smash{\hkg\mapsto\cel_\jeig(\hkg)}$ for $\smash{\hkg\in\Sset^2}$ and $\jeig=1,2,3$, such that $\smash{\cel_1\leq\cel_2<\cel_3}$ where the equality $\smash{\cel_1=\cel_2}$ holds on the acoustic axes solely. The latter are also displayed on the pseudo-transverse velocity surface plots.
We also plot on~\fref{fg:Ni-Sigma} the normalized partial scattering cross-sections $\smash{\tscati_{\indi\indj}^\#=\tscati_{\indi\indj}/\tscati_\indi}$, where the non normalized partial scattering cross sections $\smash{\tscati_{\indi\indj}}$, and total scattering cross sections $\smash{\tscati_{\indi}}$, are given by \eref{eq:tscat} as:
\textcolor{\mycolor}{\begin{equation*}
\begin{split}
\tscati_{\indi\indj}\left(\frac{\wg\hkg}{\cel_\indi(\hkg)}\right) &=2\pi\wg^2\int_{\Sset^2}\frac{1}{\cel_\indj^3(\hvq)}\dscatij_{\indi \indj}\left(\frac{\wg\hkg}{\cel_\indi(\hkg)},\frac{\wg\hvq}{\cel_\indj(\hvq)}\right)\dd\Omega(\hvq)\,, \\
\tscati_\indi\left(\frac{\wg\hkg}{\cel_\indi(\hkg)}\right) &=\sum_{\indj=1}^\Mode\tscati_{\indi\indj}\left(\frac{\wg\hkg}{\cel_\indi(\hkg)}\right)\,.
\end{split}
\end{equation*}}
Here all modes have multiplicity one so that $\Mode=3$ and the scattering cross-sections are scalars. The non-dimensional frequency parameter is $\lcor|\kg|=1$, and the correlation coefficients $\smash{\coroij_{\inda\indb}}$ of \eref{eq:Markov} are all equal for $1\leq\inda,\indb\leq 3$ (this hypothesis may be physically unrealistic but  such a discussion is for the present out of the scope of the paper). We use the product Gaussian quadrature rule studied in \emph{e.g.}~\cite{ATK82} for the computation of the integrals on the unit sphere $\Sset^2$ above.
\textcolor{\mycolor}{We may comment on these plots by noting that they have symmetries reminiscent of the underlying material symmetry. It is however hardly possible to elaborate more on this topic since to our knowledge the present results are new. The scattering cross-sections displayed here characterize the attenuation of waves by multiple scattering in the weak coupling regime. Therefore they may be used for the interpretation of ultrasonic experiments, for example. In particular, they can be directly related to some average measure of materials grain sizes in different scattering regimes, including those covered by the present theory~\cite{GOE94}}. 

\subsection{Transverse isotropy} 

Here, the elasticity tensor $\smash{\tenselas_0}$ of the background medium depends on five coefficients $\ccub_1$, $\ccub_2$, $\ccub_3$, $\ccub_4$ and $\ccub_5$ and reads:    
\begin{equation*}\label{ela-ten-cons-matr-form-trans}
\tenselas_0= \begin{bmatrix}
\ccub_1  & \ccub_2 &   \ccub_3&  0 & 0& 0\\
\ccub_2 & \ccub_1  & \ccub_3&  0 & 0& 0 \\
\ccub_3&\ccub_3  & \ccub_4 & 0 &0& 0 \\
 0  & 0   &   0&  \ccub_5  & 0& 0 \\
0 & 0 &  0&  0  &\ccub_5&0 \\
0 & 0   &   0&  0 &0& \frac{\ccub_1-\ccub_2}{2} \\
\end{bmatrix}\,.
\end{equation*}
Then the Christoffel tensor $\TA_0$ of \eref{eq:L-operators} reads:
\begin{multline}\label{disper-matr-expli-trans-cas}
\TA_0(\xg, \kg  )= \roi^{-1}(\xg)\times \\
\begin{bmatrix}
\ccub_1  \kgj_{1}^{2}  + \frac{\ccub_1-\ccub_2}{2}\kgj_{2}^{2} + \ccub_5 \kgj_{3}^{2}  & \frac{\ccub_1+\ccub_2}{2}   \kgj_{1}\kgj_{2}   & (\ccub_3 +\ccub_5 ) \kgj_{1}\kgj_{3}   \\
 \frac{\ccub_1+\ccub_2}{2}\kgj_{1}\kgj_{2}   &  \frac{\ccub_1-\ccub_2}{2}  \kgj_{1}^{2} + \ccub_1 \kgj_{2}^{2} + \ccub_5\kgj_{3}^{2}& (\ccub_3+\ccub_5)\kgj_{2}\kgj_{3}   \\
(\ccub_3 +\ccub_5 ) \kgj_{1}\kgj_{3}   &  (\ccub_3 +\ccub_5 )\kgj_{2}\kgj_{3}   &  \ccub_5(\kgj_{1}^{2} + \kgj_{2}^{2}) +   \ccub_4  \kgj_{3}^{2}  
\end{bmatrix}\,. 
\end{multline}
On the other hand, we deduce from \eref{symb-H} using again the notations of \eref{def-aver}, that:
\begin{multline*}\label{h-kq-trans}
\SyH(\xg,\kg,\vp, \vq )  = \roi^{-1}(\xg)\times\\
\displaystyle{\begin{bmatrix} \vspace*{3pt}
\begin{array}{c} \kgj_{1}\vqj_{1}\TF{\coefC}_1(\vp) + \kgj_{3}\vqj_{3}\TF{\coefC}_5(\vp) \\
+\demi\kgj_{2}\vqj_{2}(\TF{\coefC}_1(\vp)-\TF{\coefC}_2(\vp)) \end {array} & \begin{array}{c} \demi\kgj_{2}\vqj_{1}(\TF{\coefC}_1(\vp)-\TF{\coefC}_2(\vp)) \\
+\kgj_{1} \vqj_{2}\TF{\coefC}_2(\vp) \end{array} & \kgj_{1} \vqj_{3}\TF{\coefC}_3(\vp)+\kgj_{3}  \vqj_{1}\TF{\coefC}_5(\vp) \\ \vspace*{3pt}
\begin{array}{c} \demi\kgj_{1}\vqj_{2}(\TF{\coefC}_1(\vp)-\TF{\coefC}_2(\vp)) \\ +\kgj_{2} \vqj_{1}\TF{\coefC}_2(\vp) \end{array} & \begin{array}{c} \kgj_{2}\vqj_{2}\TF{\coefC}_{1}(\vp)+ \kgj_{3}\vqj_{3}\TF{\coefC}_5(\vp) \\ +\demi\kgj_{1}\vqj_{1}(\TF{\coefC}_1(\vp)-\TF{\coefC}_2(\vp)) \end{array} & \kgj_{2}\vqj_{3}\TF{\coefC}_3(\vp) + \kgj_{3}\vqj_{2}\TF{\coefC}_5(\vp) \\
 \kgj_{3} \vqj_{1}\TF{\coefC}_3(\vp)+\kgj_{1}\vqj_{3}\TF{\coefC}_5(\vp) & \kgj_{2}\vqj_{3}\TF{\coefC}_5(\vp) + \kgj_{3}\vqj_{2}\TF{\coefC}_3(\vp) & \begin{array}{c} (\kgj_{1}\vqj_{1}+\kgj_{2}\vqj_{2})\TF{\coefC}_{5}(\vp) \\ + \kgj_{3}\vqj_{3}\TF{\coefC}_4(\vp) \end{array}
\end{bmatrix}}\,.
\end{multline*}
The elasticity coefficients are related by the constraints for positive definite energy density:
\begin{displaymath}
\begin{array}{c}
\ccub_2^2<\ccub_1^2\,,\quad\ccub_3^2<\ccub_1\ccub_4\,,\\
2\ccub_1\ccub_3^2+\ccub_4\ccub_2^2-2\ccub_2\ccub_3^2<\ccub_1^2\ccub_4\,,
\end{array}
\end{displaymath}
and $\ccub_1,\ccub_4,\ccub_5>0$. There are at most either $1$ acoustic axis $\smash{\eib_3}$, or this very axis and a circular cone of axis $\smash{\eib_3}$ for this symmetry class~\cite{BOU98b} (ignoring the "special" and "pathological" cases dealt with in detail in this latter reference).

\begin{figure}
\includegraphics[scale=0.23]{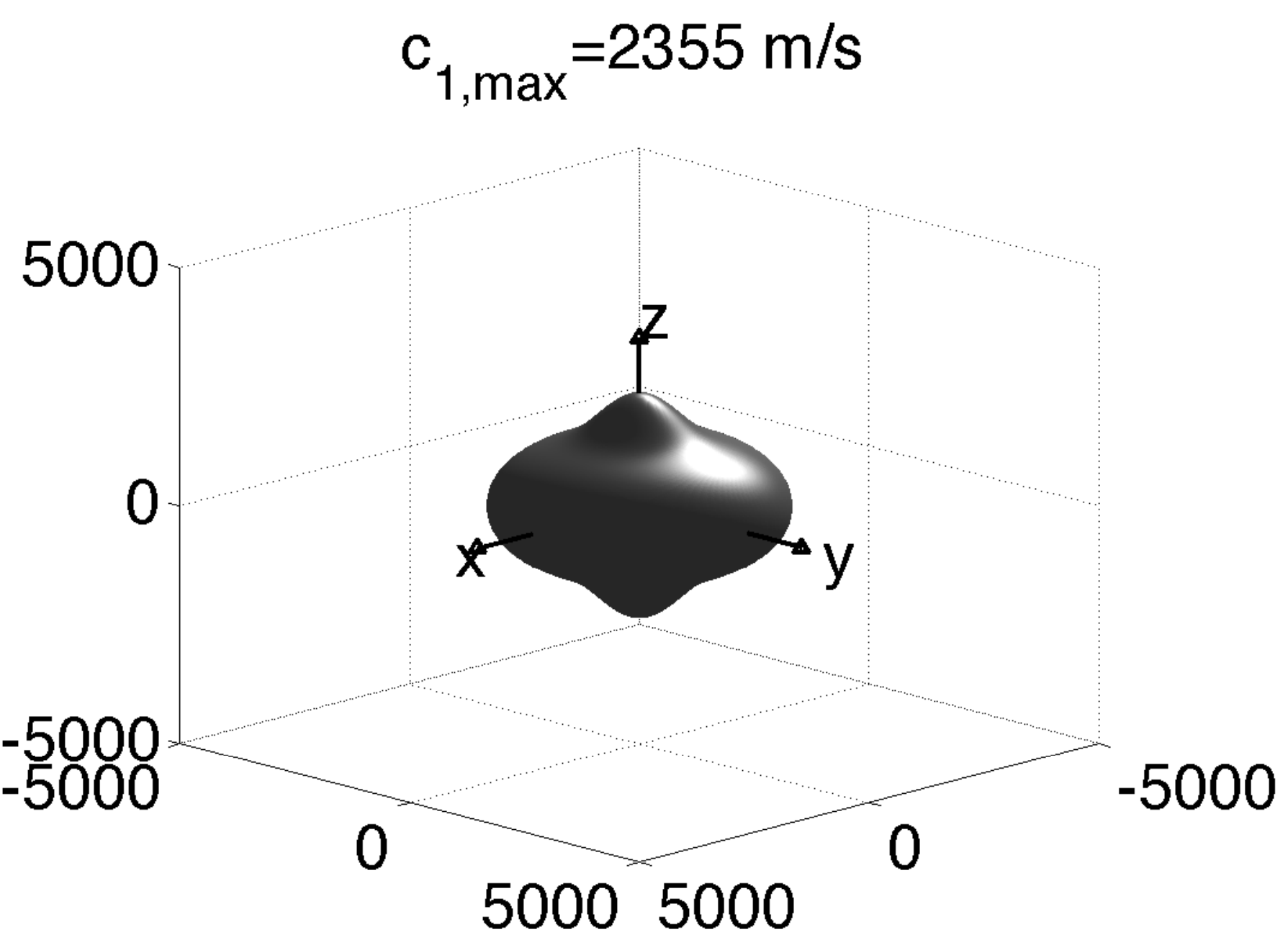}\hfill
\includegraphics[scale=0.23]{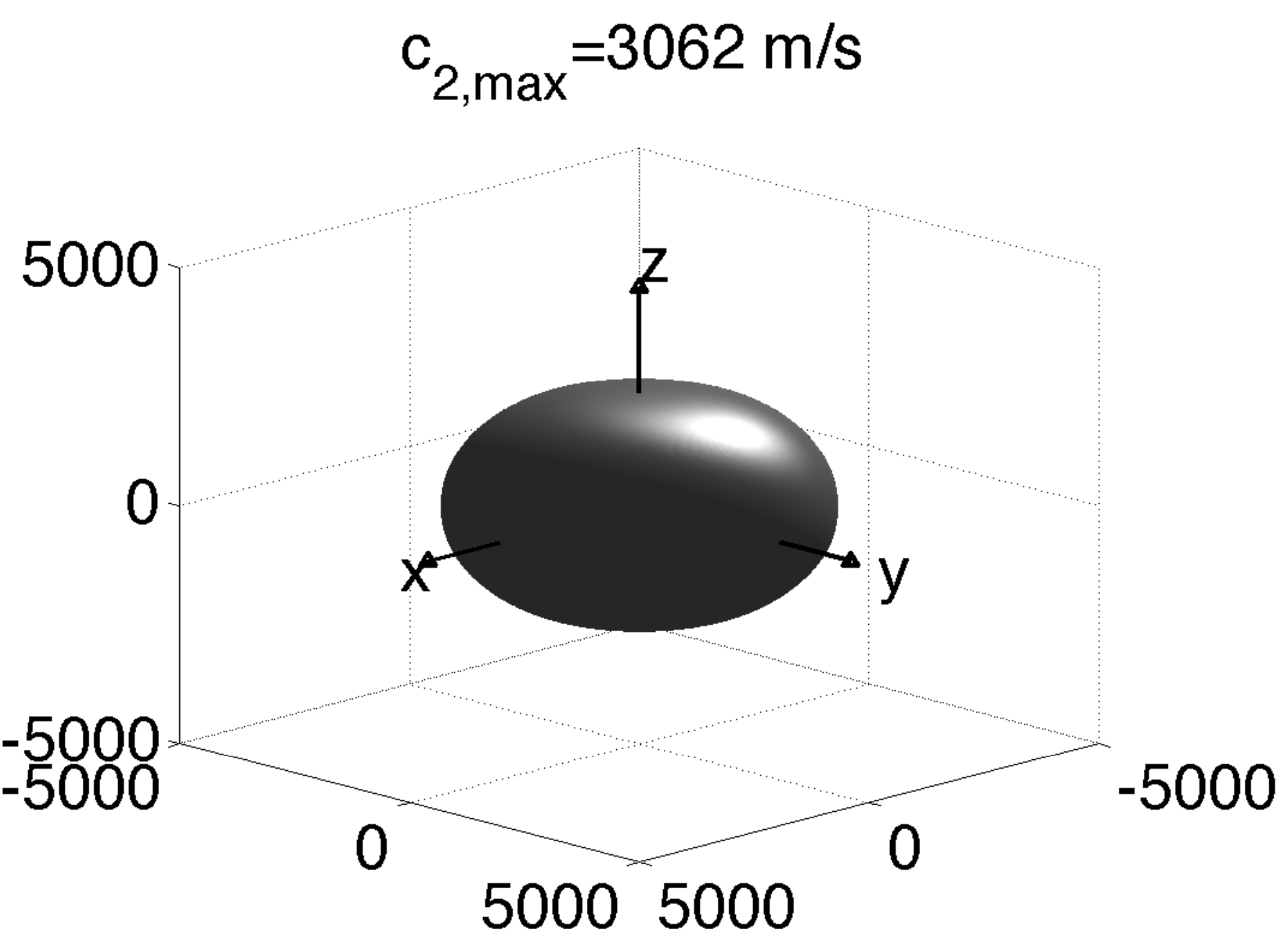}\hfill
\includegraphics[scale=0.23]{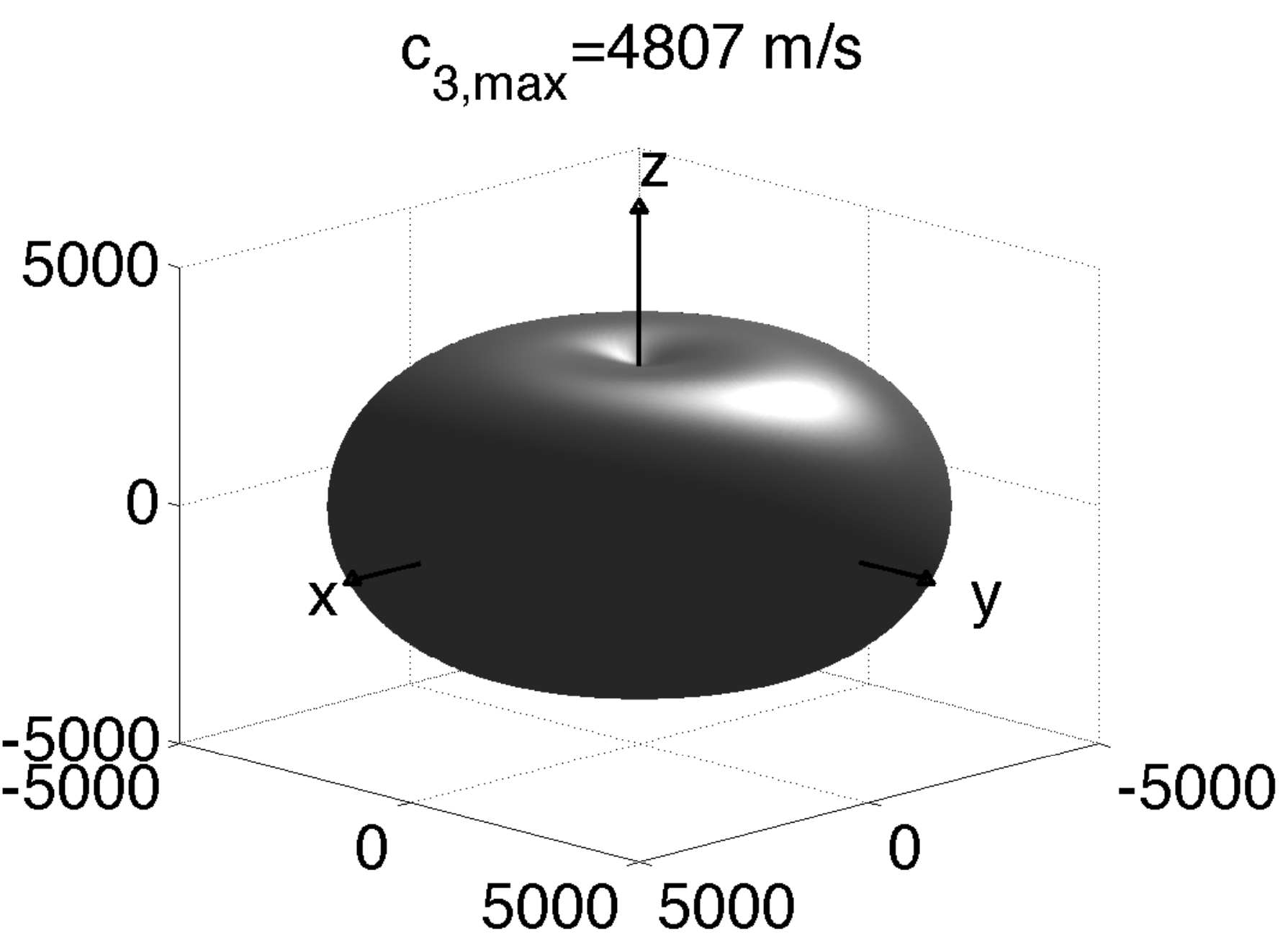}
\caption{Velocity surfaces $\smash{\hkg\mapsto\cel_\jeig(\hkg)}$ for \textcolor{\mycolor}{single crystal} zinc. Left: pseudo-transverse mode $\jeig=1$, middle: pseudo-transverse mode $\jeig=2$, right: pseudo-longitudinal mode $\jeig=3$.}\label{fg:Zn-c}
\end{figure}

\begin{figure}
\centering\includegraphics[scale=0.6]{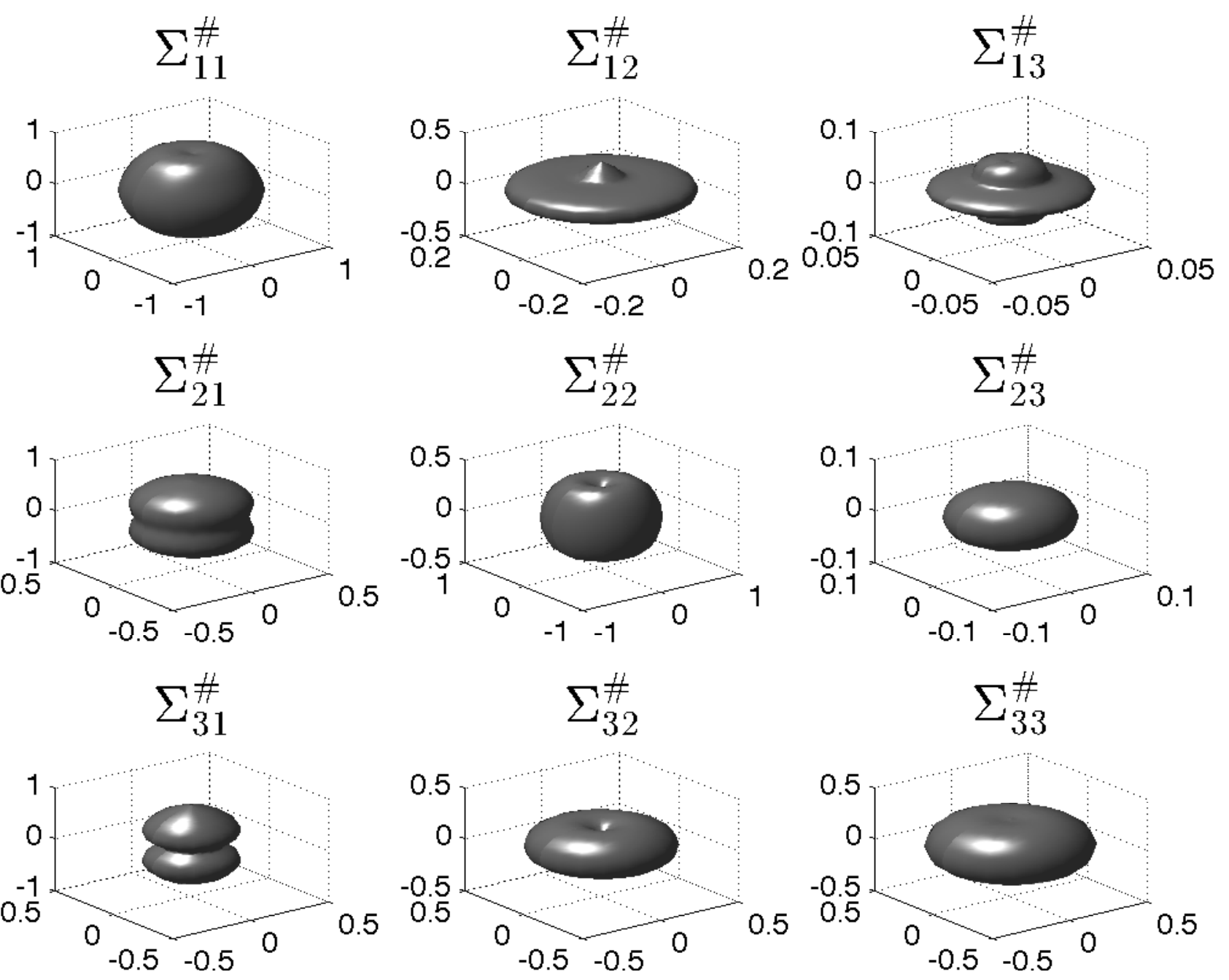}
\caption{Normalized total scattering cross-sections $\smash{\hkg\mapsto\tscati_{\indi\indj}^\#(\hkg)}$ for \textcolor{\mycolor}{single crystal} zinc with fixed frequency parameter $\lcor|\kg|=1$ and Markov model for the NCF.}\label{fg:Zn-Sigma}
\end{figure}

We apply our results to zinc (Zn), which according to the Landolt-B\"{o}rnstein database~\cite{LAN13} as cited in~\cite[Table 5]{BOU98b}, has mean elasticity constants  $\ccub_1=165$, $\ccub_2=31.1$, $\ccub_3=50.0$, $\ccub_4=61.8$, $\ccub_5=39.6$ (in GPa), and density $\roi=7140$ kg/m$^3$. The sole acoustic axis is $\smash{\eib_3}$. We first plot on~\fref{fg:Zn-c} the three velocity surfaces $\smash{\hkg\mapsto\smash{\cel_\jeig(\hkg)}}$ for $\smash{\hkg\in\Sset^2}$ and $\jeig=1,2,3$, such that $\smash{\cel_1\leq\cel_2<\cel_3}$ where the equality $\smash{\cel_1=\cel_2}$ holds on the acoustic axis solely.
We also plot on~\fref{fg:Zn-Sigma} the normalized partial total scattering cross-sections $\smash{\tscati_{\indi\indj}^\#}$ defined as in~\sref{sec:cubic}. Here all modes have multiplicity one, so that $\Mode=3$ and the scattering cross-sections are scalars as for nickel. The non-dimensional frequency parameter is $\lcor|\kg|=1$, and the correlation coefficients $\smash{\coroij_{\inda\indb}}$ are all equal for $1\leq\inda,\indb\leq 5$ (with the same reservation for this assumption as for the case of nickel).
\textcolor{\mycolor}{Regarding symmetries, the same comment as for cubic anisotropy may be done for these plots}. 

\subsection{Orthotropy} 

Here, the elasticity tensor $\smash{\tenselas_0}$ of the background medium depends on nine coefficients $\ccub_1$, $\ccub_2$, $\ccub_3$, $\ccub_4$, $\ccub_5$, $\ccub_6$, $\ccub_7$, $\ccub_8$ and $\ccub_9$ and reads:    
\begin{equation*}\label{ela-ten-cons-matr-form-ortho}
\tenselas_0= \begin{bmatrix}
\ccub_1  & \ccub_2 &   \ccub_3&  0 & 0& 0\\
\ccub_2 & \ccub_4  & \ccub_5&  0 & 0& 0 \\
\ccub_3&\ccub_5 & \ccub_6 & 0 &0& 0 \\
 0  & 0   &   0&\ccub_7  & 0& 0 \\
0 & 0 &  0&  0  &\ccub_8&0 \\
0 & 0   &   0&  0 &0&  \ccub_9\\
\end{bmatrix}\,.
\end{equation*}
Then the Christoffel tensor $\TA_0$ of \eref{eq:L-operators} reads:
 \begin{multline}\label{disper-matr-expli-ortho-cas}
 \TA_0(\xg, \kg  )= \roi^{-1}(\xg)\times \\
\begin{bmatrix}
\ccub_1  \kgj_{1}^{2}  + \ccub_9  \kgj_{2}^{2} + \ccub_8 \kgj_{3}^{2}   &(\ccub_2 +\ccub_9 ) \kgj_{1}\kgj_{2}   & (\ccub_3 +\ccub_8 ) \kgj_{1}\kgj_{3}   \\
(\ccub_2 +\ccub_9 ) \kgj_{1}\kgj_{2}   & \ccub_9 \kgj_{1}^{2} + \ccub_4 \kgj_{2}^{2} + \ccub_7  \kgj_{3}^{2}& (\ccub_5+\ccub_7 )\kgj_{2}\kgj_{3}   \\
(\ccub_3 +\ccub_8 ) \kgj_{1}\kgj_{3}   & (\ccub_5+\ccub_7 ) \kgj_{2}\kgj_{3}   &  \ccub_8 \kgj_{1}^{2} +  \ccub_7 \kgj_{2}^{2} +   \ccub_6  \kgj_{3}^{2}  
\end{bmatrix}\,.
\end{multline}
On the other hand, we deduce from \eref{symb-H} using again the notations of \eref{def-aver}, that:
\begin{multline*}\label{h-kq-ortho}
\SyH(\xg,\kg,\vp, \vq )  = \roi^{-1}(\xg)\times \\
\displaystyle{\begin{bmatrix}
\begin{array}{c} \kgj_{1}\vqj_{1}\TF{\coefC}_1(\vp)+\kgj_{2}\vqj_{2}\TF{\coefC}_9(\vp) \\ + \kgj_{3}\vqj_{3}\TF{\coefC}_8(\vp) \end{array}  & \kgj_{1} \vqj_{2} \TF{\coefC}_2(\vp)+\kgj_{2} \vqj_{1}\TF{\coefC}_9(\vp) & \kgj_{1}  \vqj_{3}\TF{\coefC}_3(\vp)+\kgj_{3}  \vqj_{1}\TF{\coefC}_8(\vp)   \\
\kgj_{1} \vqj_{2}\TF{\coefC}_9(\vp)+\kgj_{2}\vqj_{1}\TF{\coefC}_2(\vp)  &  \begin{array}{c} \kgj_{1}\vqj_{1}\TF{\coefC}_9(\vp)+\kgj_{2}\vqj_{2}\TF{\coefC}_4(\vp) \\ +\kgj_{3}\vqj_{3}\TF{\coefC}_7(\vp) \end{array} & \kgj_{2} \vqj_{3}\TF{\coefC}_5(\vp)+\kgj_{3} \vqj_{2}\TF{\coefC}_7(\vp)  \\
\kgj_{1}  \vqj_{3}\TF{\coefC}_8(\vp)+\kgj_{3}  \vqj_{1}\TF{\coefC}_3(\vp) & \kgj_{2}  \vqj_{3}\TF{\coefC}_7(\vp)+ \kgj_{3}  \vqj_{2}\TF{\coefC}_5(\vp) & \begin{array}{c} \kgj_{1}\vqj_{1}\TF{\coefC}_8(\vp) + \kgj_{2}\vqj_{2}\TF{\coefC}_7(\vp) \\ +\kgj_{3}\vqj_{3}\TF{\coefC}_6(\vp) \end{array}
\end{bmatrix}}\,.
\end{multline*}
The elasticity coefficients are related by the constraints for positive definite energy density:
\begin{displaymath}
\begin{array}{c}
\ccub_2^2<\ccub_1\ccub_4\,,\quad\ccub_3^2<\ccub_1\ccub_6\,,\quad\ccub_5^2<\ccub_4\ccub_6\,,\\
\ccub_1\ccub_5^2+\ccub_4\ccub_3^2+\ccub_6\ccub_2^2-2\ccub_2\ccub_3\ccub_5<\ccub_1\ccub_4\ccub_6\,,
\end{array}
\end{displaymath}
and $\ccub_1,\ccub_4,\ccub_6,\ccub_7,\ccub_8,\ccub_9>0$. There are at most $16$ acoustic axes for this symmetry class~\cite{BOU98b} (ignoring again the "special" and "pathological" cases dealt with in detail in this latter reference).

\begin{figure}
\includegraphics[scale=0.23]{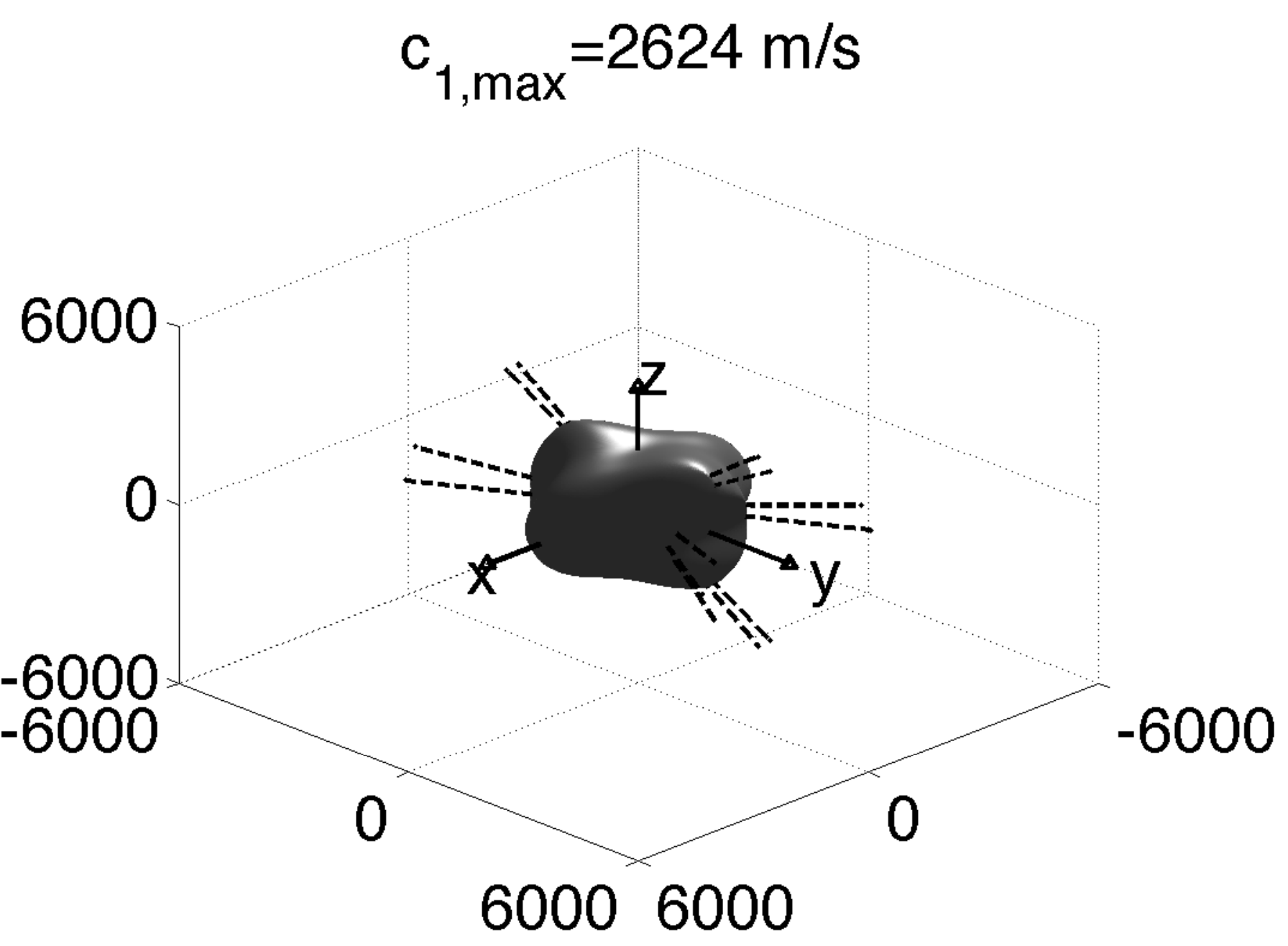}\hfill
\includegraphics[scale=0.23]{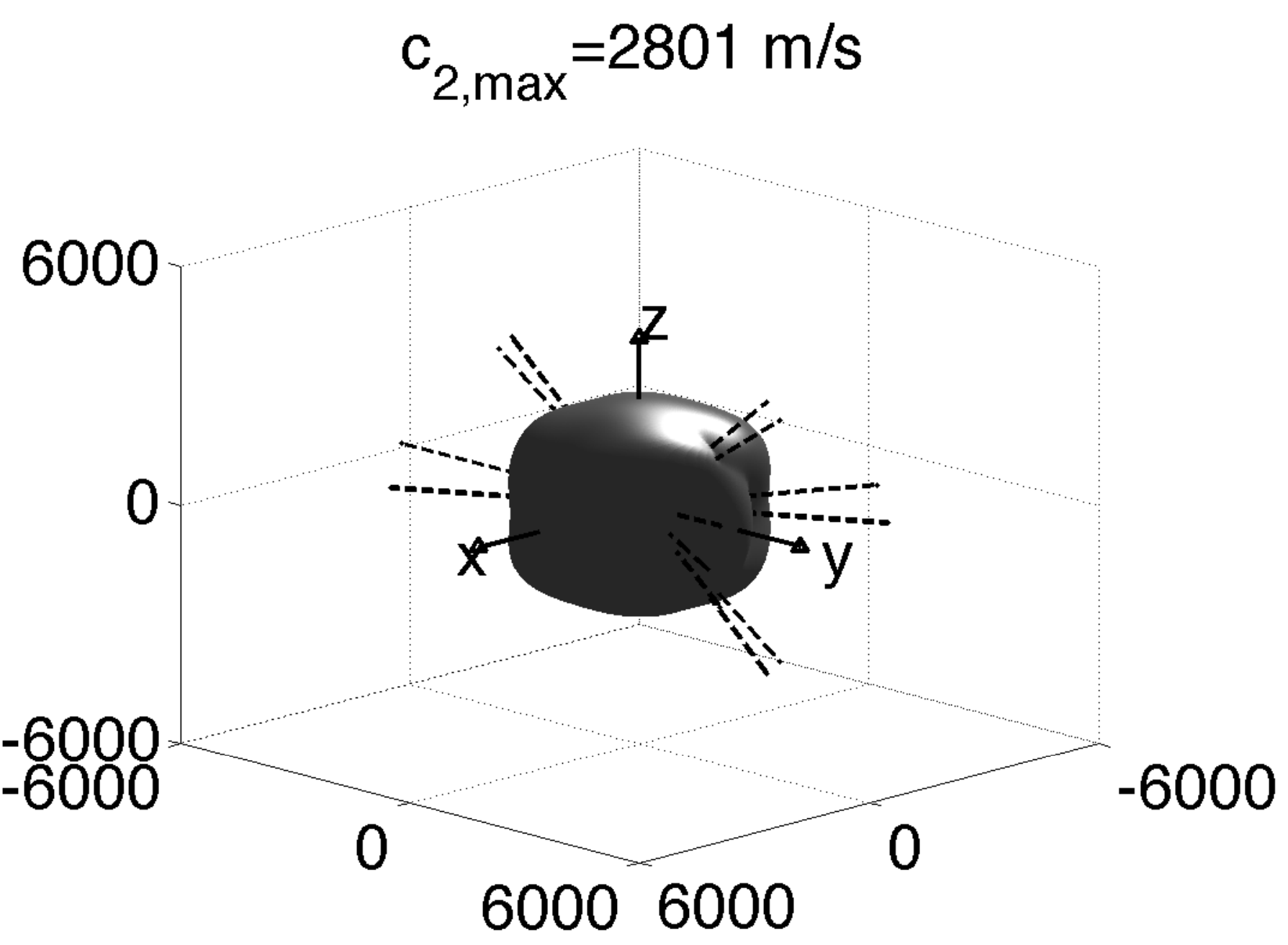}\hfill
\includegraphics[scale=0.23]{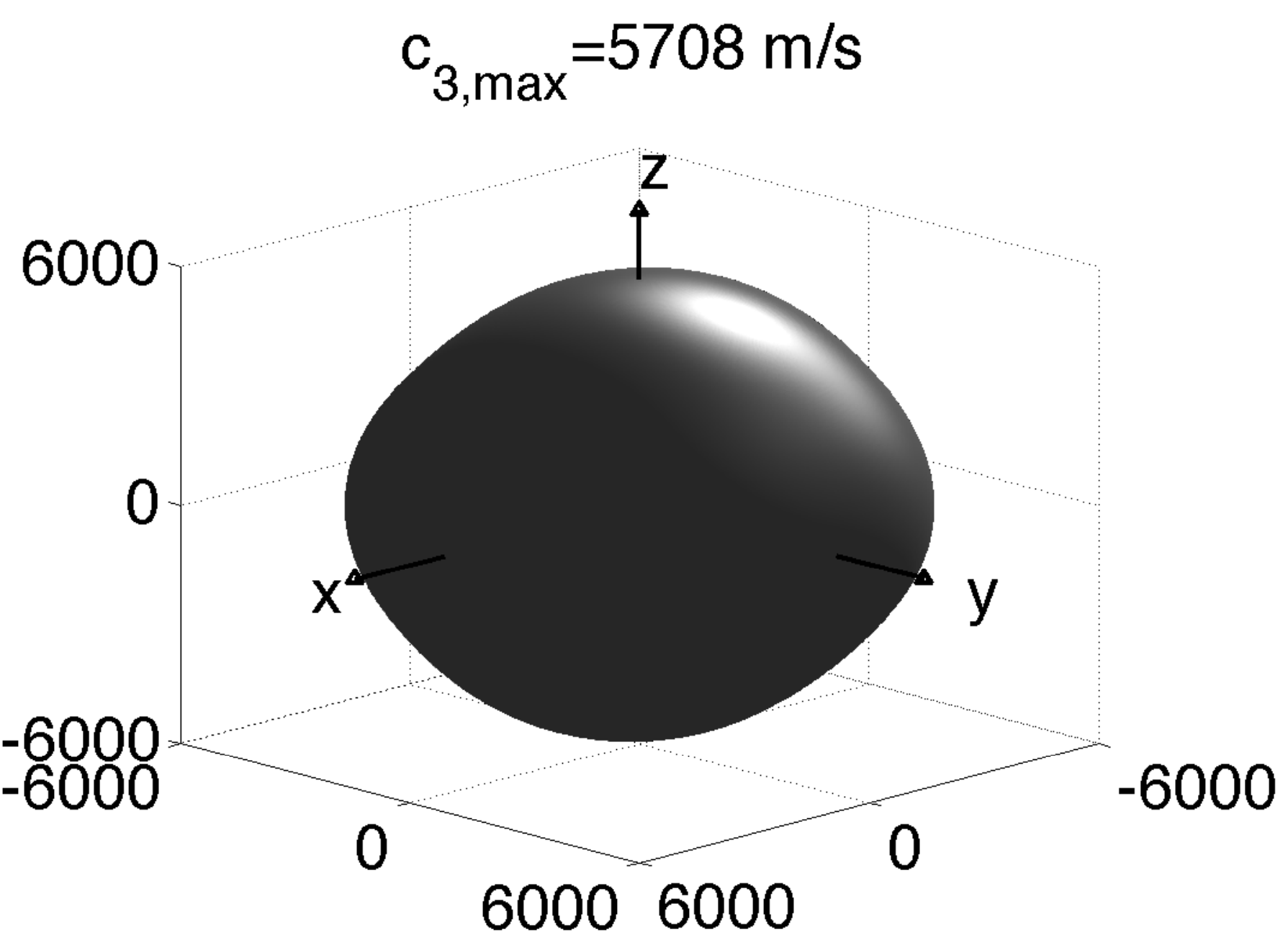}
\caption{Velocity surfaces $\smash{\hkg\mapsto\cel_\jeig(\hkg)}$ for \textcolor{\mycolor}{single crystal} celestite. Left: pseudo-transverse mode $\jeig=1$, middle: pseudo-transverse mode $\jeig=2$, right: pseudo-longitudinal mode $\jeig=3$. The dashed lines display the $10$ acoustic axes.}\label{fg:celestite-c}
\end{figure}

\begin{figure}
\centering\includegraphics[scale=0.6]{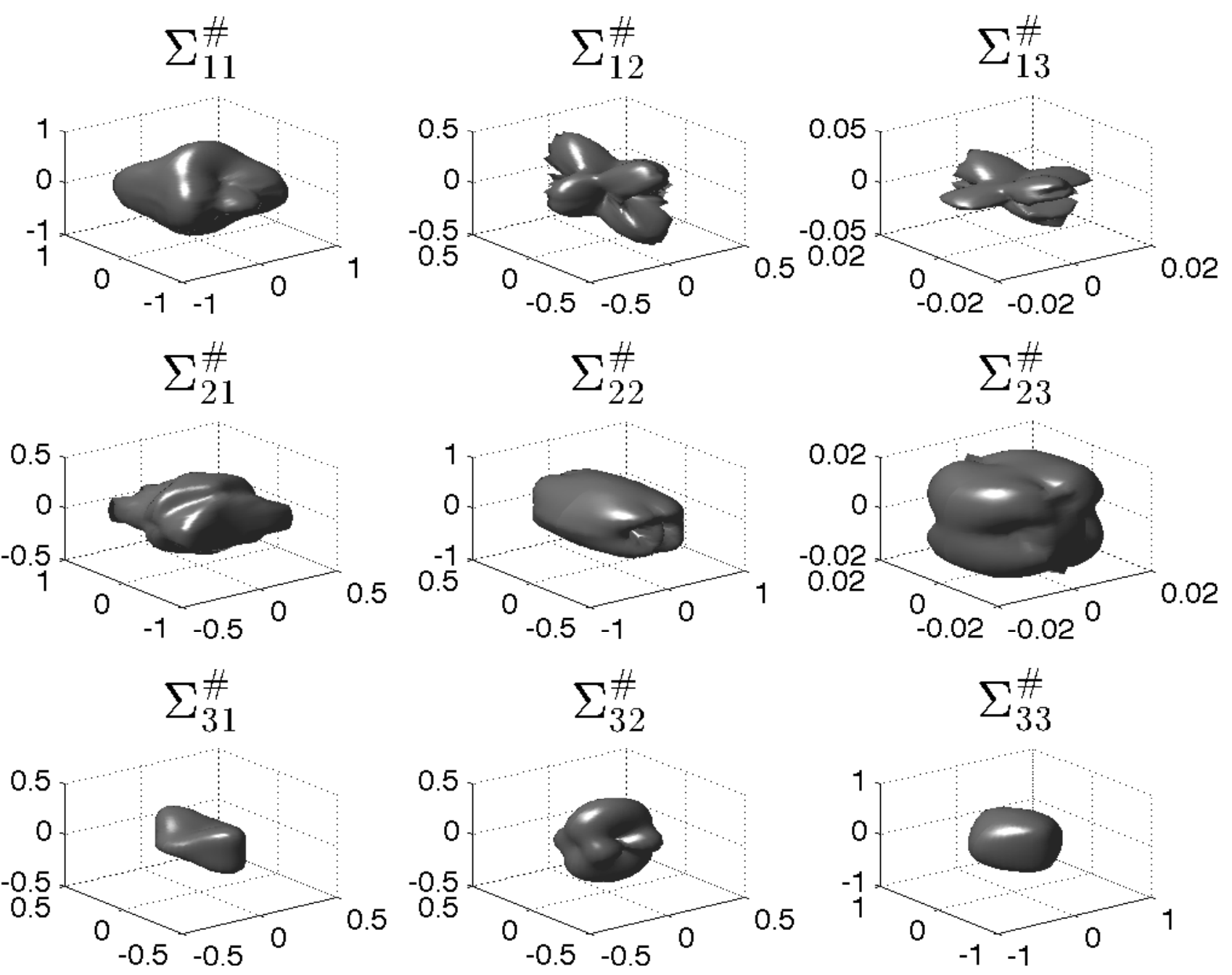}
\caption{Normalized total scattering cross-sections $\smash{\hkg\mapsto\tscati_{\indi\indj}^\#(\hkg)}$ for \textcolor{\mycolor}{single crystal} celestite with fixed frequency parameter $\lcor|\kg|=1$ and Markov model for the NCF.}\label{fg:SrSO4-Sigma}
\end{figure}

We apply our results to celestite (SrSO$_4$), which according to the Landolt-B\"{o}rnstein database~\cite{LAN13} as cited in~\cite[Table 1]{BOU98b}, has elasticity constants  $\ccub_1=104.0$, $\ccub_2=77.0$, $\ccub_3=60.0$, $\ccub_4=106.0$, $\ccub_5=62.0$, $\ccub_6=123.0$, $\ccub_7=13.9$, $\ccub_8=27.9$, $\ccub_9=26.6$ (in GPa), and density $\roi=3960$ kg/m$^3$. There are $10$ acoustic axes computed as $(0,0.77,\pm0.64)$, $(0,0.70,\pm0.71)$, $(0.49,\pm0.87,0)$, and $(\pm0.43,\pm0.88,\pm0.22)$ (with one change of sign at a time), in agreement with the data given in~\cite[Table 2]{BOU98b}. We first plot on~\fref{fg:celestite-c} the three velocity surfaces $\smash{\hkg\mapsto\cel_\jeig(\hkg)}$ for $\smash{\hkg\in\Sset^2}$ and $\jeig=1,2,3$, such that $\smash{\cel_1\leq\cel_2<\cel_3}$ where the equality $\cel_1=\cel_2$ holds on the acoustic axes solely. The latter are also displayed on the pseudo-transverse velocity surface plots.
We also plot on~\fref{fg:SrSO4-Sigma} the normalized partial total scattering cross-sections $\smash{\tscati_{\indi\indj}^\#}$ defined as in~\sref{sec:cubic}. Here again all modes have multiplicity one, so that $\Mode=3$ and the scattering cross-sections are scalars as for nickel or zinc. The non-dimensional frequency parameter is $\lcor|\kg|=1$, and the correlation coefficients $\smash{\coroij_{\inda\indb}}$ are all equal for $1\leq\inda,\indb\leq 9$ (with the same reservation for this assumption as for the cases of nickel and zinc).

\section{Conclusions}\label{sec:CL}

In this paper, the radiative transfer equations describing the propagation of high-frequency elastic (vector) waves in arbitrarily anisotropic, random media have been derived. These results generalize the models elaborated in \cite{RYZ96} for isotropic media and in \cite{BAL05} for scalar waves. In this respect they achieve the main extension identified in this latter publication for the proposed theory based on a second-order formulation of the elastic wave equation and the use of a spatio-temporal Wigner transform. It is believed that this generalization has interesting applications in the passive imaging techniques which have been developed recently in the geophysical literature, in the non destructive evaluation of heterogeneous polycrystalline materials, or in the understanding of mesoscopic phenomena such as the enhanced coherent back-scattering effect or the refocusing properties of time-reversed waves in random media. An immediate perspective of the present work consists in deriving the diffusion limit of the radiative transfer equations applicable to anisotropic media~\cite{MAR06}. Another direction is to consider the influence of the correlation structures of the random inhomogeneities on the shape of the scattering cross-sections in view of possibly develop composite materials with particular directional properties. Such correlation features may also be enriched by the random matrix models studied in a different context~\cite{TA10,GUI13}. These extensions are the subject of ongoing investigations. 

\section*{Acknowledgement}

The authors wish to thank the CNRS Federation "Francilienne de M\'ecanique, Mat\'eriaux, Structures et Proc\'ed\'es" (F2M CNRS-FR2609)
for financial support in this study.


\bibliographystyle{plain}
\bibliography{article-postdoc}







\end{document}